# Bayesian nonparametric spectral analysis of locally stationary processes

Yifu Tang*, Claudia Kirch† Jeong Eun Lee‡ and Renate Meyer§

**Abstract.** Based on a novel dynamic Whittle likelihood approximation for locally stationary processes, a Bayesian nonparametric approach to estimating the time-varying spectral density is proposed. This dynamic frequency-domain based likelihood approximation is able to depict the time-frequency evolution of the process by utilizing the moving periodogram previously introduced in the bootstrap literature. The posterior distribution is obtained by updating a bivariate extension of the Bernstein-Dirichlet process prior with the dynamic Whittle likelihood. Asymptotic properties such as sup-norm posterior consistency and $L_2$-norm posterior contraction rates are presented. Additionally, this methodology enables model selection between stationarity and non-stationarity based on the Bayes factor. The finite-sample performance of the method is investigated in simulation studies and applications to real-life data-sets are presented.

## 1 Introduction

Statistical inference based on a finite number of real-valued observations $X_t, t = 1, \dots, T$, that have been sequentially observed over time, i.e., a finite stretch of a time series $X_t, t \in \mathbb{Z}$, requires some assumptions about the extent to which this finite section is representative of the whole time series. One of the most often made assumption is that the underlying time series is stationary. For a stationary time series, the second order dependence structure is uniquely characterized by its (time-invariant) spectral density, the Fourier transform of the autocovariance function. This dependence structure is often analyzed in the frequency domain using the Whittle likelihood approximation (Whittle, 1957) as it simplifies the likelihood evaluation because the Fourier transformation normalizes the Fourier coefficients and effectively diagonalizes their covariance matrix. Thereby, spectral density estimation via the Whittle likelihood avoids numerical inversion of a high-dimensional covariance matrix and in contrast to the exact likelihood, depends directly on the spectral density instead of indirectly via an inverse Fourier transform. It has been used extensively in a wide range of applications as for instance those given in Rao and Yang (2020). Asymptotic independence and normality hold in a certain sense for a large class of non-Gaussian stationary time series (Hannan, 1973; Shao and Wu, 2007). Methods for spectral density estimation based on the Whittle likelihood with proven asymptotic properties have been developed both within the frequentist as well as the Bayesian framework. Some well-known studies include Franke and Härdle (1992); Giraitis and Robinson (2001); Kreiss et al. (2003); Gangopadhyay et al. (1999) among others.

*Department of Statistics, The University of Auckland, Auckland, New Zealand, ytan580@aucklanduni.ac.nz

†Institute for Mathematical Stochastics, Department of Mathematics, Otto-von-Guericke University, Magdeburg, Germany, claudia.kirch@ovgu.de

‡Department of Statistics, The University of Auckland, Auckland, New Zealand, kate.lee@auckland.ac.nz

§Department of Statistics, The University of Auckland, Auckland, New Zealand, renate.meyer@auckland.ac.nz

However, in many practical applications, the assumption of stationarity may not hold. Instead of being strictly stationary, many time series exhibit a slowly time-varying dependence structure, for instance, the variability of the electrical activity of the heart or the brain in biomedical studies (Fiecas and Ombao, 2016), hydrological and environmental time series (Ramachandra Rao et al., 2003), exchange rates and stock indices in financial and economic studies (Holan et al., 2012) and interferometer noise in gravitational wave measurements (Digman and Cornish, 2022). Their stochastic behavior changes slowly with time where often locally, i.e., in a small neighborhood of each time point $t = 1, \ldots, T$ of the time series, the behavior is very close to stationary. Hence, locally, the stochastic (2nd order) behavior of the time series can well be described by the spectral density of the stationary approximation at each rescaled time point $u \approx t/T$ making the time-varying spectral density $f(u, \lambda)$ a (smooth) function of both (rescaled) time $0 \leqslant u \leqslant 1$ and (rescaled) frequency $0 \leqslant \lambda \leqslant 1$. Thus, time-varying spectral density estimation essentially comes down to a two-dimensional surface estimation problem under suitable smoothness assumptions.

A rigorous mathematical treatment of asymptotic concepts such as consistency and asymptotic normality for locally stationary time series has been made possible with the introduction of infill asymptotics based on a rescaling of time to the unit interval by Dahlhaus (1997). In particular, this formulation covers the class of autoregressive processes with time-varying parameters, see Dahlhaus (2012) for an extensive survey and Dahlhaus et al. (2017) for an extension to non-linear processes. There is an alternative framework for local stationarity called locally stationary wavelet processes (Nason and von Sachs, 1999; Nason et al., 2000) which is not considered of this paper.

Parametric approaches for the class of locally stationary time series have been developed both in the frequentist and Bayesian framework (e.g., Roueff and Sánchez-Pérez (2018); Yang et al. (2016)). While parametric approaches are efficient when correctly specified, they make strong assumptions about the data-generating process and can give biased results when these are not satisfied. In the interest of developing more flexible nonparametric approaches for locally stationary time series, many authors partition the time series into approximately stationary segments. This is motivated by the fact that a locally stationary process can be approximated by a stationary process within a neighborhood of a fixed (rescaled) time point. Frequentist methods using piecewise stationary processes are based on the usual spectral density estimation techniques for stationary time series such as smoothing the periodogram or fitting piecewise autoregressive processes (Adak, 1998; Bruce et al., 2020). Bayesian nonparametric approaches for piecewise stationary time series such as Rosen et al. (2009, 2012) use reversible jump type algorithms for a data-driven determination of the number of stationary segments, a mixture of smoothing splines to model the log spectrum of each segment and time varying mixing weights to allow for non-stationarity across segments. However, piecewise stationary processes are only approximations of locally stationary processes and fail to accurately represent the smooth time-varying nature of locally stationary processes as will be demonstrated in Section 4.2. Another approach is based on a joint modeling of the temporal and frequency components of the time-varying spectral density in the time-frequency domain. Examples include data-adaptive kernel smoothing (van Delft and Eichler, 2019), the smoothing spline ANOVA (Guo et al., 2003; Qin and Wang, 2008) in the frequentist framework and online Bayesian inference under the state-space formulation (Wolfe et al., 2004; Everitt et al., 2013).

In general, any Bayesian analysis for the time-varying spectral density requires the specification of a suitable (pseudo-) likelihood for the time series in addition to a prior on the space of bivariate functions. In practice, the estimation of the time-varying spectrum is often based on local periodograms which are computed in the same way as the usual periodograms but depend only on a short window of the data. Local periodograms of overlapping windows result in periodogram ordinates that are highly dependent which makes the definition of a corresponding (pseudo-)likelihood a formidable task. Using the local periodograms of non-overlapping segments results in yet another piecewise stationary approximation which we aim to avoid for the above reasons. Therefore, we propose a *dynamic Whittle likelihood* approximation based on *moving* versions of the local periodograms that avoid dependencies across windows by a judicious selection of the periodogram ordinates. These have been shown to have favorable asymptotic properties by Häfner and Kirch (2017).

For correctly specified parametric models, the Bernstein-von Mises theorem guarantees that frequentist and Bayesian inference procedures are asymptotically equal to the first order of approximation as long as the true parameter value is in the support of the prior. For infinite-dimensional parameter spaces, however, this does not hold in this generality and for any new Bayesian nonparametric technique it is necessary to prove that the posterior distribution in fact contracts around the true parameter (Kleijn and van der Vaart, 2006). None of the previous Bayesian nonparametric approaches for locally stationary time series demonstrate such elemental and indispensable theoretical properties as posterior consistency and contraction rates. In this respect, our approach is unique in considering and providing theoretical guarantees of asymptotic properties. Furthermore, we demonstrate very competitive finite sample performance when compared to both frequentist and Bayesian state-of-the-art alternative methods for locally stationary time series. For reproducibility of results and ease of uptake, our implementation of the proposed approach is publicly available as part of the R package `beyondWhittle` (Meier et al., 2017).

The rest of the article is organized as follows. Moving periodogram and the dynamic Whittle likelihoods are introduced in Section 2. In Section 3, we introduce the bivariate Bernstein-Dirichlet process prior as well as the resulting asymptotic results such as sup-norm posterior consistency and $L_2$-norm posterior contraction rate. Important findings of a larger simulation study in addition to a case study based on a time series of weekly egg prices are presented in Section 4. A discussion is given in Section 5. Additional material is provided in the supplementary material including additional information on computational issues and illustrative R code, additional simulation results and data analyses as well as the mathematical proofs.

## 2 Dynamic Whittle likelihood

A major motivation for spectral analysis of stationary time series is the fact that the Fourier transformation has a decorrelating (and normalizing) effect on the data. Indeed, in some appropriate asymptotic sense the periodogram ordinates are independent and exponentially distributed with expectation proportional to the spectral density for a wide range of stationary time series (see e.g. Shao and Wu (2007)). This observation also motivated the famous likelihood approximation proposed by Whittle (1957) which has become very popular both in frequentist and in Bayesian statistics, and for parametric as well as non-parametric inference (see e.g. Hannan 1973; Poetscher 1987; Pawitan and O'Sullivan 1994; Choudhuri et al. 2004; Kirch et al. 2019). Even for parametric inference for stationary Gaussian time series the Whittle approximation has computational advantages over the true likelihood because it does not require matrix inversion. But for parametric models, Contreras-Cristán et al. (2006) have shown that the goodness of approximation and efficiency depends on the amount of autocorrelation as well as deviations from Gaussianity for small/medium sample sizes.

Locally stationary time series can be thought of as behaving approximately stationary in small neighbourhoods with slowly changing spectral density $f(u, \lambda)$ at rescaled time point $u$ and rescaled frequency $\lambda$. To prove posterior consistency and contraction rates of this time-varying spectral density we will adopt the theoretical framework of locally stationary processes according to Dahlhaus (1997) where roughly $X_t \approx \sum_{j=-\infty}^{\infty} a(u, j)\epsilon_{t-j}$ for standardized i.i.d. innovations $\{\epsilon_t\}$ having fourth moments and sufficiently smooth filter functions $a(\cdot, j) : (0, 1] \to \mathbb{R}$, $j \in \mathbb{R}$. For a mathematically rigorous definition we refer to Section A of the supplementary material. In this framework the time-varying (power) spectral density (tv-PSD) is given by

$$f(u, \lambda) = \frac{1}{2\pi} \left| \sum_{j \in \mathbb{Z}} a(u, j) \exp(-i\pi\lambda j) \right|^2, \quad (u, \lambda) \in [0, 1]^2. \tag{2.1}$$

In contrast to the stationary situation, this requires the estimation of a bivariate function ideally by means of established periodogram-based methods. While global periodograms are meaningless in the context of locally-stationary time series, the decorrelating property still holds if a periodogram is calculated locally from a short stretch of a much longer locally stationary time series (Sergidis and Paparoditis, 2008, Lemma 4). However, the dependence structure between such local periodograms obtained from overlapping stretches of data is very complicated, rendering the construction of a corresponding likelihood difficult, if not impossible. Instead of using non-overlapping data stretches corresponding to a piecewise stationary approximation, the proposed *dynamic Whittle likelihood* approximation is based on moving periodograms which were introduced by Häfner and Kirch (2017) in the context of bootstrapping. These moving periodograms also possess the decorrelating (and normalizing) property of the usual periodograms (in a suitable sense) which can be utilized to obtain a Whittle approximation for locally stationary time series.

To take the time varying structure into account, the construction of the moving periodograms is based on local windows of width $2m+1$ . Using an odd window length avoids the need for a special treatment of the periodogram at frequency $\pi$ (or 1 after rescaling). Also note that this will not yield additional computational complexities since fast algorithms such as the fast Fourier transform can only be used in the context of global but not moving periodograms (Häfner, 2014, Chapter 10). Then the window is shifted to the right to calculate the next (shifted) local periodogram ordinate at the next Fourier frequency (where after reaching Fourier frequency $\lambda_m$ we move back to frequency $\lambda_1$ in a circular fashion, further details can be found in the following Definition 1).

For ease of notation assume that we have observed $X_t = X_{t,T}$ at times $t = -m + 1, \cdots, T+m$.

**Definition 1** (Moving periodogram ordinates)**.** The *moving periodogram ordinates* $\mathrm{MI}_t$ (of order $m$) at time point $t = 1, \cdots, T$ of $X_{-m+1}, \cdots, X_{T+m}$ are defined by

$$\mathrm{MI}_t = \frac{1}{2\pi(2m+1)} \left| \sum_{\nu=0}^{2m} X_{\nu+t-m} \exp(-i\pi\nu\lambda_{\mathrm{mod}(t)}) \right|^2, \tag{2.2}$$

where $\lambda_j = \frac{2j}{2m+1}$, $j = 1, 2, \cdots, m$ are the Fourier frequencies and $\mathrm{mod}(t) = 1 + ((t-1) \mod m)$.

In particular, the moving periodogram at time point $t$ is given by the local periodogram centered at that time point taken at the Fourier frequency $\lambda_{\mathrm{mod}(t)}$.

The key observation to define a meaningful likelihood approximation is that the moving periodogram ordinates are also asymptotically independent and exponentially distributed in some sense. The independence holds with the exception of a small number of periodogram ordinates but this exception does not influence the asymptotic behavior. For more details, we refer to Theorem 2.2 in Häfner and Kirch (2017); see also Proposition P.1 in Section E of the supplementary material.

Based on this observation we define the following Whittle approximation for locally stationary time series which assumes that all moving periodograms are independent and exponentially distributed with expectation equal to the corresponding time-varying spectral density. We name this likelihood approximation the *dynamic Whittle likelihood* as it captures the time-varying structure of the process dynamically by using the moving periodograms.

**Definition 2** (Dynamic Whittle likelihood)**.** The *dynamic Whittle likelihood* based on the moving periodogram ordinates is defined by

$$L_{DW}^{(1)}(f) = \prod_{t=1}^{T} \frac{1}{f(t/T, \lambda_{\mathrm{mod}(t)})} \exp\left( -\frac{\mathrm{MI}_t}{f(t/T, \lambda_{\mathrm{mod}(t)})} \right). \tag{2.3}$$

The number of periodogram ordinates for Whittle-type likelihoods and periodogram-based estimators in the literature (Choudhuri et al., 2004; Sergidis and Paparoditis, 2008; Dette et al., 2011) is about half of the number of observations whereas for the dynamic Whittle likelihood $L_{DW}^{(1)}$ it is close to the sample size $T$. This feature may result in a heavy computational burden for large time series as each evaluation of $L_{DW}^{(1)}(f)$ requires computing the value of $f$ at $T$ points. To mitigate this problem, we introduce the following thinned versions of $L_{DW}^{(1)}$ where an evaluation of the thinned likelihood with thinning factor $i$ only requires the calculation of $f$ at $T/i$ points.

**Definition 3** (Thinned dynamic Whittle likelihood)**.** The *thinned dynamic Whittle likelihood* with thinning factor $i = 2, 3$ is given by

$$L_{DW}^{(i)}(f) = \prod_{l=1}^{B_i} \prod_{j=1}^{m} \frac{1}{f(u_{j,l,i}, \lambda_j)} \exp\left( -\frac{\mathrm{MI}_{(u_{j,l,i}T)}}{f(u_{j,l,i}, \lambda_j)} \right), \tag{2.4}$$

where $B_i = \lceil \frac{T-m}{im} \rceil$ and $u_{j,l,i} = \frac{i(l-1)m+j}{T}$.

For $i = 1$ the definition in (2.4) yields the original dynamic Whittle likelihood as defined in (2.3) if $T/m \in \mathbb{N}$. Otherwise, the original dynamic Whittle likelihood contains $(T \mod m)$ additional factors, which can be taken into account in (2.4) by adding factors with $l = \lceil T/m \rceil$ and $j$ running only from 1 to $(T \mod m)$. A similar adaptation can also be made for the thinned versions if $\lceil T/(im) \rceil$ differs from $B_i$. Such adaptations were made in the simulation study to ensure that the likelihood contains terms for all observations. A thinning factor larger than 3 is not recommended in general as as it would periodically omit stretches of data and not make use of all available data points. The finite-sample performances of $L_{DW}^{(i)}$ for different thinning factors are compared in Section 4.2.

The parameter $m$ in the dynamic Whittle approximation corresponds to the width $2m+1$ of the window used to calculate the moving periodograms. The larger $m$ is the finer the resolution of the spectrum (at each time point) but at the cost of having a coarser grid on the time scale. As such, the choice of $m$ is effectively a trade-off between precision in time and precision in frequency. The $L_2$-norm contraction rate obtained in Theorem 3.2 below becomes minimal for a choice of $m \propto T^{1/3}$ but the result may not be optimal and does not provide any information about the proportionality constant. Thus, we still need a heuristic choice of $m$ which should be based on the expected smoothness of the spectral time variation.

To obtain asymptotic results concerning the posterior in the context of locally stationary processes, we need the following relation between $m$ and the sample size $T$.

**Assumption $\mathcal{A}$.1.** *It holds that $m \to \infty$ (for $T \to \infty$), but $\frac{m^3}{T^2} = O(1)$.*

In particular, this assumption is satisfied by $m \propto T^{1/3}$ which minimizes the $L_2$-norm contraction rate in Theorem 3.2. Furthermore, we need that the observed time series is locally stationary in the sense of Definition S.1 as given in Section A of the supplementary material in addition to the following assumptions on the smoothness of the underlying time-varying spectral density $f_0$ with filter function $a_0$:

**Assumption $\mathcal{A}$.2.** *(a)* $\sup_{u\in[0,1]} \sum_{h\in\mathbb{Z}} |h\, a_0(u,h)| < \infty$.

*(b)* $\delta_0 := \inf_{(u,\lambda)\in[0,1]^2} f_0(u,\lambda) > 0$, *where* $f_0$ *is the underlying time-varying spectral density.*

Assumption $\mathcal{A}$.2 (a) implies Assumption A.1 of Häfner and Kirch (2017) as well as their additional assumption that $\sup_u \sum_{h\in\mathbb{Z}} |a_0(u,h)| \sqrt{|h|} < \infty$. Moreover, part (a) of Assumption $\mathcal{A}$.2 in combination with Definition S.1 (c) (iii) guarantees that the time-varying spectral density is continuously differentiable (i.e., $f_0 \in C^1([0,1]^2)$). Versions of both Assumptions $\mathcal{A}$.2 (a) and (b) are common in the context of spectral analysis of (locally) stationary time series (see, e.g., Shao and Wu 2007; Liu and Wu 2010; Kreiss and Paparoditis 2015; Yang and Zhou 2022).

# 3 Bayesian inference for the tv-PSD: the BDP-DW method

Apart from the likelihood, Bayesian inference for the time-varying spectral density requires the specification of a prior distribution on a suitable parameter space. To avoid potential biases when using parametric models in situations where the parametric model assumptions do not hold or cannot be verified and to make the analysis flexible and robust, we define a nonparametric prior on the space of time-varying spectral density functions. The definition based on the bivariate Bernstein-Dirichlet process prior is given in Section 3.1 before defining the posterior distribution obtained by updating the bivariate Bernstein-Dirichlet process prior with the dynamic Whittle likelihood in Section 3.2. The Bayesian procedure involving the Bernstein-Dirichlet process prior and the dynamic Whittle likelihood is referred to as the *BDP-DW* method. Sup-norm posterior consistency and $L_2$-norm contraction rates of the BDP-DW posterior are stated in Section 3.2 with corresponding proofs in Section E of the supplementary material. Section 3.3 discusses the computation of the Bayes factor for making a choice between stationarity and non-stationarity within the framework of local stationarity.

## 3.1 Bivariate Bernstein-Dirichlet process prior

By Assumption $\mathcal{A}$.2 and Definition S.1, the true time-varying spectrum $f_0$ is bounded away from 0 and continuously differentiable. This motivates the following definition of the parameter space

$$\Theta = \left\{ f \in C^1\left([0,1]^2\right) : f \geqslant \delta, \|f\|_\infty \leqslant M_0, \|\partial_1 f\|_\infty \leqslant M_1, \|\partial_2 f\|_\infty \leqslant M_2 \right\}, \quad (3.1)$$

where $\partial_i f$ denotes the partial derivative of $f$ with respect to the $i$-th variable. Here, we require that $\delta$ is chosen small enough and $M_j$, $j = 0,1,2$, large enough so that $f_0 \in \Theta$ and all four inequalities are strict. Then, we equip $\Theta$ with the uniform metric $d_\infty$ on $C\left([0,1]^2\right)$ such that $(\Theta, d_\infty)$ is a metric space. The $\sigma$-field equipped on $\Theta$ is the Borel-$\sigma$-field of $C\left([0,1]^2\right)$ restricted to $\Theta$.

The (univariate) Bernstein-Dirichlet process prior, originally introduced by Petrone (1999b) as a nonparametric prior on the space of distribution functions on $[0, 1]$, has been successfully applied to the estimation of probability density functions (Petrone, 1999a; Kruijer, 2008; Zhao et al., 2013; Barrientos et al., 2015), the spectral density functions of stationary time series (Choudhuri et al., 2004; Kirch et al., 2019) and random fields (Zheng et al., 2010), to name but a few. Here we generalize the univariate Bernstein-Dirichlet process prior to the bivariate case. To elaborate, we put a prior on $\Theta$ using a mixture of bivariate Bernstein polynomials defined as

$$b\left(u, \lambda; k_1, k_2, G\right) = \sum_{j_1=1}^{k_1} \sum_{j_2=1}^{k_2} w_{k_1,k_2}(G)(j_1, j_2)\beta\left(u; j_1, k_1 - j_1 + 1\right)\beta\left(\lambda; j_2, k_2 - j_2 + 1\right), \tag{3.2}$$

where $G$ is a probability measure on $[0, 1]^2$, $w_{k_1,k_2}(G)(j_1, j_2) = G\left(\left(\frac{j_1-1}{k_1}, \frac{j_1}{k_1}\right] \times \left(\frac{j_2-1}{k_2}, \frac{j_2}{k_2}\right]\right)$, and $\beta(x; a, b) = \frac{\Gamma(a+b)}{\Gamma(a)\Gamma(b)} x^{a-1}(1-x)^{b-1}$ is the beta density with parameters $a$ and $b$. The first index $j_1$ corresponds to time while the second one corresponds to frequency. Therefore, (3.2) with $k_1 = 1$ models a spectral density that is constant over time, i.e., the spectral density of a stationary process. In Section 3.3 we suggest using the posterior probability function of $k_1$ to make a choice between stationarity and non-stationarity.

The bivariate Bernstein-Dirichlet process prior is specified by the following hierarchical model:

**Assumption $\mathcal{A}$.3** (The bivariate Bernstein-Dirichlet process prior)**.**

*(a)* $f(u, \lambda) = \tau b\left(u, \lambda; k_1, k_2, G\right)$ *with the function* $b$ *defined in* (3.2).

*(b)* $G$ *follows a Dirichlet process, i.e., for any partition* $\{A_1, A_2, \cdots, A_n\}$ *of* $[0, 1]^2$, $\{G(A_j)\}_{j=1}^n$ *jointly follow a Dirichlet distribution*

$$\left(G(A_1), \cdots, G(A_n)\right) \sim Dir\{MG_0(A_1), \cdots, MG_0(A_n)\},$$

*where* $M$ *is a weight parameter and* $G_0$ *is a base measure on* $[0, 1]^2$ *(Müller and Quintana, 2004). It is assumed that* $G_0$ *has a continuous probability density* $g_0$ *with full support.*

*(c)* $k_i$ *has probability mass function,* $\rho_i(k_i) > 0$, *for* $k_i = 1, 2, \cdots$, $i = 1, 2$. *Moreover, we assume* $-\ln \rho_i(k_i) = O\left(k_i^2 \ln k_i\right)$ *when deriving the posterior contraction rate.*

*(d) The distribution of* $\tau$ *has continuous Lebesgue density* $\pi_\tau$ *on* $(0, \infty)$ *having full support.*

*(e)* $G$, $k_1$, $k_2$ *and* $\tau$ *are a priori independent.*

This hierarchical prior model defines a probability measure on an infinite-dimensional function space which is convenient for the formulation and theoretical proofs. When it comes to practical computation, however, one may need to seek a finite-dimensional approximation. We will elaborate on such an approximation in Section B.2 of the supplementary material.

Let $\Pi_{BD}$ denote the law induced by the above prior model. The following theorem states the prior positivity of neighbourhoods of the true spectral density. The proof of the theorem can be found in Section E.3 of the supplementary material.

**Theorem 3.1.** *For $r > 0$ let $U_\infty(r) = \{f \in \Theta : \|f - f_0\|_\infty < r\}$ denote a sup-norm neighborhood of the true underlying time-varying spectral density $f_0$. Under Assumption $\mathcal{A}.3$, we have $\Pi_{BD}(U_\infty(r)) > 0$ for any $r > 0$. Moreover, for sufficiently small $r$, it holds that*

$$-\ln \Pi_{BD}(U_\infty(r)) = O\left(-\frac{\ln r}{r^4}\right).$$

By the above theorem $\Pi_{BD}(\Theta) > 0$ such that we can now define the prior (to be used in the proofs) as

$$\Pi_0(A) = \frac{\Pi_{BD}(A)}{\Pi_{BD}(\Theta)}$$

for any $A \in \mathcal{B}\left(C\left([0,1]^2\right)\right) \cap \Theta$.

## 3.2 Posterior consistency and contraction rate

With the bivariate Bernstein-Dirichlet process prior and the dynamic Whittle likelihood at hand, we can now introduce the posterior:

$$\Pi_T^{(i)}(A) = \frac{\int_A L_{DW}^{(i)}(f)\Pi_0(\mathrm{d}f)}{\int_\Theta L_{DW}^{(i)}(f)\Pi_0(\mathrm{d}f)}$$

for any $A \in \mathcal{B}\left(C\left([0,1]^2\right)\right) \cap \Theta$ and a given thinning factor $i \in \{1,2,3\}$. The following theorem shows that this posterior concentrates around the underlying time-varying spectrum $f_0$ and gives the corresponding contraction rate. The proof is contained in Section E.4 of the supplementary material.

**Theorem 3.2** (sup-norm posterior consistency and $L_2$-norm contraction rate)**.** *Under Assumptions $\mathcal{A}.1$, $\mathcal{A}.2$ and $\mathcal{A}.3$, for $i = 1,2,3$:*

*(a) For any $r > 0$, we have $\Pi_T^{(i)}(U_\infty^c(r)) \to 0$ in $\mathrm{P}$-probability when $T \to \infty$.*

*(b) Moreover, for any positive divergent sequence $M_T \to \infty$, we have $\Pi_T^{(i)}\left(U_2^c\left(M_T \max\left\{m^{-\frac{1}{4}}, (m/T)^{\frac{1}{8}}\right\}\right)\right) \to 0$ in $\mathrm{P}$-probability when $T \to \infty$, where*

$$U_2(r) = \left\{f \in \Theta : \|f - f_0\|_2 \coloneqq \sqrt{\int_0^1 \int_0^1 |f(x,y) - f_0(x,y)|^2 \,\mathrm{d}x\,\mathrm{d}y} < r\right\}. \quad (3.3)$$

*Remark.* The posterior consistency is with respect to the uniform metric $d_\infty$ which is stronger than the metric induced by $L_2$-norm. The above $L_2$-norm contraction rate becomes minimal for a choice of $m \propto T^{1/3}$ resulting in a contraction rate of $T^{-1/12}$.

### 3.3 Model selection: Stationarity versus local stationarity

It is known that a locally stationary time series, which possesses a continuous tv-PSD $f(u,\lambda)$, is stationary if and only if $f(u,\lambda)$ is independent of $u$ (Dette et al., 2011). This feature, in conjunction with the BDP-DW procedure, suggests a novel way to make a choice between stationarity and non-stationarity within the framework of local stationarity. For a mixture of Bernstein polynomials of the form (3.2) where the first variable is the rescaled time, the polynomial does not vary with rescaled time if $k_1 = 1$. To see whether stationarity is plausible, one may calculate the Bayes factor of the stationary model $\{k_1 = 1\}$ versus the general locally stationary model $\{k_1 \in \mathbb{N}\}$. If the time series is indeed stationary, then the Bayes factor will be larger than 1. Due to the discrete nature of $k_1$ and the fact that the stationary model is nested in the general locally stationary model, we employ the celebrated Savage-Dickey estimator of the Bayes factor (Dickey, 1971; Verdinelli and Wasserman, 1995) in our study. To be specific, the Savage-Dickey estimator for stationary model $\{k_1 = 1\}$ is the ratio of its posterior probability to its prior probability. i.e., $\hat{B}_{01} = \frac{\hat{\Pi}_T^{(i)}(\{k_1=1\})}{\Pi_0(\{k_1=1\})}$. It is worth mentioning that an upperbound of $\hat{B}_{01}$ is $1/\Pi_0\left(\{k_1 = 1\}\right)$. A simulation study to illustrate the properties of this model selection procedures is conducted in Section 4.3.

## 4 Simulation and case study

In this section we illustrate the empirical performance of the BDP-DW method by means of a simulation study and a case study of real data on weekly egg prices. Results of additional simulation studies and data examples (daily particulate matter (PM) 2.5 and gravitational wave data) are presented in Section D of the supplementary material.

In practice, the moving periodograms are not available for the first and last $m$ observations so that the time-varying spectral density estimates are provided on the time interval $[m/T, 1-m/T]$ (without introducing any additional biases). On the remaining time end intervals, we use a constant extension based on the estimator at time $m/T$ and $1-m/T$, respectively. While this does not affect the asymptotic behavior, this results in worse performance in practice, in particular for faster-varying spectral densities, due to boundary effects.

Details of the prior specifications for simulation and case studies are given in Section B.1 of the supplementary material. All simulation results in Sections 4.2 and 4.3 are based on 1000 repetitions of the respective data-generating process (DGP). The MCMC algorithm is run for $1.1 \times 10^5$ iterations with a burn-in period of $6 \times 10^4$ iterations and a thinning interval of 5.

### 4.1 Computational aspects

Since the posterior distribution, proportional to the product of the bivariate Bernstein-Dirichlet process prior and the dynamic Whittle likelihood, is analytically intractable, we use a simulation-based approach to posterior computation. This is based on a blocked

Metropolis-Hastings (MH) sampler, similar to the Gibbs sampling strategy described in Choudhuri et al. (2004), adopting the Sethuraman representation of the Dirichlet process $G$ in (3.2). Full details are given in Section B.2 of the supplementary material. The algorithm is implemented in the R package `beyondWhittle` and Section F provides illustrative R code for generating posterior samples of the time-varying spectral density for one of the simulation studies as well as for summarizing and plotting the results.

## 4.2 Simulation study

The simulations in this section are based on 1000 repetitions of the following locally stationary time series with $T = 1500$ and i.i.d. standard normally distributed innovations $\{w_t\}$:

LS1a. $X_{t,T} = w_t + 1.122\left(1 - 1.718\sin\left(\frac{\pi}{2}\frac{t}{T}\right)\right)w_{t-1} - 0.81w_{t-2}, \quad t = 1, 2, \cdots, T;$

LS2a. $X_{t,T} = w_t + 1.1\cos\left(1.5 - \cos\left(4\pi\frac{t}{T}\right)\right)w_{t-1}, \quad t = 1, 2, \cdots, T;$

LS3a. $X_{t,T} = \left(1.2\frac{t}{T} - 0.6\right)X_{t-1,T} + w_t, \quad t = 1, 2, \cdots, T.$

The first two DGPs are examples of time-varying moving average processes while the last one is an example of a time-varying autoregressive process. For illustrative purposes, realizations for each of these time series can be found in Section C.1 of the supplementary material. The first figure in Figure 1 shows the true time-varying spectral density for LS2a, while the corresponding plots for the other two examples can be found in the first row of Figure II in Section C.3 of the supplementary material. The first time series LS1 has a much slower varying spectral density than LS2, while LS3 is also slowly varying but has a peak at two of the corners which is particularly difficult to estimate due to boundary effects both in the time and frequency direction.

We compare the precision for different methods in terms of the median and interquartile range of the average square error (ASE) of the logarithm of a tv-PSD estimate calculated as follows:

$$ASE = \frac{1}{T(K+1)}\sum_{t=1}^{T}\sum_{j=0}^{K}\left(\ln\hat{f}\left(\frac{t}{T},\frac{j}{K}\right) - \ln f_0\left(\frac{t}{T},\frac{j}{K}\right)\right)^2, \tag{4.1}$$

where $\hat{f}$ is the given estimate and $K$ is set to 99 (i.e., finer scaled than $m$). When it comes to a Bayesian method, the posterior mean is used as an estimate of the tv-PSD.

All computations in Section 4.2 are performed on a virtual machine with 64GB RAM, 16 VCPUs, and an Ubuntu Linux operating system.

### Choice of tuning parameters

We have introduced thinned likelihoods to speed up the computations: The results from the simulations (presented in Section C.2 of the supplementary material) indeed indicate that the run-time decreases with increased thinning at the cost of precision. For example, the median ASE for LS1a with no thinning is at 0.10, it increases to 0.12 for thinning factor 2 and even to 0.14 for thinning factor 3. At the same time the median run-time decreases from 5.48 minutes to 4.37 and 4.06 minutes. Unless otherwise stated in the following we will use the thinning factor 2.

The window size $m$ is a trade-off between time and frequency resolution: Indeed for the slower time-varying LS1a spectrum, a choice of $m = 50$ (or even $m = 75$) with a higher resolution in frequency yields best results (median ASE of 0.12 (rounded) for both choices versus a median ASE of 0.14 for $m = 25$). For the faster time-varying LS2a spectrum a choice of $m = 25$ corresponding to a higher time resolution is best with a median ASE of 0.15 as compared to 0.21 and 0.38 for $m = 50$ and $m = 75$. For this particular example the error occuring due to boundary effects in the time direction is quite large. For example, the ASE for $m = 25$ improves from 0.15 to 0.13 if only calculated for $t = m, \ldots, T - m$.

### Comparison with alternative state-of-the-art methods

In this section, we will compare the BDP-DW methodology based on a window size of $m = 50$ and thinning factor $i = 2$ (i.e. not optimally tuned) with the following three competing methods from the literature: The nonparametric smoothing spline ANOVA approach (Smooth ANOVA) combined with the DGML criterion of Qin and Wang (2008) with pre-specified time-frequency grid $(32, 32)$ and the same specifications as in Section 6.5.3 of Wang (2011), the multitaper spectral estimate proposed by Bruce et al. (2020) with their suggested parameter values and the Adaptspec method (Rosen et al., 2012; Bertolacci et al., 2022) with 20000 iterations, a burn-in period of 12000 iterations, the number of basis functions set to 10, and the maximum number of segments set to 10.

The Smooth ANOVA approach and the BDP-DW method assume that the underlying time series is locally stationary and model the tv-PSD using a continuous function while the Adaptspec method and the multitaper estimator utilize a piece-wise stationary approximation to a locally stationary time series and treat the tv-PSD as a piecewise constant along the time axis. To give a better understanding of this effect, besides the locally stationary processes LS1–LS3 defined in Section 4.2, we also include the following piecewise stationary time series in this simulation study

PS1. $X_{t,T} = a_{t,T} X_{t-1,T} + w_t, \quad a_{t,T} = -0.5\, \mathrm{I}\left\{t \leqslant \frac{T}{2}\right\} + 0.5\, \mathrm{I}\left\{t > \frac{T}{2}\right\} \quad t = 1, 2, \cdots, T.$

For all four models we no longer only consider innovations $\{w_t\}$ with (a) i.i.d. standard normal distribution but also with (b) a standardized Student's t distribution with degrees of freedom 3 and with (c) standardized Pareto distribution with scale 1 and shape 4. This gives an impression how sensitive the procedures are with respect to deviations from Gaussianity in particular in the case of heavier tailed distributions (as (b) and (c) have second but not fourth moments) and skewness (as (c) is right-skewed).

For PS1 we use the correctly specified number and lengths of segments for the multitaper methodology.

Table 1: Median and IQR (in brackets) of average square error of the tv-PSD estimates obtained through the BDP-DW method, smoothing spline ANOVA (Smooth ANOVA), Adaptspec method and multitaper estimator. The smallest median ASE in each row is bold.

| DGP | BDP-DW | Smooth ANOVA | Adaptspec | Multitaper |
|---|---|---|---|---|
| LS1a | **0.1186**(0.0265) | 0.2381(0.0521) | 0.1762(0.0814) | 0.3837(0.0423) |
| LS1b | **0.1817**(0.0746) | 0.4473(0.1614) | 0.2791(0.1190) | 0.5687(0.1205) |
| LS1c | **0.2877**(0.1468) | 0.7336(0.2293) | 0.4962(0.2378) | 0.8144(0.1962) |
| LS2a | **0.2147**(0.0448) | 0.2257(0.0424) | 0.3104(0.0577) | 0.2816(0.0348) |
| LS2b | **0.2829**(0.0892) | 0.4618(0.1619) | 0.4076(0.1322) | 0.4928(0.1220) |
| LS2c | **0.3823**(0.1301) | 0.7390(0.2349) | 0.6950(0.1788) | 0.7491(0.1789) |
| LS3a | **0.0252**(0.0124) | 0.0270(0.0132) | 0.0603(0.0283) | 0.1652(0.0294) |
| LS3b | **0.0842**(0.0798) | 0.2586(0.1521) | 0.1636(0.1085) | 0.4039(0.1096) |
| LS3c | **0.2029**(0.1601) | 0.5595(0.2293) | 0.4819(0.2331) | 0.6720(0.1842) |
| PS1a | 0.0848(0.0166) | 0.0779(0.0288) | **0.0213**(0.0092) | 0.1397(0.0269) |
| PS1b | 0.1414(0.0732) | 0.3342(0.1449) | **0.1243**(0.1420) | 0.1809(0.0667) |
| PS1c | 0.2399(0.1285) | 0.5857(0.2297) | 0.3928(0.2599) | **0.1995**(0.0934) |

Table 1 contains the median (IQR) of the ASEs for all realizations. Clearly, the two methodologies based on locally stationary time series (BDP-DW and smooth ANOVA) outperform the two competitors based on piecewise stationary approximations for the three locally stationary time series while the latter have a better behavior for the piecewise stationary time series. Plots of all four estimators (based on the same randomly selected realization) for LS2a and PS1a are given in Figure 1. Corresponding plots for LS1a and LS3a can be found in Figure II in Section C.3 of the supplementary material. These plots clearly show the difference in estimation where the BDP-DW and Smooth ANOVA obtain smooth estimates while the other two methods obtain piecewise stationary estimates.

For truly locally stationary time series the dynamic Whittle likelihood clearly outperforms the other methods in terms of having the smallest median ASE and most of the time a significantly smaller IQR. For non-Gaussian time series the difference is even more pronounced, showing that the BDP-DW methodology is most robust with respect to deviations from Gaussianity. This is true despite the fact that we used a thinning factor of 2 (and not the better full dynamic Whittle likelihood) and a window size of $m = 50$ which is not optimal for LS2. The results in the table are complemented by plots in Section C.3 of the supplementary material showing the pointwise median for each grid point in addition to the pointwise IQR giving an indication of where precision is lost.

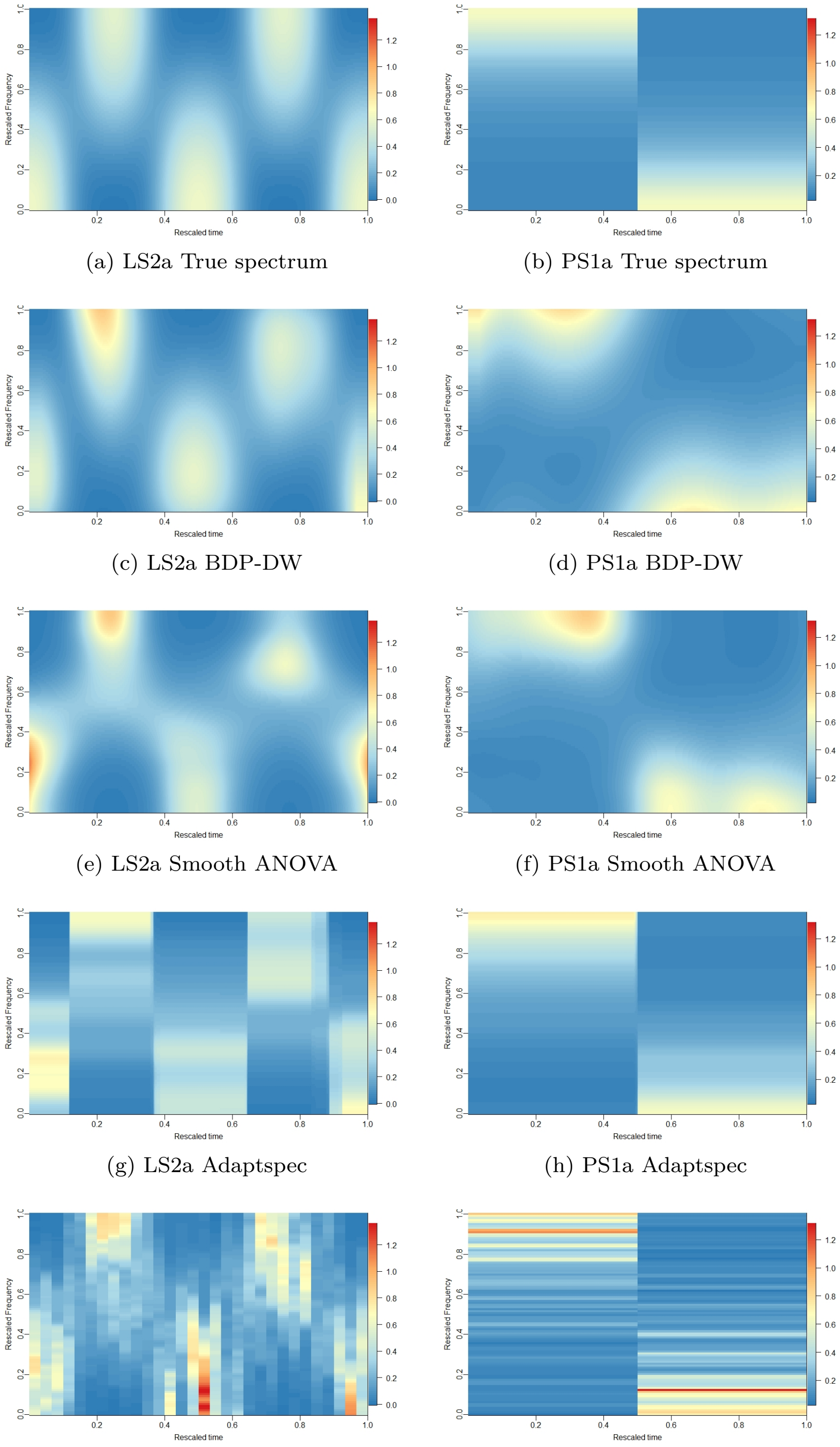

(a) LS2a True spectrum (b) PS1a True spectrum

(c) LS2a BDP-DW (d) PS1a BDP-DW

(e) LS2a Smooth ANOVA (f) PS1a Smooth ANOVA

(g) LS2a Adaptspec (h) PS1a Adaptspec

(i) LS2a Multitaper (j) PS1a Multitaper

Figure 1: The true and estimated time-varying spectral density functions for DGPs LS2a and PS1a. All plots in the same columns share the same color scale and correspond to the same realized time series (as displayed in Section C.1 of the supplementary material).

### 4.3 Choosing between stationarity and non-stationarity

In this section we investigate empirically the performance of using the BDP-DW estimator for model selection as detailed in Section 3.3 (again based on window size $m = 50$ and thinning factor $i = 2$). In addition to the non-stationary processes that we previously considered, we also simulate realizations of the following two stationary processes, again with $T = 1500$ and i.i.d. standard normal innovations $\{w_t\}$:

S1 $X_t = 0.75X_{t-1} + w_t + 0.8w_{t-1}, \quad t = 1, 2, \cdots, T;$

S2 $X_t = w_t - 0.36w_{t-1} + 0.85w_{t-2}, \quad t = 1, 2, \cdots, T.$

The results for the Savage-Dickey estimates $\hat{B}_{01}$ of the Bayes factor are summarized in Table 2.

Table 2: Summary statistics of the Savage-Dickey estimates of the Bayes factor. The first column contains the percentage of the cases where the estimated Bayes factor is greater than 1. The rest of the 5 columns contains the lower 2.5, 5, 10, 25 and 50 percentiles of the estimated Bayes factor.

| DGP | $\hat{B}_{01} > 1$ | 2.5 perc. | 5 perc. | 10 perc. | 25 perc. | 50 perc. |
|---|---|---|---|---|---|---|
| LS1a | 0% | 0 | 0 | 0 | 0 | 0 |
| LS2a | 0.6% | 0 | 0 | 0 | 0 | 0 |
| LS3a | 0% | 0 | 0 | 0 | 0 | 0 |
| PS1a | 0% | 0 | 0 | 0 | 0 | 0 |
| S1 | 95.8% | 0.0981 | 1.6115 | 14.3824 | 26.6070 | 27.2208 |
| S2 | 99.8% | 24.9118 | 26.7625 | 27.1706 | 27.2699 | 27.2808 |

Recall that $\hat{B}_{01} > 1$ indicates that the posterior is in favor of stationarity where confidence in stationarity increases with increasing values of the Bayes factor. Clearly, the Bayes factor based on the BDP-DW estimate does an excellent job in making the right choice between stationarity and non-stationarity (for both the locally stationary and the piecewise stationary cases). Furthermore, a clearer distinction is achieved for S2 in comparison to S1. This is not surprising given that S1 is closer to unit-root-behavior than S2 making it more difficult to classify as stationary.

With the current prior (specified in Section B.1 of the supplementary material), the theoretical maximum of the estimated Bayes factor, as given in Section 3.3, is 27.2808.

Further illustrations of the Bayes factor using real data can be found in Section D of the supplementary material. These include a stretch of noise measurements by the Laser Interferometer Gravitational-Wave Observatory collected during the third LIGO/Virgo observation run. The aim was to check whether the instrumental noise remains stationary over a longer duration compared to the brief period of a fraction of a second that a gravitational wave signal emitted by a binary black hole merger typically lasts. If stationary, this would provide a justification for using the Welch method of averaging over periodograms of overlapping noise segments to obtain a consistent estimate of the noise spectral density that is then used for the purpose of estimating the gravitational wave signal parameters.

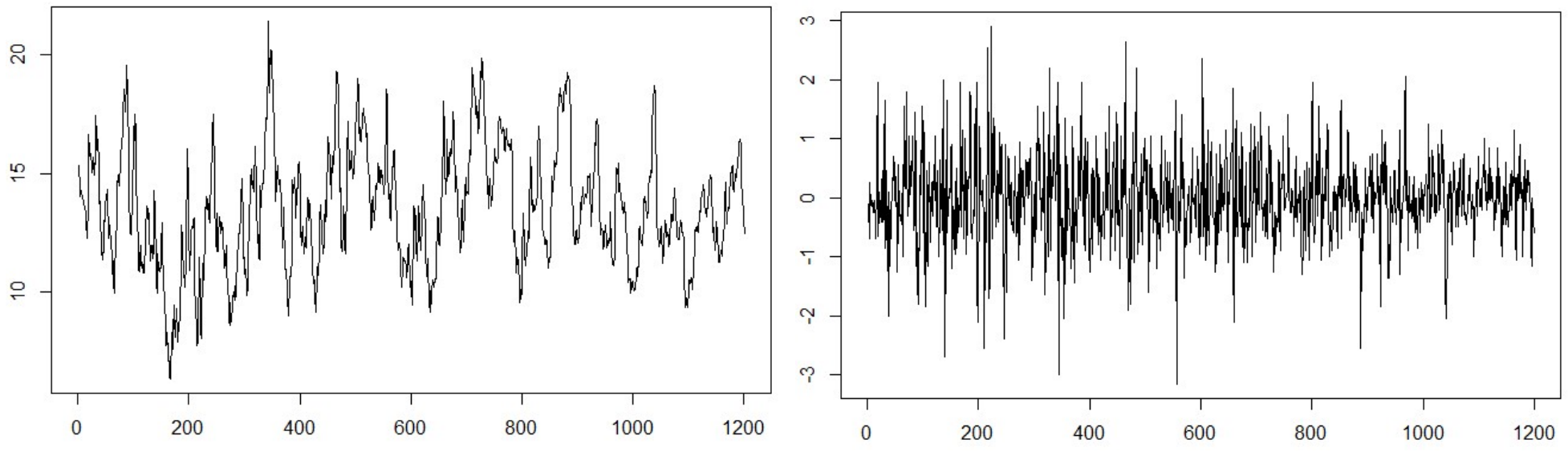

(a) Original egg prices time series

(b) First-order difference

Figure 2: Time series plots of the weekly egg price data.

### 4.4 Case study: Weekly egg prices

In this section, we consider 1201 observations of weekly egg prices at a German agriculture market between April 1967 and May 1990. The data set is provided in Fan and Yao (2003) and has been studied extensively in the literature. Following Paparoditis (2010), the first-order differences of the original time series are analyzed. Both the original and the differenced time series are displayed in Figure 2. We choose a thinning factor $i = 1$ and window size $m = 50$ for the BDP-DW method. The estimated time-varying spectral density is displayed in Figure 3.

The estimated time-varying spectral density exhibits changes along the time direction, which indicate that the transformed time series might have non-stationary behavior. This is further corroborated by the result that the estimated Savage-Dickey Bayes factor equals 0.

## 5 Discussion and outlook

In conclusion, the proposed combination of the bivariate Bernstein-Dirichlet process prior and the dynamic Whittle likelihood based on the moving periodogram is capable of modeling the temporal and spectral components of the time-varying spectral density simultaneously. Not only can we give theoretical guarantees of asymptotic robustness in terms of posterior consistency and contraction rates but also demonstrate very competitive performance in simulations when compared to both the true spectral density and the estimates produced by alternative state-of-the-art methods for locally stationary time series. Moreover, we have made an R package (i.e., the `beyondwhittle` package) readily available for general use. However, for extremely long time series (such as 30 seconds of real Advanced LIGO data with a sampling rate of 16384 Hz), utilizing the proposed sampling scheme that relies on Markov Chain Monte Carlo algorithms could be impractical due to the potentially lengthy amount of time required to complete the process.

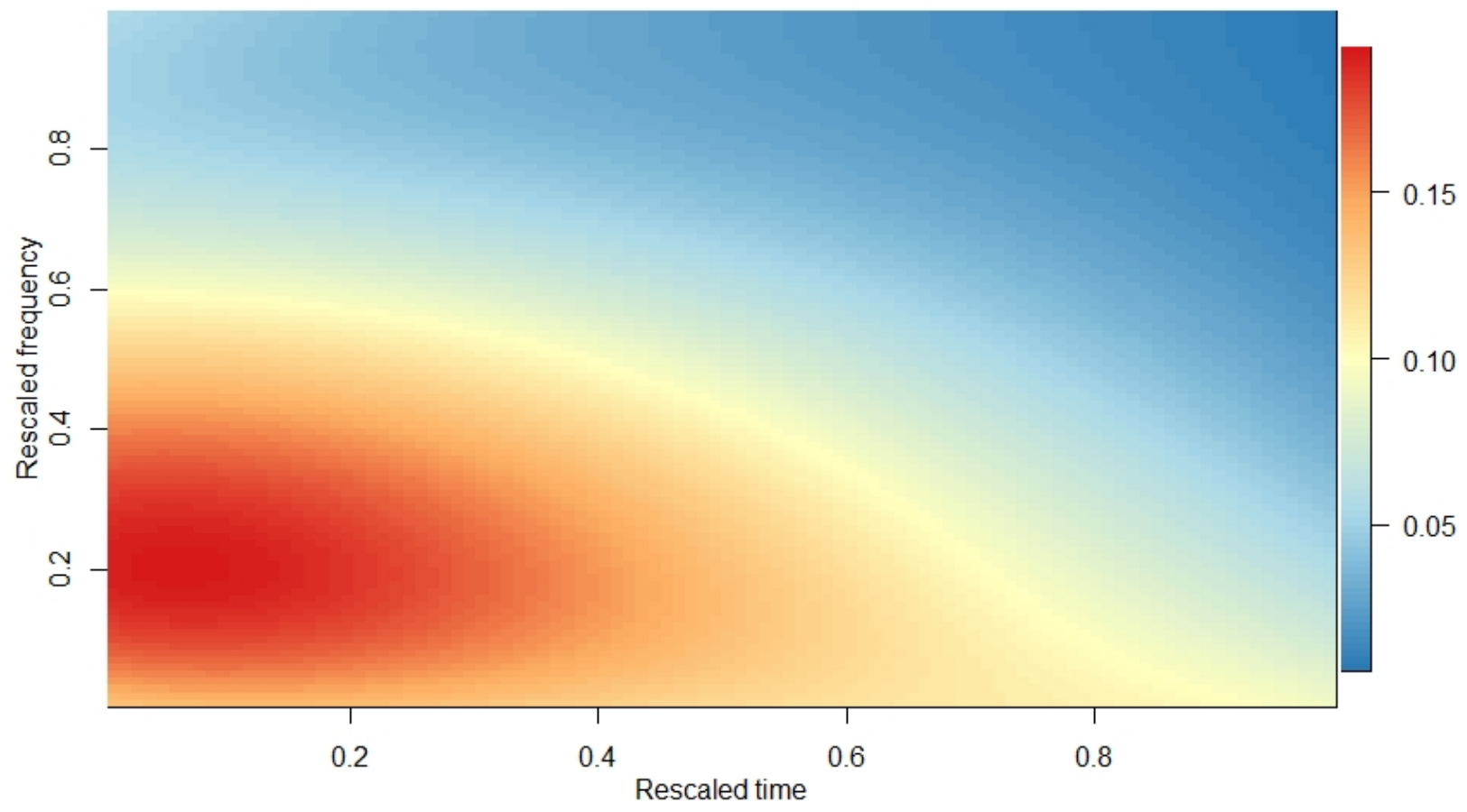

Figure 3: Estimated time-varying spectral density of the first-order difference of the weekly egg price data shown in Fig. 2(b).

Regarding future directions, the innovative approach utilizing a dynamic Whittle likelihood presents several opportunities for expanding upon the proposed methodology. These potential extensions are highlighted in the following paragraphs and will be the primary areas of focus for future research.

First, one potential enhancement to the posterior computation when using the dynamic Whittle likelihood for long time series could involve variational Bayes techniques (e.g. Hu and Prado (2023)).

Secondly, there are various approaches proposed in the literature to deal with the time-varying spectral density matrix of multivariate locally stationary time series (Ombao et al., 2005; Guo and Dai, 2006; Li and Krafty, 2019; Bertolacci et al., 2022; Chau and von Sachs, 2022). The Whittle-based Bayesian methodology for stationary multivariate time series as in Meier et al. (2020) can be used to extend the BDP-DW approach to multivariate time series.

A parametric correction to the dynamic Whittle likelihood similar to Kirch et al. (2019) is worth investigating to further improve the finite-sample performance in particular at peaks in the spectrum. Similarly, other types of basis functions, such as B-spline basis and wavelet basis functions (Shen and Ghosal, 2015), which exhibit faster rates of convergence, could be explored.

Häfner and Kirch (2017) also define a moving inverse Fourier transform which reconstructs a locally stationary time series in the time domain from independent normal variates (with appropriate variances) in the frequency domain. Such an approximation could also be used to construct a likelihood for a locally stationary time series in the time domain. While this offers no advantage if the aim of the analysis is the estimation of the time-varying spectrum, it can be combined with time-domain methods such as in a Bayesian regression context.

Finally, some real-world time series may exhibit time-varying long-memory behavior which is not considered in this paper. Although there are some frequentist methods which aim to model the long-memory time-varying spectral density (Roueff and von Sachs, 2011; Lu and Guegan, 2011; Wang, 2019; Chan and Palma, 2020), no Bayesian method is currently available for locally stationary long-memory time series. The BDP-DW method has the potential to be advanced even further to accommodate long memory by incorporating a parametric component to model the pole at zero frequency.

## Acknowledgements

C.K. and R.M. acknowledge funding by DFG Grant KI 1443/3-2 and Y.T. by a the University of Auckland Doctoral Scholarship. R.M. gratefully acknowledges support by the Marsden grant MFP-UOA2131 from New Zealand Government funding, administered by the Royal Society Te Apārangi. Y.T., J.L and R.M. thank the Centre for eResearch at the University of Auckland for their technical support. We thank Dr. Patricio Maturana-Russel for valuable discussions during the initial stages of the project.

# Supplementary materials for "Bayesian nonparametric spectral analysis of locally stationary processes"

In this file, a mathematically rigorous definition of locally stationary processes is given in Section A. In Section B, the details of prior specification and posterior sampling scheme are given. Sections D and C contain further case studies and simulation results, respectively. The proofs of the asymptotic results which are listed in Section 3.2 of the main text and relevant auxiliary propositions are given in Section E. R code demonstrating the use of the R package `beyondWhittle` for locally stationary time series is in Section F.

## A Locally stationary time series

In this section, we state the definition of locally stationary time series as introduced by Dahlhaus (1997, 2003). To this end, we regard the process $\{X_t\}$ as part of a sequence of processes $\{X_{t,T}\}$ with the following properties:

**Definition S.1** (Locally Stationary Processes)**.** Let $(\Omega, \mathcal{P}, \mathrm{P})$ be a probability space. A collection of random variables $\{X_{t,T}\}$ defined on $(\Omega, \mathcal{P})$ is called *locally stationary* if

$$X_{t,T} = \sum_{j=-\infty}^{\infty} a_{t,T}(j)\, \varepsilon_{t-j},\ t = 1, 2, \cdots, T,\ T \in \mathbb{N}$$

and the following conditions are satisfied:

(a) $\{\varepsilon_t,\ t \in \mathbb{Z}\}$ are independent and identically distributed with $\mathrm{E}\varepsilon_1 = 0$, $\mathrm{E}\varepsilon_1^2 = 1$ and $\mathrm{E}\varepsilon_1^4 < \infty$.

(b) $\sup_t |a_{t,T}(j)| \leqslant \frac{K}{l(j)}$, where $K > 0$ is a constant not depending on $T$, $l(j) = 1$ when $|j| \leqslant 1$ and $l(j) = |j| \ln^{1+\kappa} |j|$ otherwise for some $\kappa > 0$.

(c) There exists functions $a(\cdot, j) : (0, 1] \to \mathbb{R}$, $j \in \mathbb{Z}$ satisfying

  (i) $\sup_t \left|a_{t,T}(j) - a\left(\frac{t}{T}, j\right)\right| \leqslant \frac{K}{Tl(j)}$;

  (ii) $|a(u, j) - a(v, j)| \leqslant \frac{K|u-v|}{l(j)}$;

  (iii) $\sup_u \left|\frac{\partial^i}{\partial u^i} a(u, j)\right| \leqslant \frac{K}{l(j)}$, $i = 0, 1, 2, 3$.

The above definition guarantees in particular, that $X_{t,T}$ is close to the stationary linear process $\widetilde{X}_t(u) = \sum_{j=-\infty}^{\infty} a(u, j)\, \varepsilon_{t-j}$ as long as $t/T$ is close to $u$ (for a mathematical rigorous formulation we refer e.g. to (1.1.19) in Sergides (2008)). This formalizes the notion that the locally stationary process behaves locally (i.e., in a small neighbourhood of any rescaled time point $u$) like a stationary process.

The time-varying (power) spectral density (tv-PSD) is then described within this model as

$$f(u,\lambda) = \frac{1}{2\pi}\left|\sum_{j\in\mathbb{Z}} a(u,j)\exp\left(-i\pi\lambda j\right)\right|^2, \quad (u,\lambda)\in[0,1]^2,$$

and $c(u,h) = 2\pi\int_0^1 f(u,\lambda)\exp(i\pi\lambda h)\,\mathrm{d}\lambda$ is the time-varying covariance of lag $h$, $h\in\mathbb{Z}$ (at rescaled time $u$). To build intuition we note that $f(u,\cdot)$ is the spectral density of the stationary approximation $\{\widetilde{X}_t(u)\}$ in the (rescaled) time point $u$.

# B Computational aspects of the BDP-DW method

Before describing the Markov chain Monte Carlo (MCMC) algorithm to sample from the posterior distribution, we address some details of prior specification that are of practical importance.

## B.1 Details of the prior specification

Since the parameter space $\Theta$ defined in (3.1) in the main text is only for theoretical proofs, there is no need to choose the associated bounds in practice. Using the set of beta densities $\beta(\cdot;j,k-j)$ to define the mixture of Bernstein polynomials (in (3.2) in the main text) has the undesirable effect that close to either endpoint of the unit interval the value of the mixture is mainly determined by just one basis function each w.r.t. time and frequency. This yields a prior density that exhibits a strange behavior at the boundaries of the unit square (see e.g. Section 5.1 of Meier (2018) for an illustration in the univariate case) that deviates from a constant plane (as expected for a noninformative prior) at the boundaries. This in turn induces an undesirable sensitivity of the posterior computational procedures and estimators to the weights corresponding to these (four) basis functions.

Therefore, to achieve a more stable mixture behavior on the boundaries of $[0,1]^2$ the beta densities $\beta(\cdot;j,k-j)$ are replaced by their truncated and dilated counterparts:

$$\beta_{\xi_l}^{\xi_r}(x;j,k-j) = c_{r,l}(j,k)\beta\left(\xi_l + x\left(\xi_r-\xi_l\right);j,k-j\right)$$

where $0<\xi_l<\xi_r<1$, $x\in[0,1]$ and $c_{r,:l}(j,k)$ is the normalizing constant such that the integral of $\beta_{\xi_l}^{\xi_r}(\cdot;j,k-j)$ over $[0,1]$ is 1. In the empirical studies, we employ the truncated Bernstein polynomial basis with $\xi_l = 0.1$ and $\xi_r = 0.9$. Unlike the standard beta densities, they have the advantage that the resulting polynomial mixture at either endpoint of the unit interval is determined by several basis functions rather than just one basis function.

The prior of the polynomial degree $k_i$ is $\rho_i(k_i)\propto\exp\left(-0.01k_i\ln k_i\right)$, $i=1,2$. To ease the computational burden, the values of $k_i$ are thresholded at $k_{i,\max}=100$. In practice, one may first set the threshold to 100 and increase it should the estimated posterior probability of $\{k_i = 100\}$ be large. As for the Dirichlet process, we choose $M = 1$ and $G_0$ to be the uniform distribution over $[0,1]^2$. The prior for $\tau$ is chosen to be an Inverse-Gamma distribution with shape and scale parameters all being 0.001.

## B.2 Posterior computation

Since the posterior distribution is analytically intractable, we conduct the posterior computation via a simulation-based MCMC method. In order to facilitate the computation, we utilize the Sethuraman representation for a Dirichlet process (Sethuraman, 1994). The random function $G$ is represented as $G = \sum_{l=1}^{\infty} p_l \delta_{W_l}$, where $\delta$ denotes a point mass function, $p_1 = V_1$, $p_l = (1-V_1)\cdots(1-V_{l-1})V_l$ for $l \geqslant 2$, $V_l \sim \text{beta}(1, M)$, $W_l = (W_{1l}, W_{2l}) \sim G_0$ and $(W_1, W_2, \cdots, V_1, V_2, \cdots)$ are all independent. To achieve a finite series representation, we truncate the series at a large $L$ and represent $G$ as $G = \sum_{l=0}^{L} p_l \delta_{W_l}$, where $p_0 = 1 - \sum_{l=1}^{L} p_l$, $W_0 = (W_{10}, W_{20}) \sim G_0$. The additional $p_0\delta_{W_0}$ is to make sure that $G$ is still a probability measure even after truncation (see also Choudhuri et al. 2004; Zheng et al. 2010). A choice of $L = L_i = \max\{20, (mB_i)^{\frac{1}{3}}\}$ is recommended by Zheng et al. (2010), where $B_i$ is defined in Definition 3 in the main text. Then the random function can be written as

$$
\begin{aligned}
f(u_1, u_2; k_1, k_2, G) &= \tau b(u_1, u_2; k_1, k_2, G) \\
&= \tau \sum_{j_1=1}^{k_1} \sum_{j_2=1}^{k_2} \left[ \sum_{l=0}^{L} p_l \, \mathrm{I}\left\{ \frac{j_1 - 1}{k_1} < W_{1l} \leqslant \frac{j_1}{k_1}, \frac{j_2 - 1}{k_2} < W_{2l} \leqslant \frac{j_2}{k_2} \right\} \right. \\
&\qquad\qquad \left. \cdot \beta(u_1; j_1, k_1 - j_1 + 1)\, \beta(u_2; j_2, k_2 - j_2 + 1) \right] \\
&= \tau \sum_{l=0}^{L} p_l \beta(u_1; \lceil k_1 W_{1l} \rceil, k_1 - \lceil k_1 W_{1l} \rceil + 1)\, \beta(u_2; \lceil k_2 W_{2l} \rceil, k_2 - \lceil k_2 W_{2l} \rceil + 1),
\end{aligned}
$$

where $\mathrm{I}\{\cdot\}$ denotes an indicator function and the last equation is due to the fact that the indicator $\mathrm{I}\left\{ \frac{j_1-1}{k_1} < W_1 \leqslant \frac{j_1}{k_1}, \frac{j_2-1}{k_2} < W_2 \leqslant \frac{j_2}{k_2} \right\} = 1$ if and only if $j_l = \lceil k_l W_l \rceil$, $l = 1, 2$. Hence, the prior density has the representation

$$
\left[ \prod_{l=1}^{L} M(1 - V_l)^{M-1} \right] \left[ \prod_{l=0}^{L} g_0(W_{1l}, W_{2l}) \right] \rho_1(k_1)\rho_2(k_2)\pi(\tau)
$$

and the posterior is proportional to the product of $L_{DW}^{(i)}(f)$ and the prior density. Since we have $3L+5$ parameters $(\tau, k_1, k_2, W_{10}, \cdots, W_{1L}, W_{20}, \cdots, W_{2L}, V_1, V_2, \cdots, V_L)$ and the full conditionals are not available in a closed form, we employ a block-wise Metropolis-Hastings (MH) algorithm to sample from the posterior. The proposal for $k_l$, $l = 1, 2$ is chosen as a random walk scheme with symmetrized Poisson distribution increments. Since $W_{1j}, W_{2j}, V_j$ are all in $(0, 1)$, we employ logit transformation so that the transformed parameters $\tilde{W}_{1j}, \tilde{W}_{2j}, \tilde{V}_j$ take values in $\mathbb{R}$. We then update the transformed parameters by dividing them into three blocks $\{\tilde{W}_{1j}\}_{j=0}^{L}$, $\{\tilde{W}_{2j}\}_{j=0}^{L}$ and $\{\tilde{V}_j\}_{j=1}^{L}$ and applying adaptive block-wise MH steps with normal random walk proposals. For example, given the value $\{\tilde{V}_j^{(i)}\}_{j=1}^L$ of block $\{\tilde{V}_j\}_{j=1}^L$ in iteration $i$ of the Markov Chain, a proposal for $\{\tilde{V}_j^{(i+1)}\}_{j=1}^L$ is generated by a $L$-dimensional normal distribution with mean $\{\tilde{V}_j^{(i)}\}_{j=1}^L$ and a covariance matrix determined adaptively in the same manner as Section 2 of Roberts and Rosenthal (2009). Finally, the proposal for $\ln \tau$ is a uniform distribution centered at the old value with width determined adaptively during burn-in.

# C Additional simulation results

## C.1 Plots of the time series used in the simulation study in Section 4.2 of the main text and corresponding estimators

In this section, we provide time series plots (Figure I) based on one single simulation for each of the three locally stationary and the piecewise stationary time series that were used to generate a sample estimate in Figure 1 in the main text.

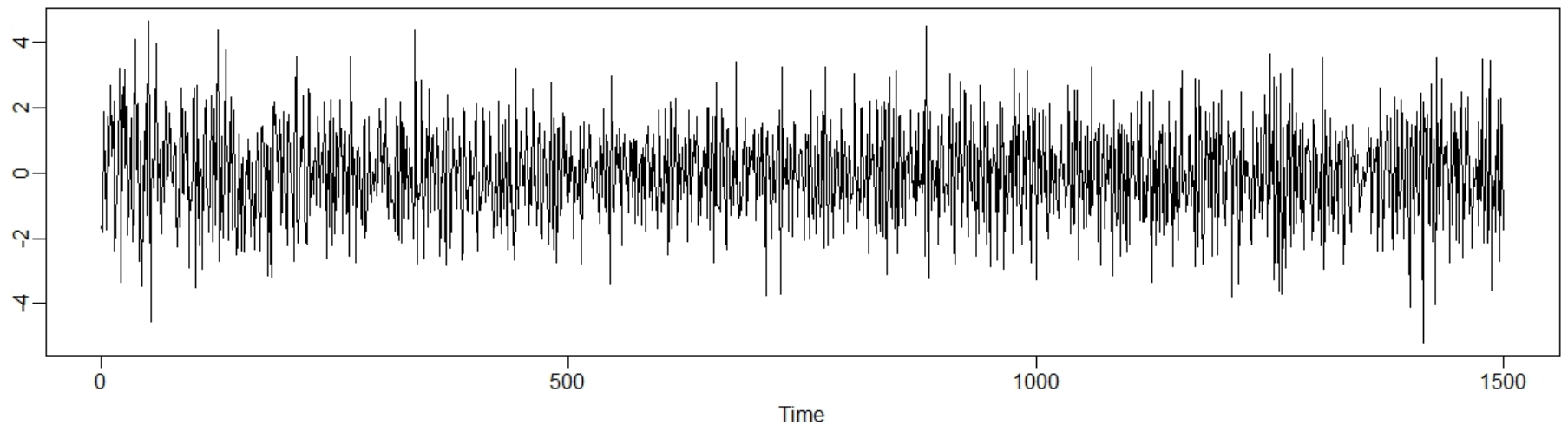

LS1a

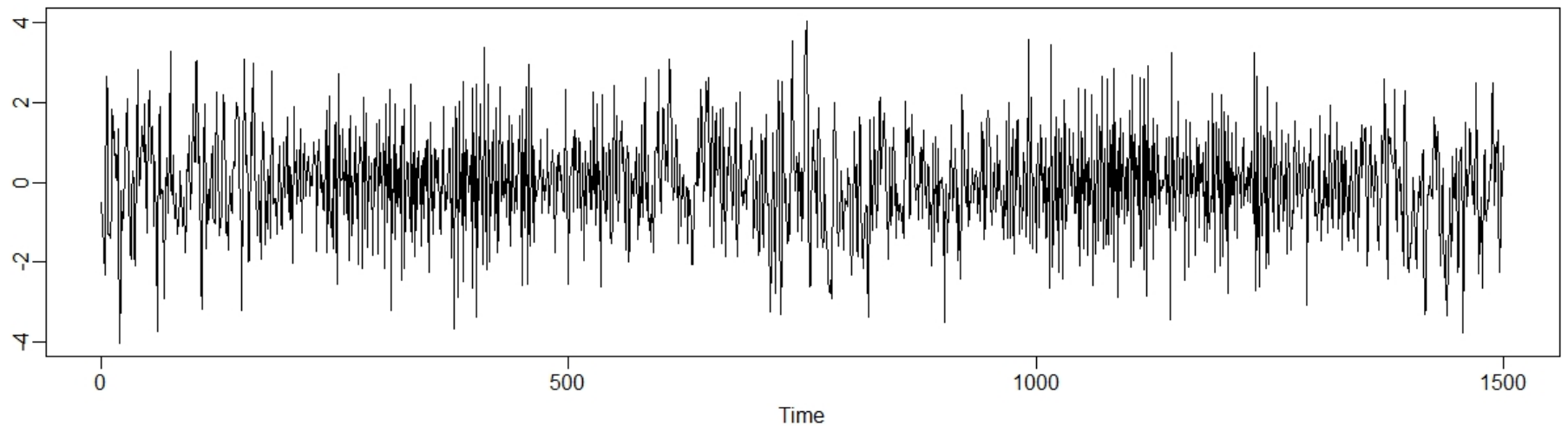

LS2a

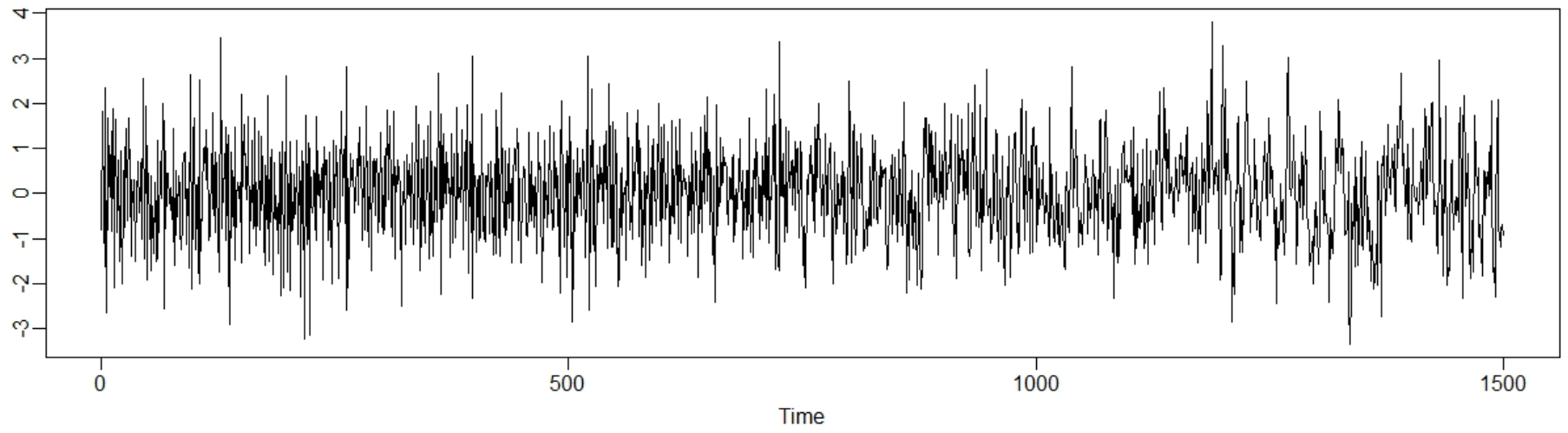

LS3a

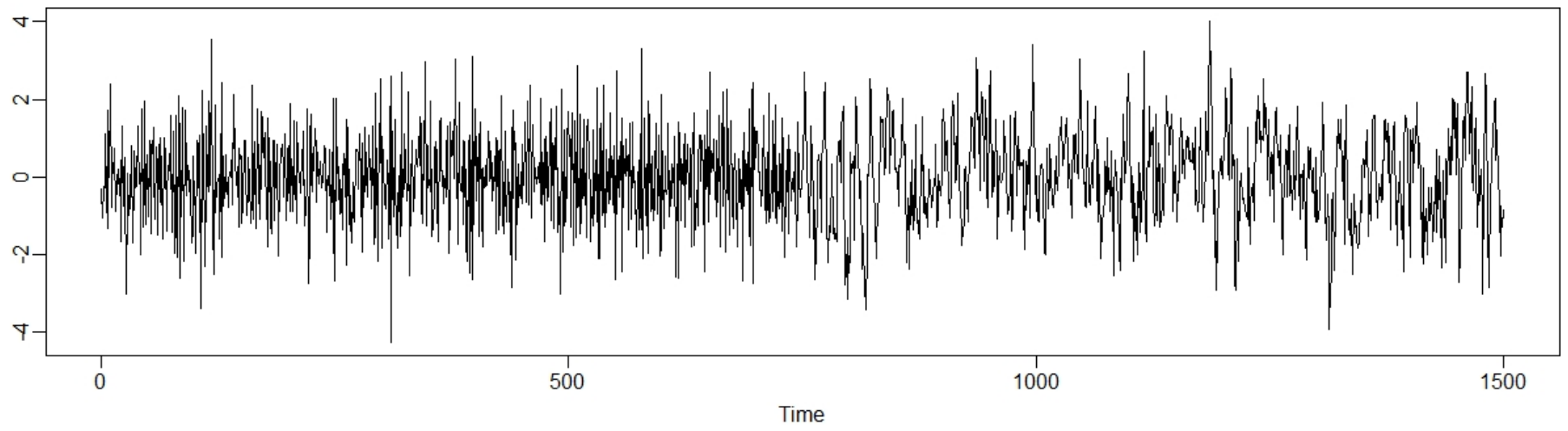

PS1a

Figure I: Randomly selected realizations of DGPs LS1a, LS2a, LS3a and PS1a.

## C.2 Complete simulations: Choice of tuning parameters

The proposed methodology requires the specification of the window size which acts as a tuning parameter. As mentioned before, its choice implies a trade-off between the resolution in time and frequency. Additionally, we have introduced thinned likelihoods to speed up the computations. In this section, we will compare these versions of the methodology to get an impression of how sensitive the methodology is with respect to the choice of window size $m$ and thinning factor $i$. All computations are performed on a virtual machine with 64GB RAM, 16 VCPUs, and an Ubuntu Linux operating system.

The results are summarized in Table I. As expected, there is a trade-off between the accuracy of the estimation and the computation time. A smaller thinning factor leads to a better estimate at the expense of run time. Despite the better empirical performance without thinning, we used a thinning factor of 2 in all other simulation studies.

Table I: Median and IQR (in brackets) of ASE and run-time pertaining to BDP-DW approach with different thinning factors and window size $m = 50$. The unit of run-time is minutes. The thinning factor $i$ is denoted by upper superscript.

| DGP | Type | $L_{DW}^{(1)}$ | $L_{DW}^{(2)}$ | $L_{DW}^{(3)}$ |
|---|---|---|---|---|
| LS1a | ASE | 0.1037(0.0210) | 0.1186(0.0265) | 0.1402(0.0362) |
| | Run-time | 5.475(0.142) | 4.365(0.088) | 4.059(0.093) |
| LS2a | ASE | 0.2076(0.0423) | 0.2147(0.0488) | 0.2361(0.0527) |
| | Run-time | 5.790(0.173) | 4.526(0.109) | 4.150(0.105) |
| LS3a | ASE | 0.0236(0.0121) | 0.0252(0.0124) | 0.0311(0.0155) |
| | Run-time | 5.423(0.127) | 4.409(0.106) | 4.064(0.091) |

Next, we investigate how sensitive the BDP-DW approach is with respect to the choice of the window size $m$. Three window sizes ($m = 25, 50, 75$) are chosen and the BDP-DW procedure with thinning factor 2 is applied to the simulated data described above. The results are summarized in Table II. The results clearly show the trade-off between time and frequency resolution: While for the slower time-varying LS1a spectrum, a choice of $m = 50$ (or even $m = 75$) with a higher resolution in frequency yields best results, a choice of $m = 25$ corresponding to a higher time resolution is best for the faster time-varying LS2a spectrum. For LS2a, longer window sizes yield significantly worse ASEs. For this particular example, the error occurring due to boundary effects in the time direction is quite large. For example, the ASE for $m = 25$ improves from 0.1479 to 0.1324 if calculated only for $t = m, \ldots, T - m$. For LS3a all three window sizes show comparable performance.

Table II: Median and IQR (in brackets) of of ASE pertaining to the BDP-DW estimates with thinning factor $i = 2$ and different window sizes $m = 25, 50, 75$.

| DGP | $m = 25$ | $m = 50$ | $m = 75$ |
|---|---|---|---|
| LS1a | 0.1392 (0.0346) | 0.1186 (0.0265) | 0.1157 (0.0272) |
| LS2a | 0.1479 (0.0304) | 0.2147 (0.0488) | 0.3796 (0.0667) |
| LS3a | 0.0238 (0.0107) | 0.0252 (0.0124) | 0.0279 (0.0153) |

## C.3 Additional plots for the comparison with the state-of-the-art in Section 4.2 of the main text

Figure II shows heat maps of the tv-PSD estimates based on one single simulated time series for DGPs LS1a and LS3a.

Additionally, we give complementary plots (Figures III–VI) to the summary statistics in Table 1 in the main text: More precisely, we plot the pointwise median and pointwise IQR of all 1000 realizations to provide insight into the estimation accuracy. Not surprisingly, BDP-DW and Smooth ANOVA clearly suffer from boundary effects (in time direction) in terms of having more variable estimates at the boundaries where the effect is severer for smooth ANOVA than for BDP-DW. Furthermore, the estimate is more volatile close to peaks which is also not surprising. The multitaper method has larger variability also outside boundaries and peaks, in particular the variability of the estimates are proportional to the values of the spectral density estimates.

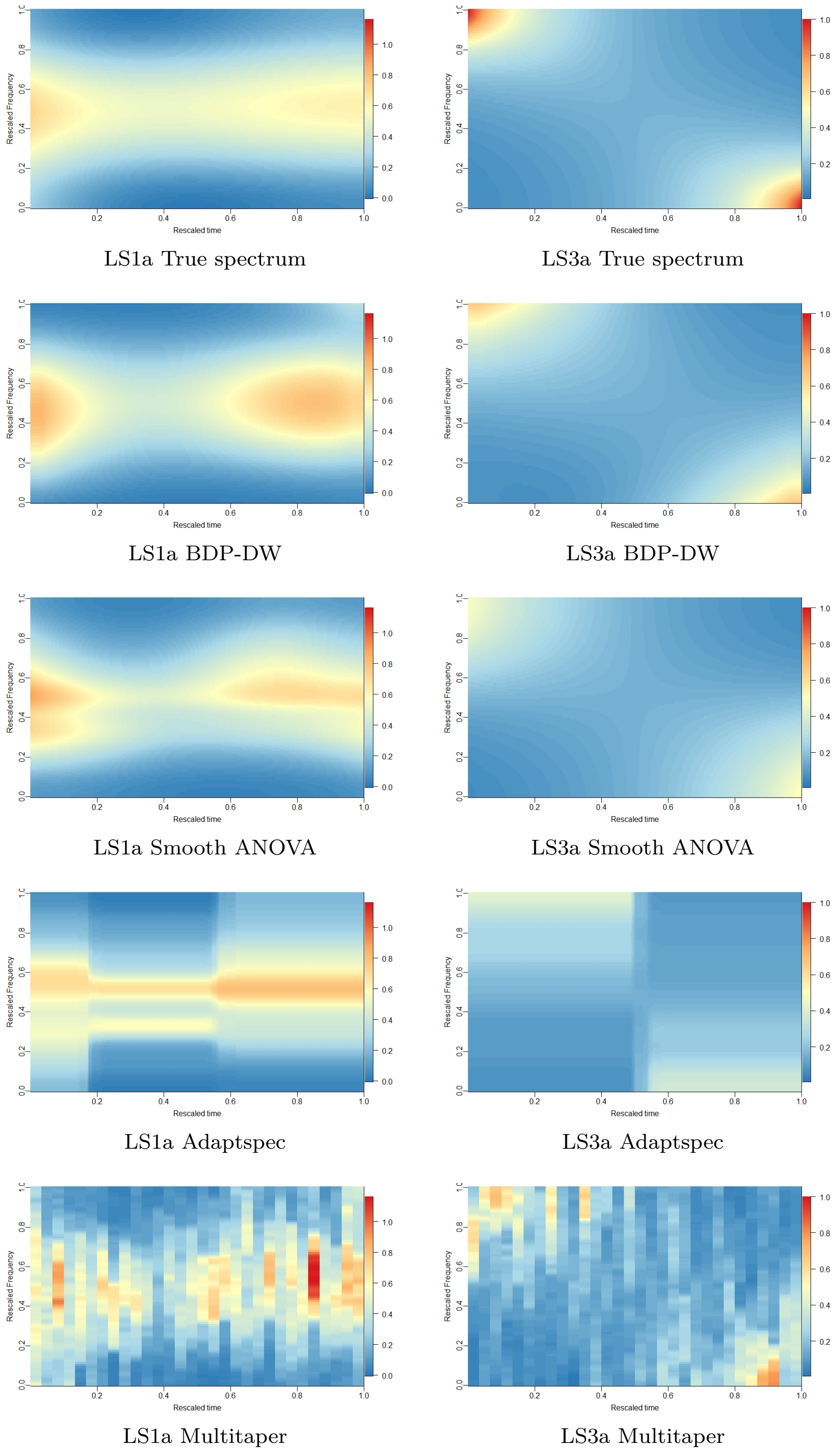

Figure II: The true and estimated time-varying spectral density functions for DGPs LS1a and LS3a. The plots in the same column correspond to the same realized time series displayed in Figure I and also share the same color scale.

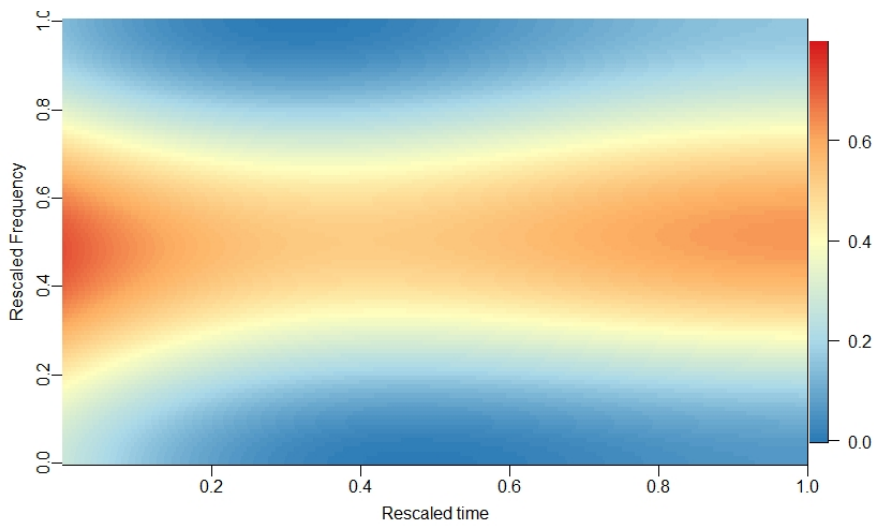

LS1a True spectrum

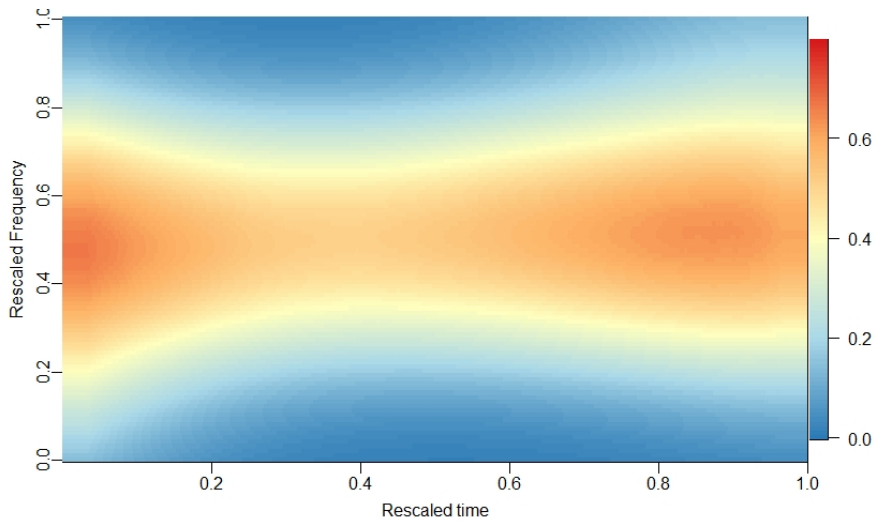

LS1a BDP-DW

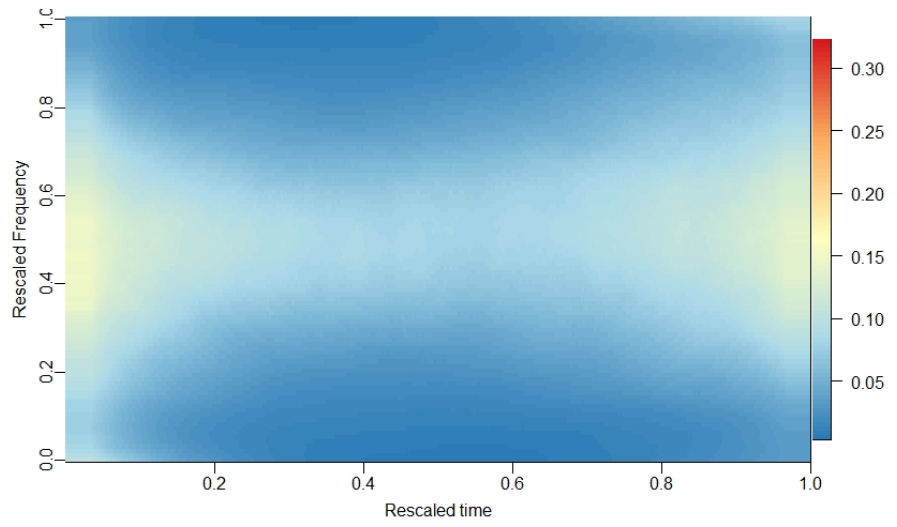

LS1a BDP-DW

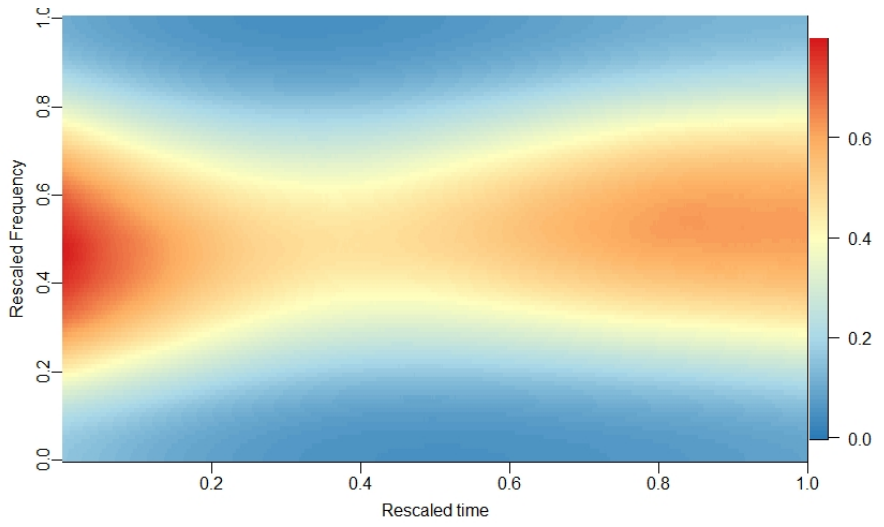

LS1a Smooth ANOVA

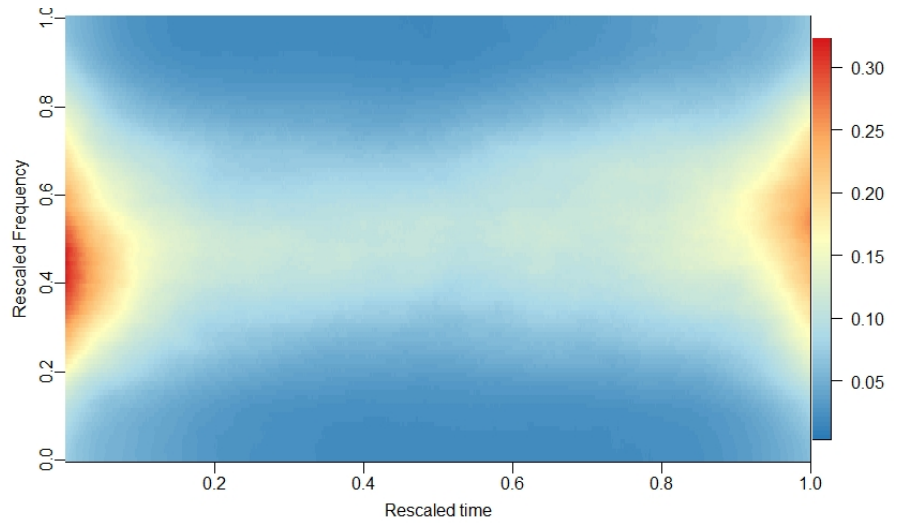

LS1a Smooth ANOVA

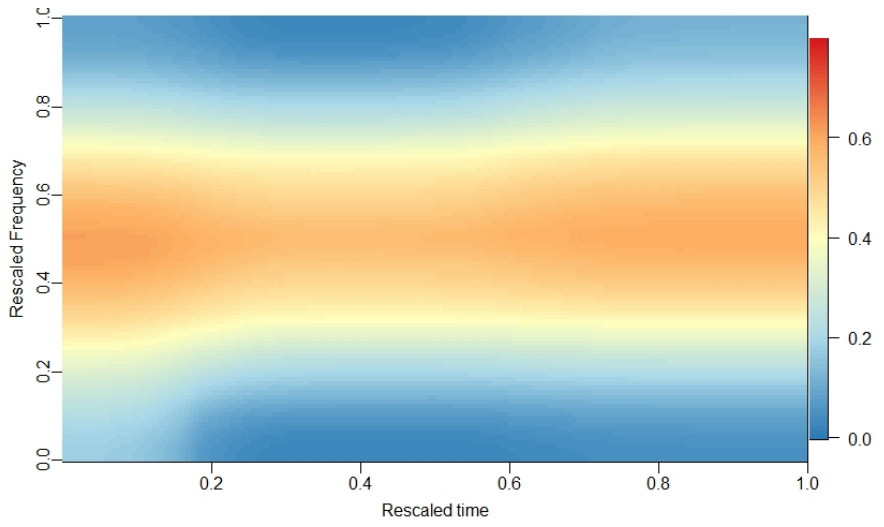

LS1a Adaptspec

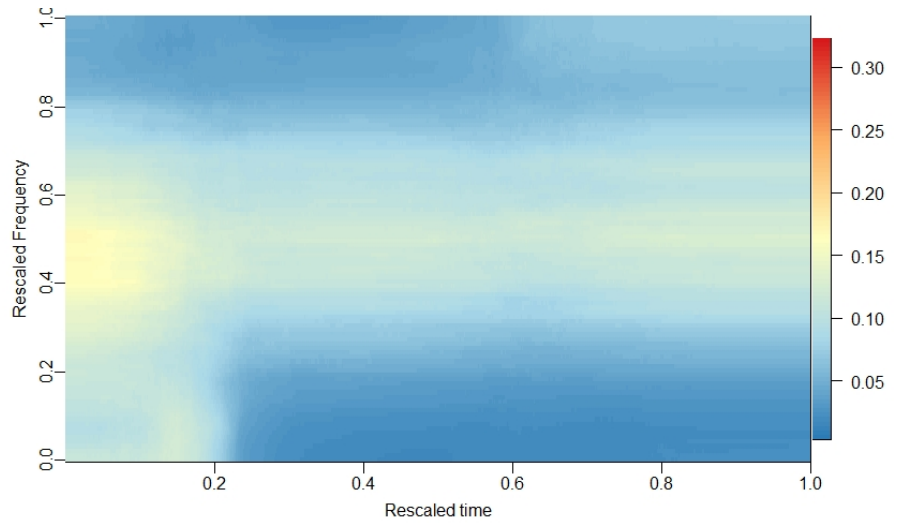

LS1a Adaptspec

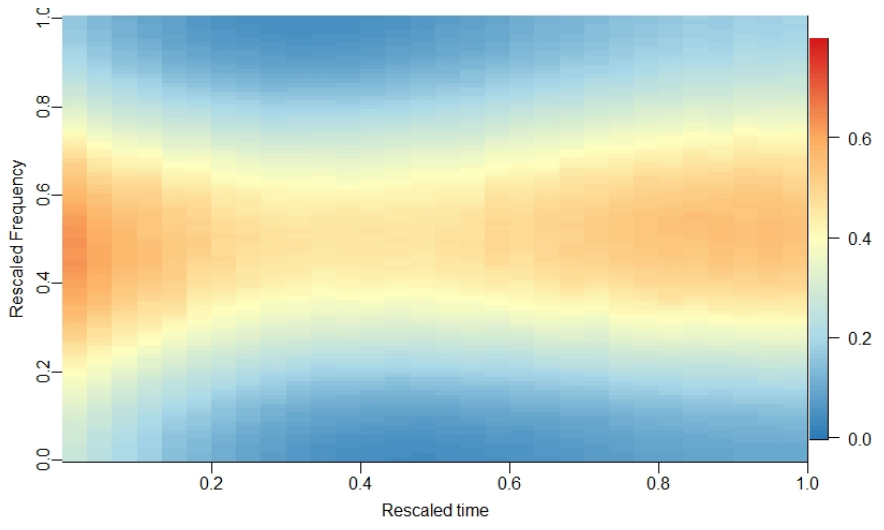

LS1a Multitaper

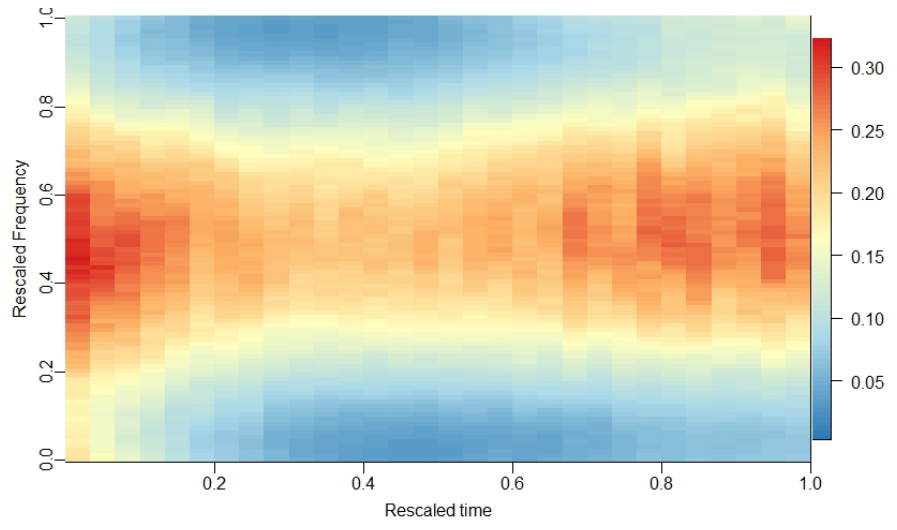

LS1a Multitaper

Figure III: Left column: The pointwise median tv-PSD estimates of all 1000 realizations of LS1a. Right column: The pointwise IQR. For each plot, the x-axis is rescaled time and the y-axis is rescaled frequency. The plots in the same column share the same color scale.

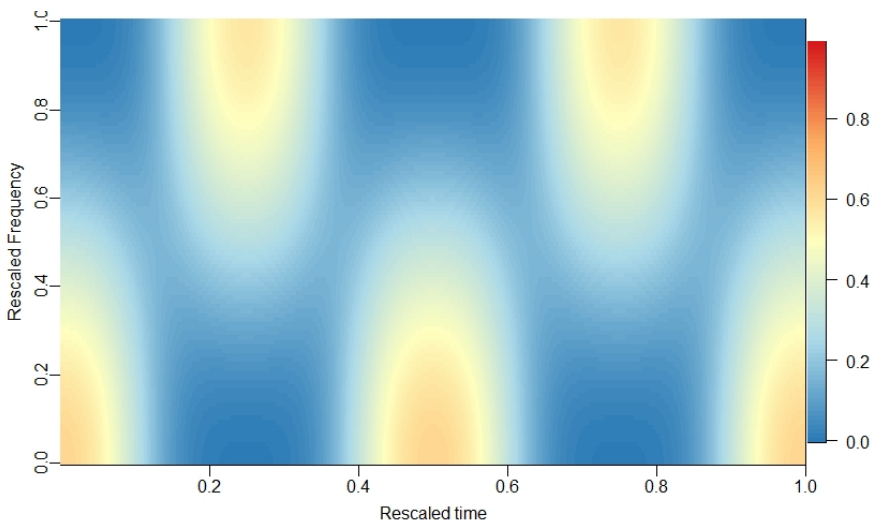

LS2a True spectrum

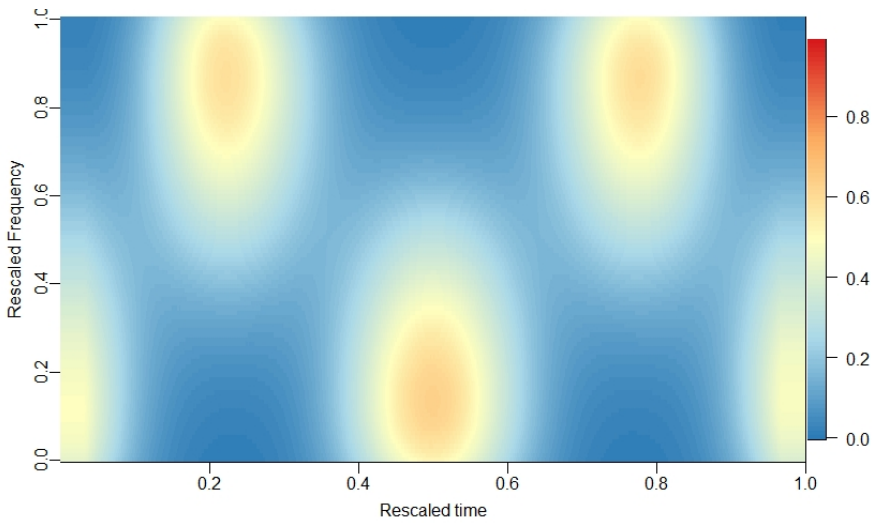

LS2a BDP-DW

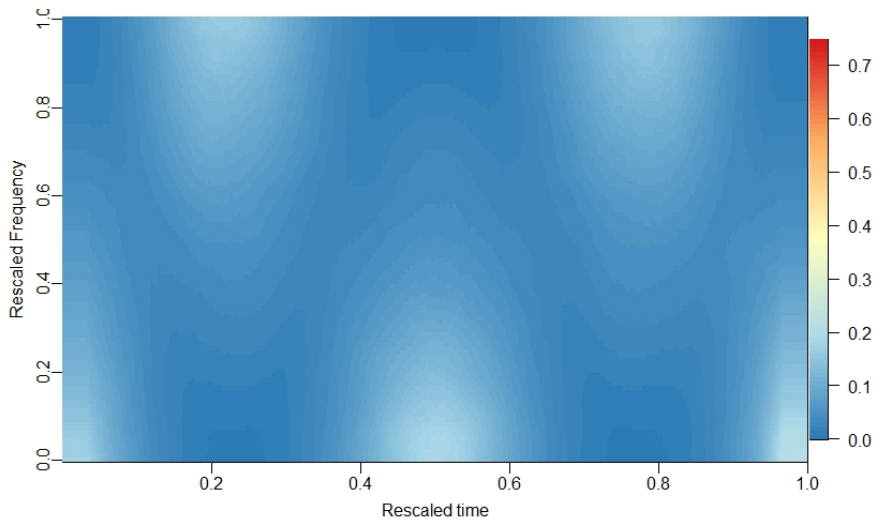

LS2a BDP-DW

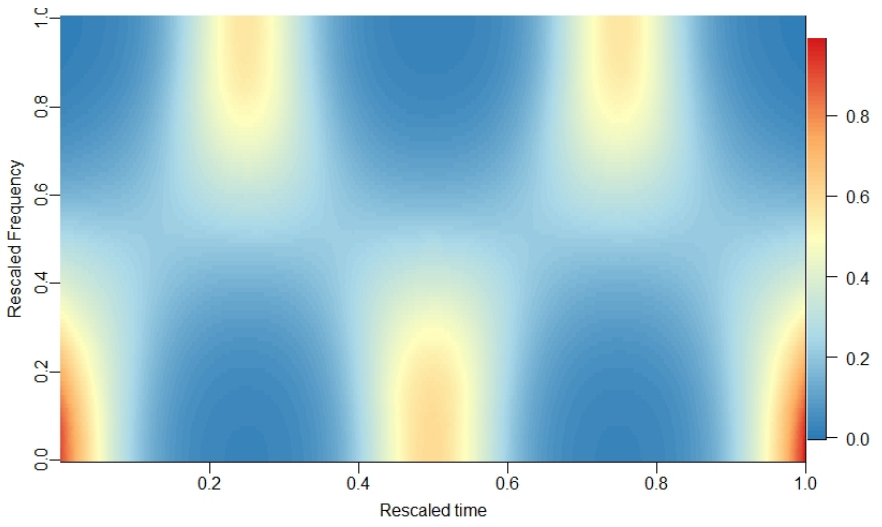

LS2a Smooth ANOVA

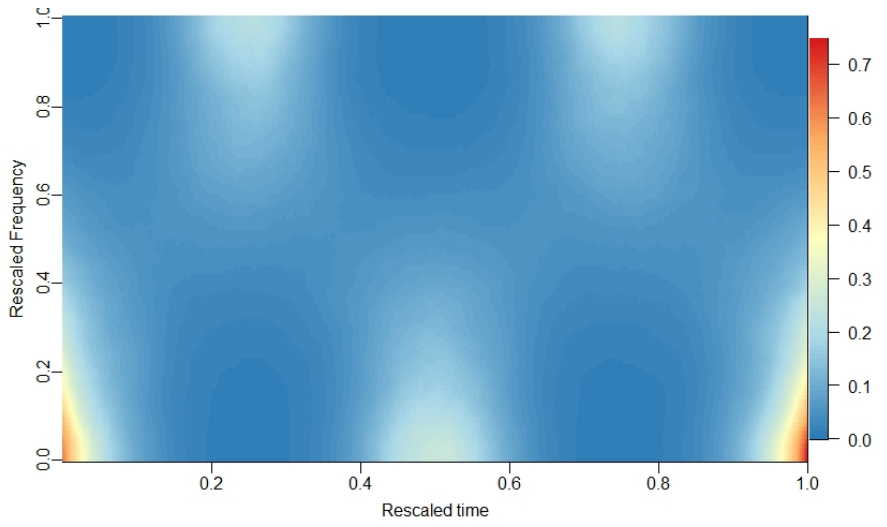

LS2a Smooth ANOVA

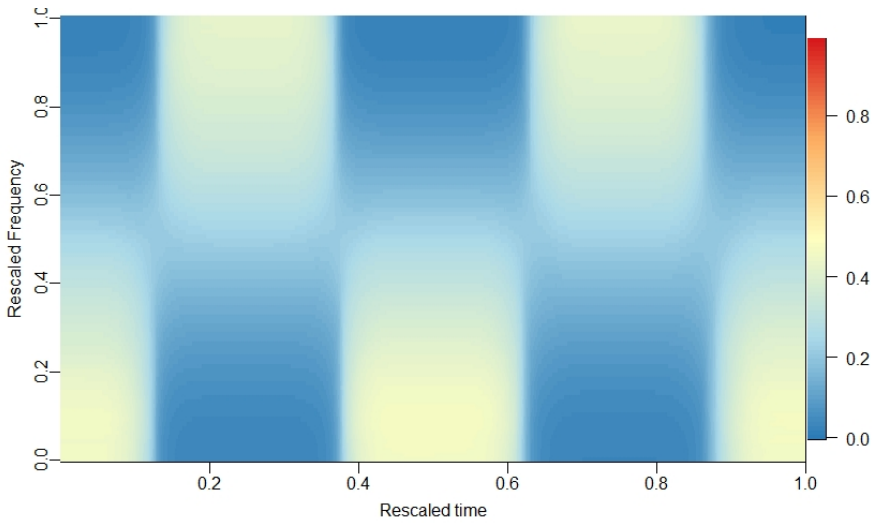

LS2a Adaptspec

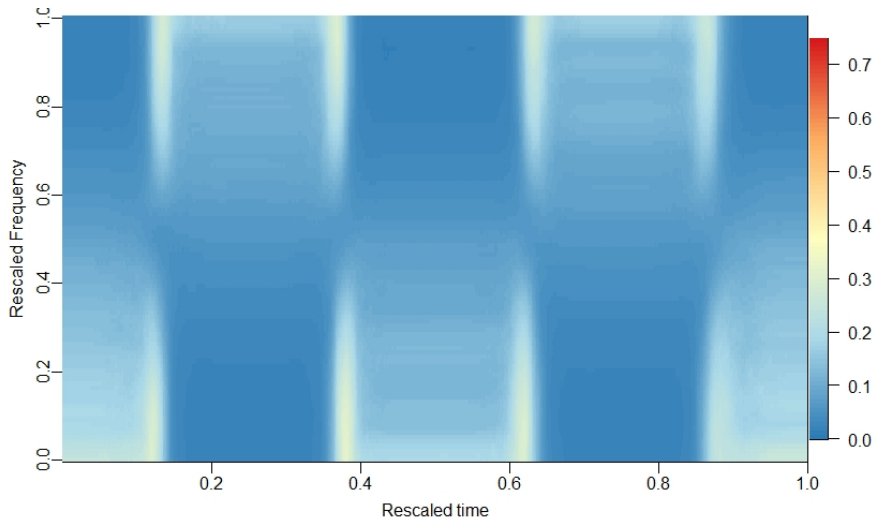

LS2a Adaptspec

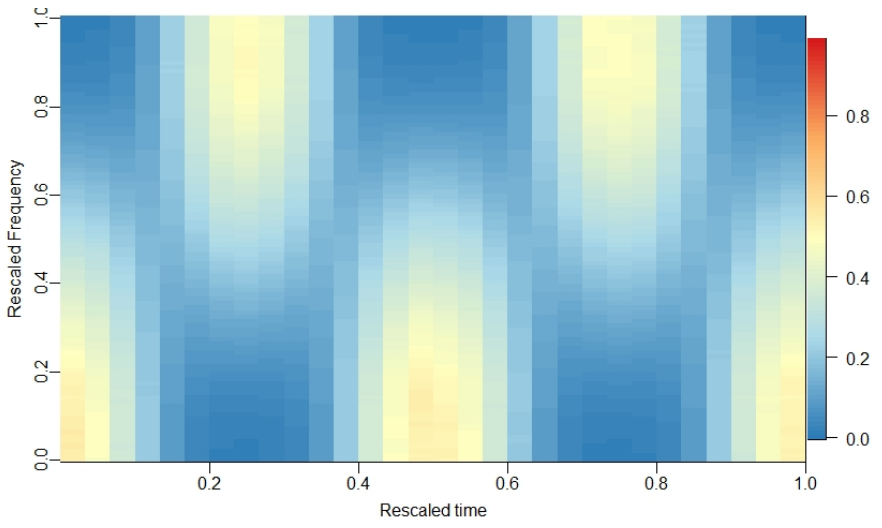

LS2a Multitaper

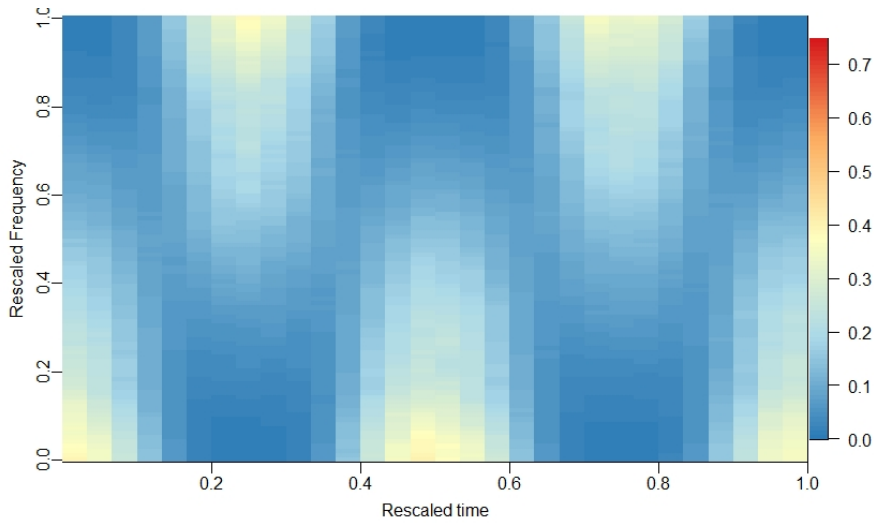

LS2a Multitaper

Figure IV: Left column: The pointwise median tv-PSD estimates of all 1000 realizations of LS2a. Right column: The pointwise IQR. For each plot, the x-axis is rescaled time and the y-axis is rescaled frequency. The plots in the same column share the same color scale.

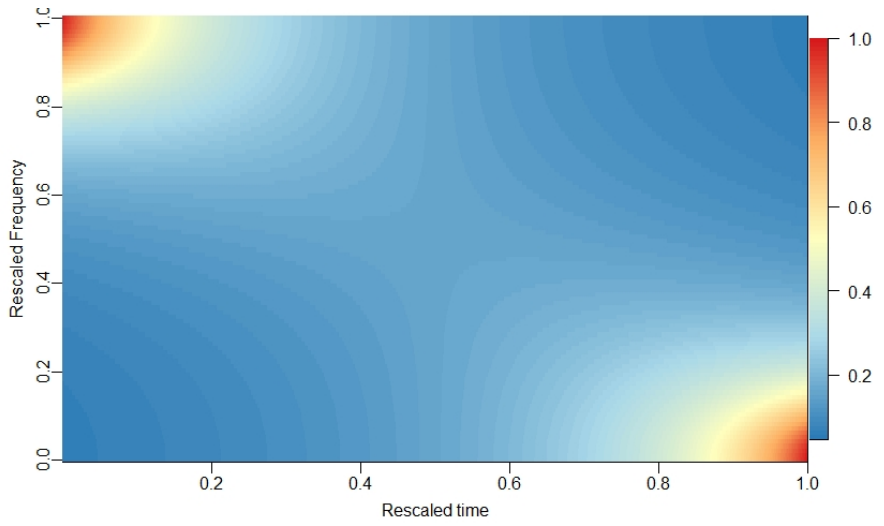

LS3a True spectrum

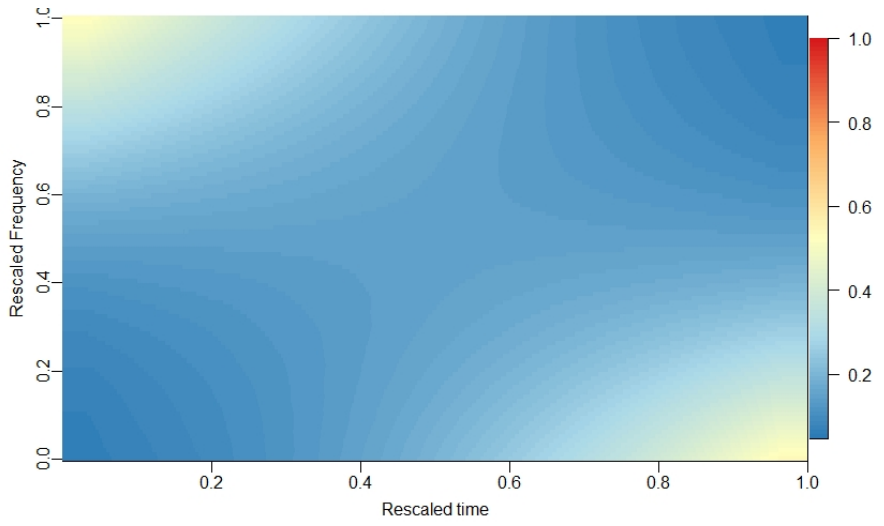

LS3a BDP-DW

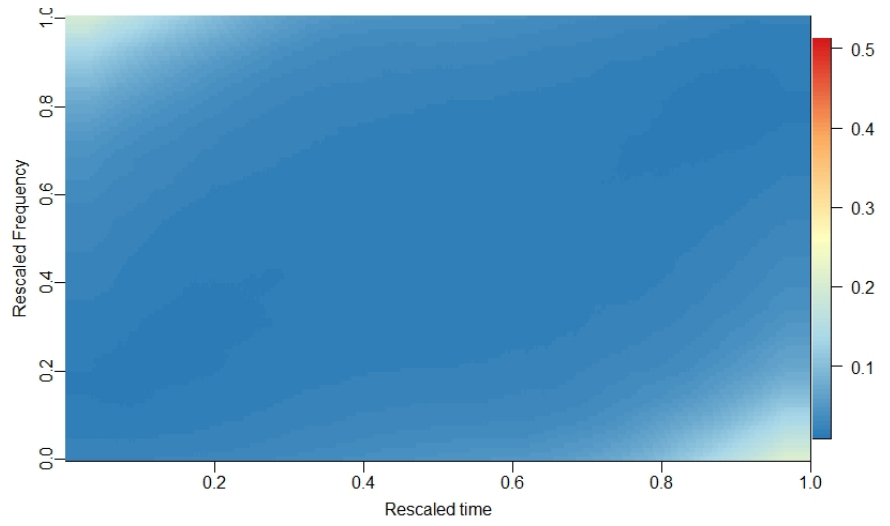

LS3a BDP-DW

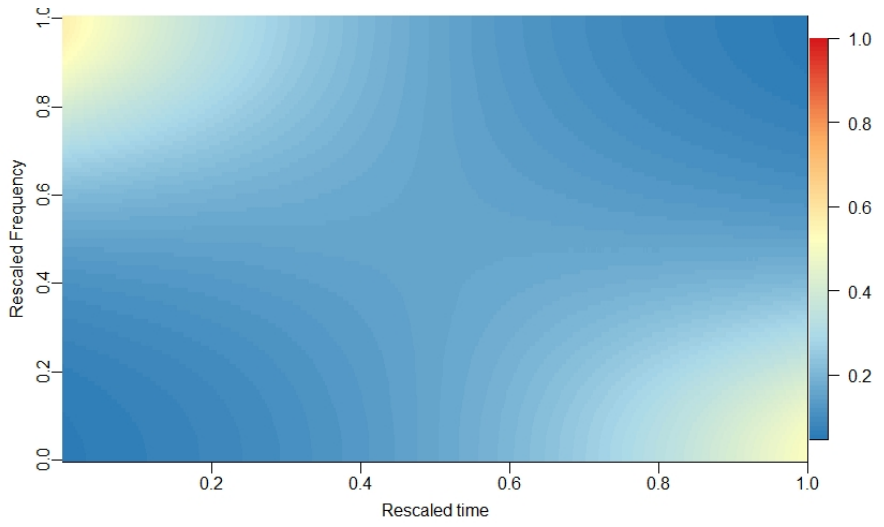

LS3a Smooth ANOVA

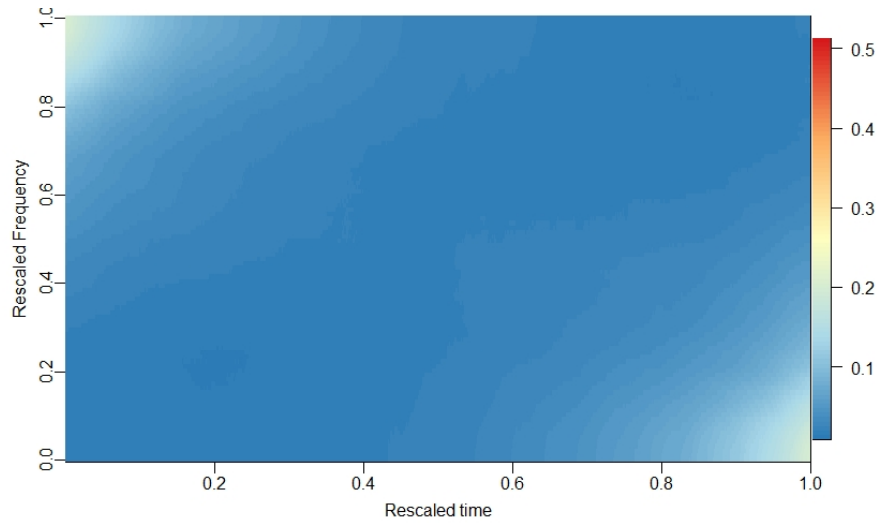

LS3a Smooth ANOVA

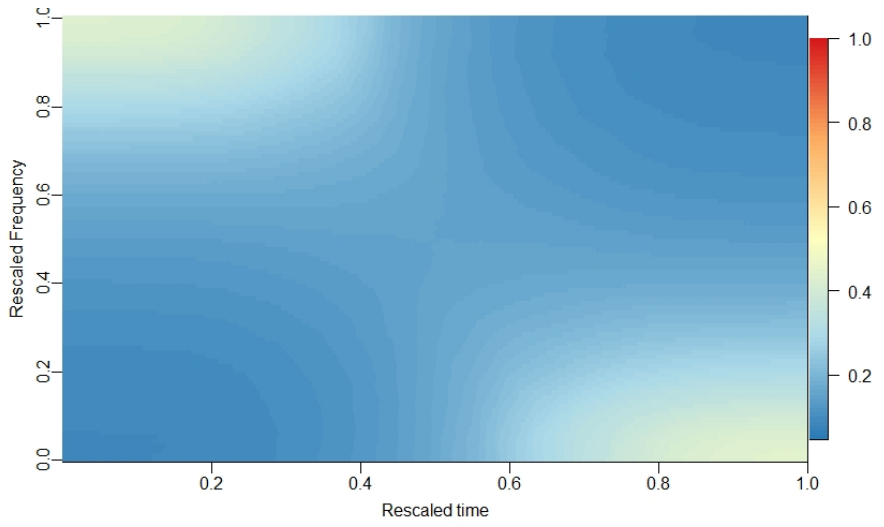

LS3a Adaptspec

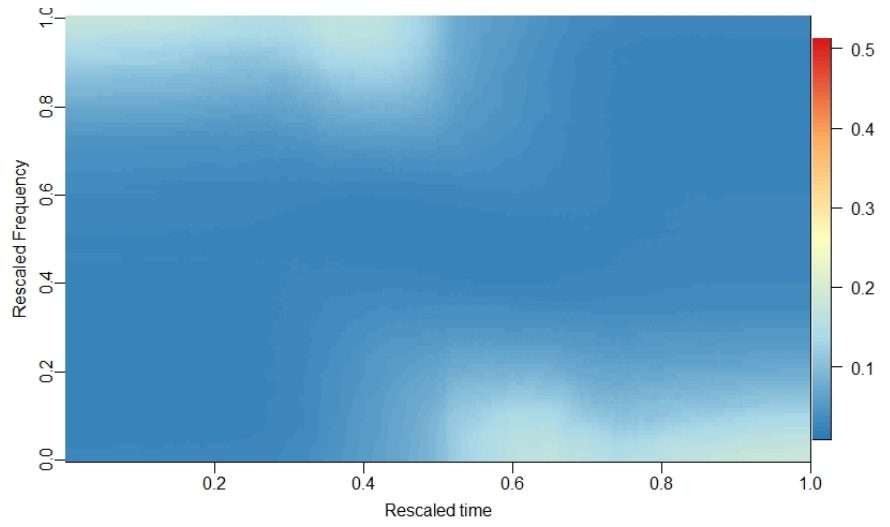

LS3a Adaptspec

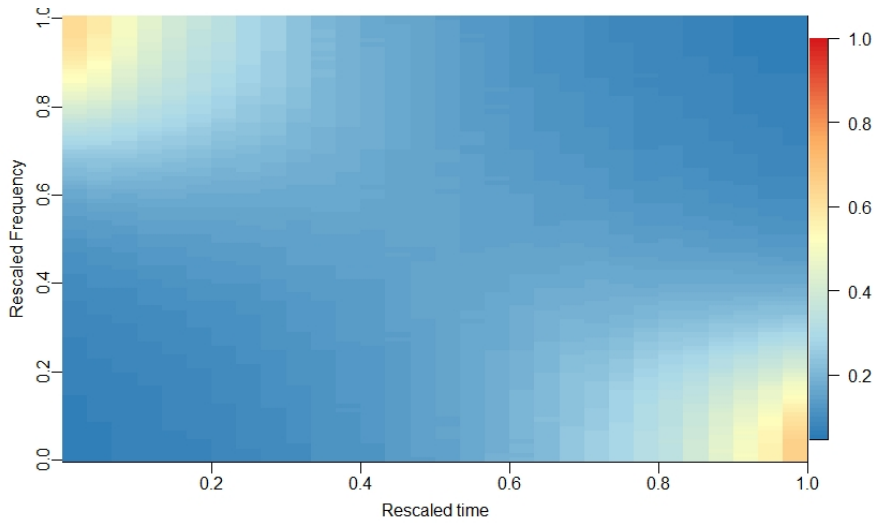

LS3a Multitaper

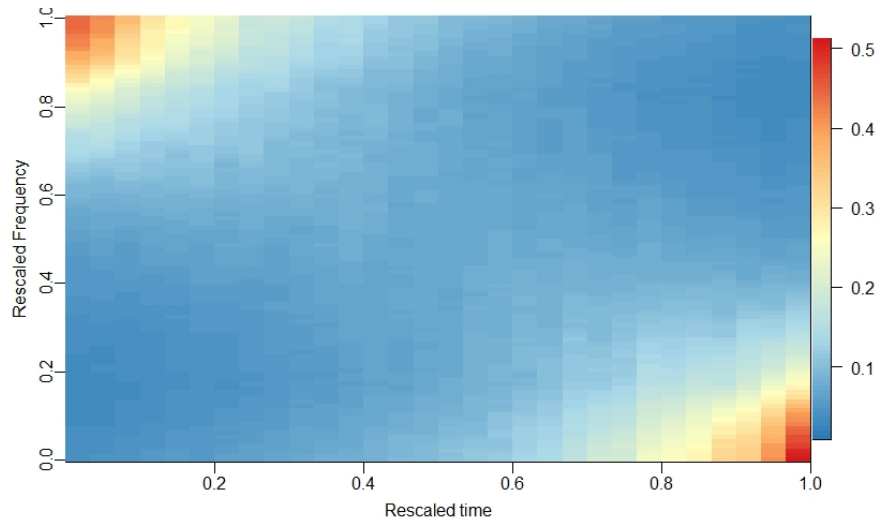

LS3a Multitaper

Figure V: Left column: The pointwise median tv-PSD estimates of all 1000 realizations of LS3a. Right column: The pointwise IQR. For each plot, the x-axis is rescaled time and the y-axis is rescaled frequency. The plots in the same column share the same color scale.

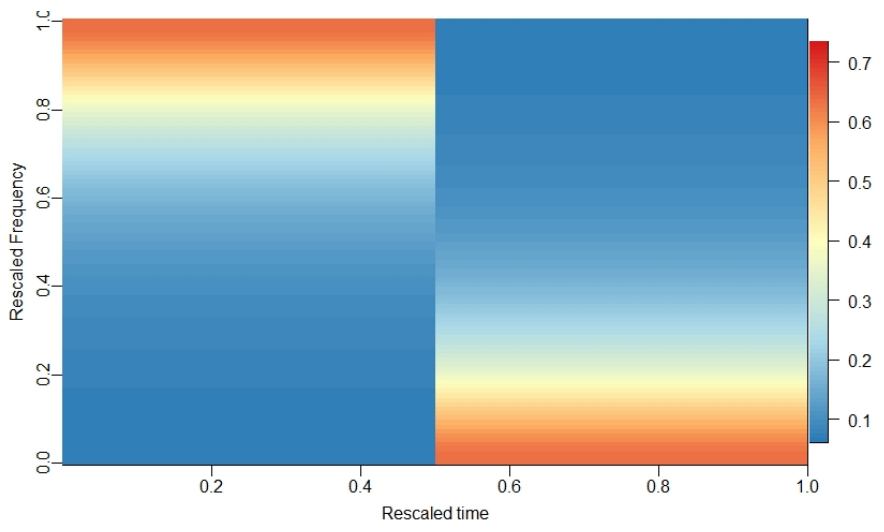

PS1a True spectrum

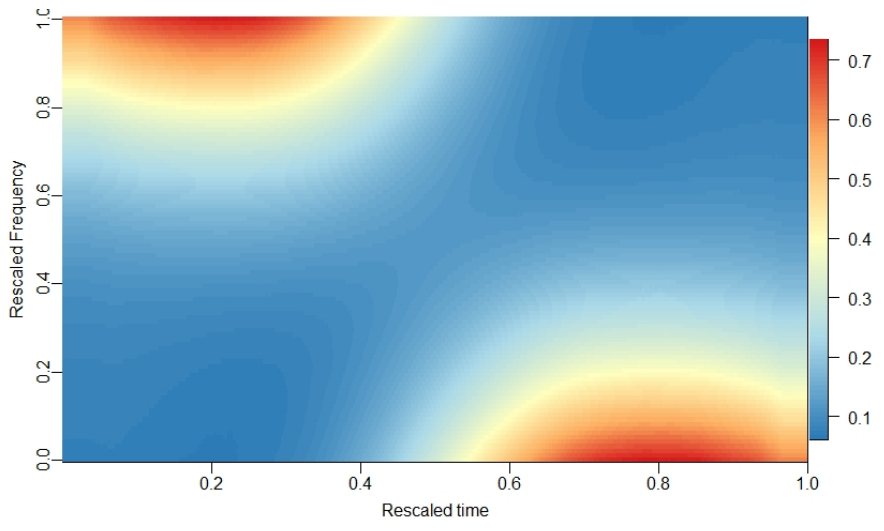

PS1a BDP-DW

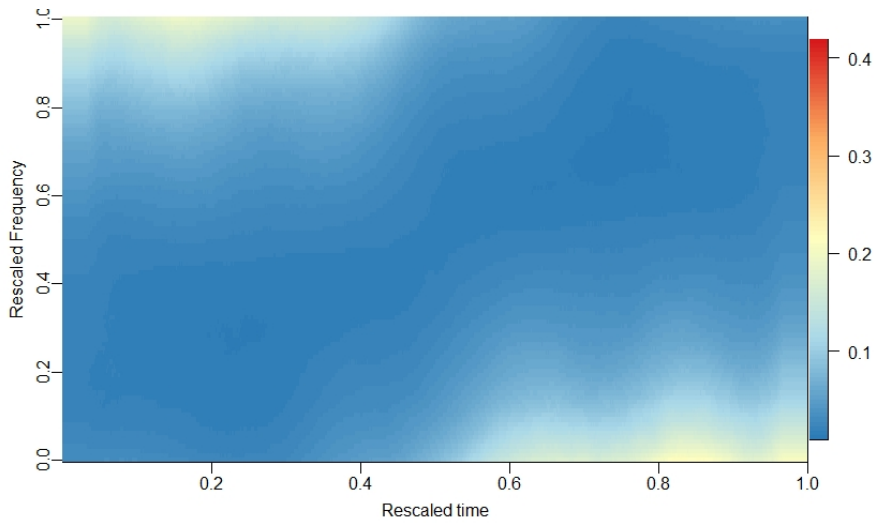

PS1a BDP-DW

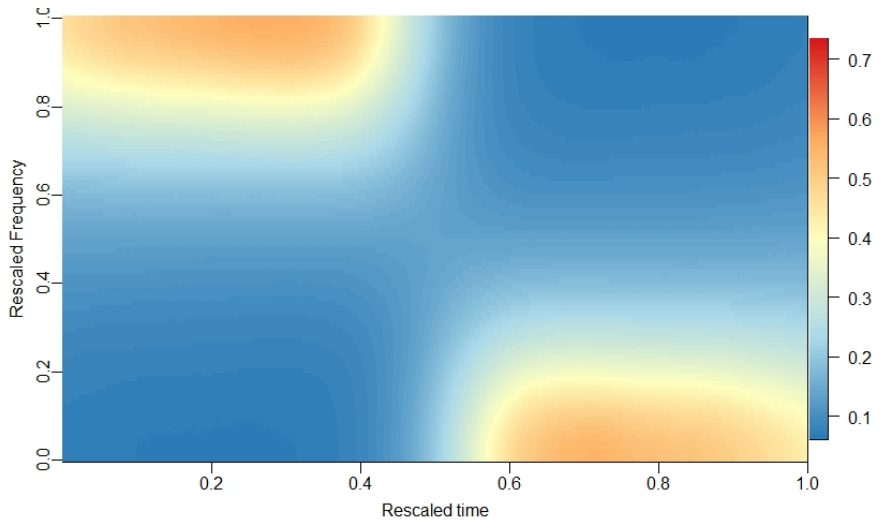

PS1a Smooth ANOVA

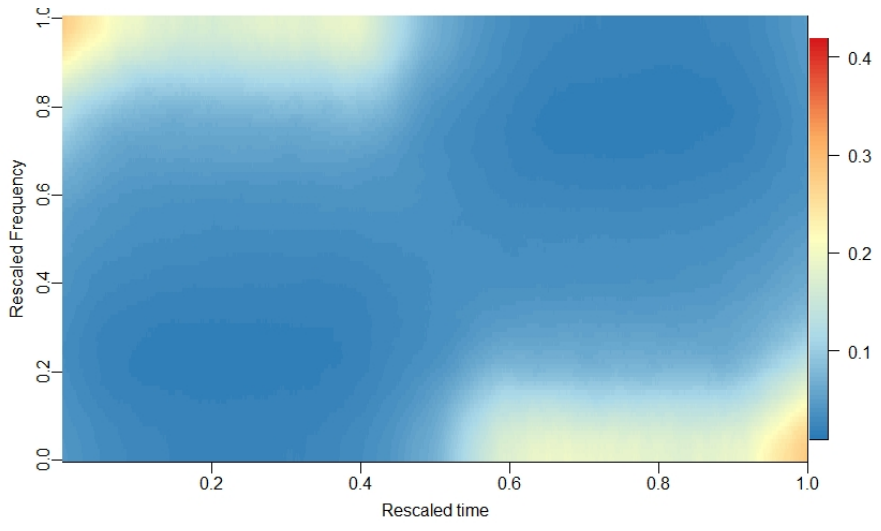

PS1a Smooth ANOVA

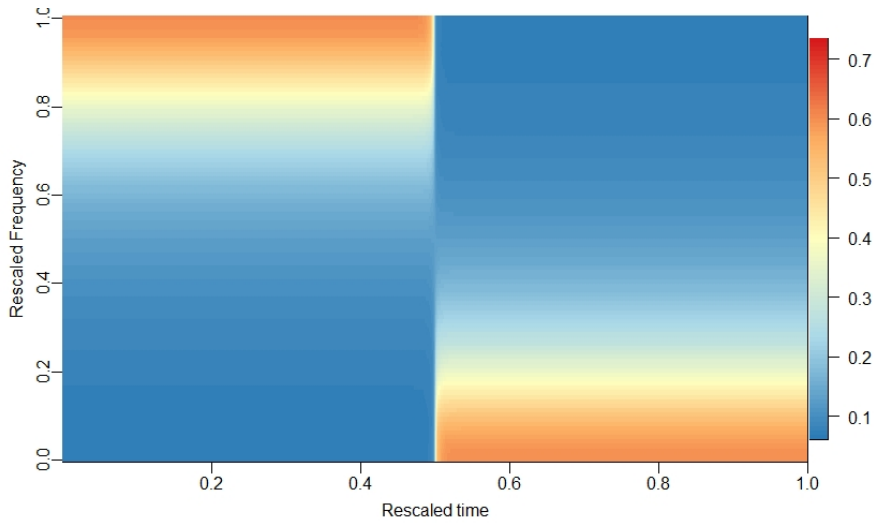

PS1a Adaptspec

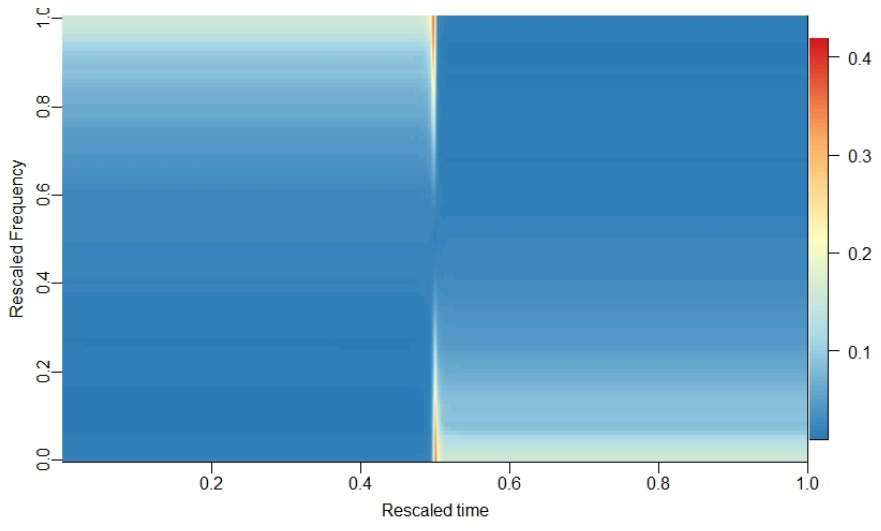

PS1a Adaptspec

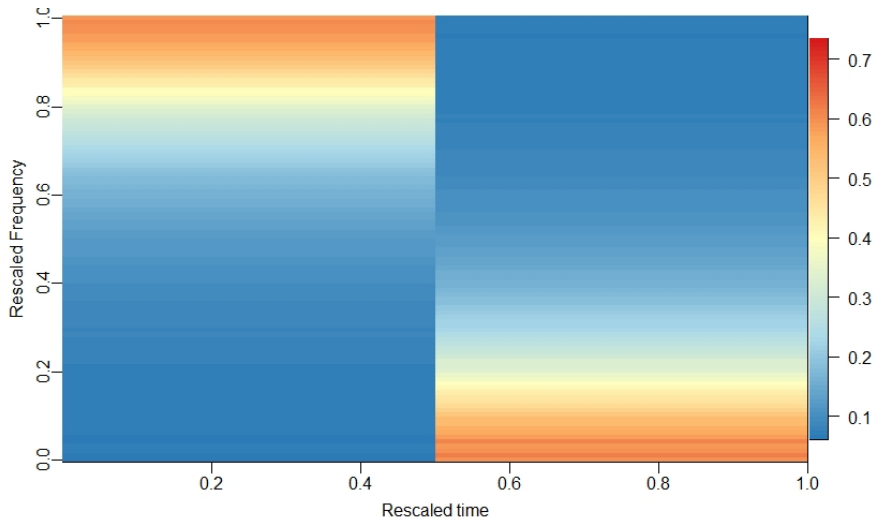

PS1a Multitaper

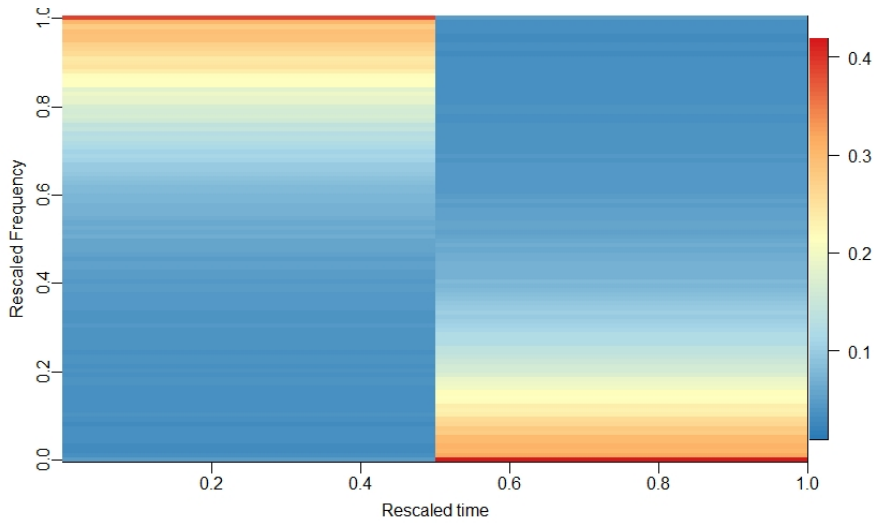

PS1a Multitaper

Figure VI: Left column: The pointwise median tv-PSD estimates of all 1000 realizations of PS1a. Right column: The pointwise IQR. For each plot, the x-axis is rescaled time and the y-axis is rescaled frequency. The plots in the same column share the same color scale.

# D Further case studies

## D.1 Gravitational wave data

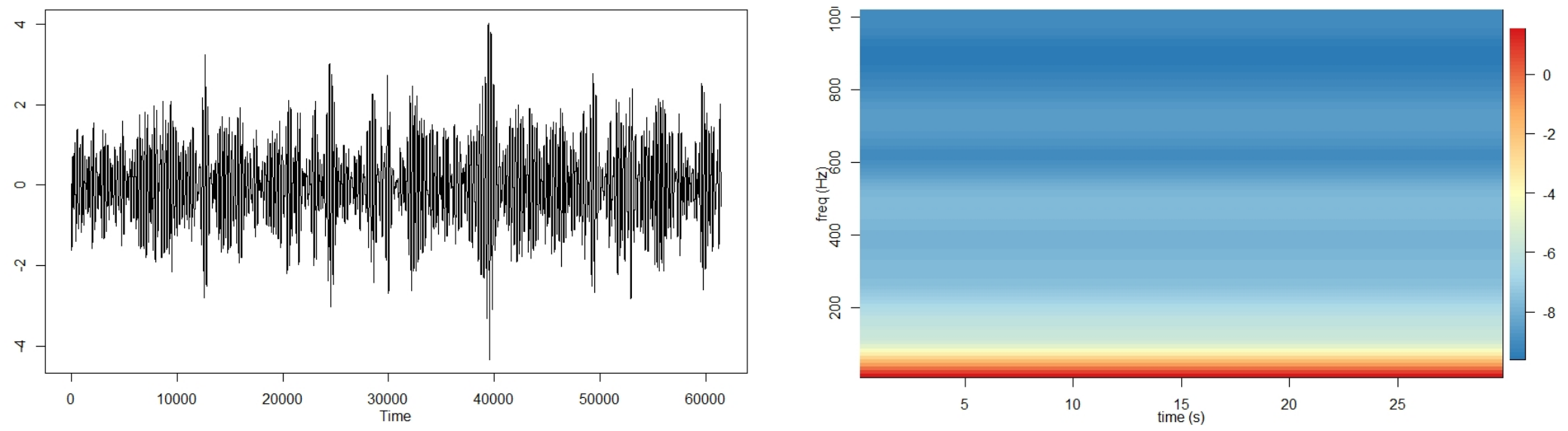

(a) LIGO data (down-sampled and standardized)

(b) Estimated log time-varying spectral density for LIGO data

Figure VII: LIGO time series and estimated time-varying spectral density

The Laser Interferometer Gravitational-Wave Observatory (LIGO) is designed to detect gravitational waves, deformations of the space-time metric that propagate at the speed of light. Predicted by Einstein's General Theory of Relativity (see Einstein (1916)) already over a hundred years ago, the actual detection of gravitational waves required the construction of extremely precise interferometers to detect relative changes in the interferometer arm lengths of the order of $10^{-21}$. For the very first detection of gravitational waves, GW150914, emitted by the inspiral of two stellar mass black holes in 2015 (Abbott et al., 2016), the founders of LIGO were awarded the Nobel Prize in Physics in 2017. Here, we consider 30 seconds of real LIGO data from detector H1 starting GPS time 1256669184 with the sampling rate at 4096 Hz collected during the third LIGO/Virgo observation run (O3), made publicly available in the Gravitational Wave Open Science Center `https://www.gw-openscience.org/`. This particular stretch of data does not contain a gravitational wave signal but only instrumental noise. LIGO instruments are impacted by various noise sources such as quantum sensing, seismic, thermal suspension, and gravity gradient noise in addition to transient anthropogenic noise. An accurate characterization of the interferometer noise in terms of its spectral density is critical for an accurate and robust extraction and estimation of signal waveforms (Christensen and Meyer, 2022). For the purpose of signal parameter estimation, it is usually assumed that the interferometer noise is stationary. While this is a reasonable assumptions for short burst signals such as those from binary black hole mergers that last only a fraction of a second, neutron star mergers can last for several minutes and it is expected that LIGO interferometer noise is slowly time-varying over periods of more than one minute but stationary for shorter stretches (Abbott et al., 2020).

We follow the usual procedures by first applying a low-pass Butterworth filter to the data and then down-sampling to 2048 Hz. Furthermore, the data are standardized before being analyzed to avoid potential numerical issues. The dynamic Whittle likelihood with thinning factor $i = 3$ and window size $m = 400$ is used in the analysis. The estimated Savage-Dickey Bayes factor of $\{k_1 = 1\}$ is 27.2802, which is far greater than 1, indicating that this segment of data may be covariance stationary. The estimated time-varying spectral density is shown in Figure VII. It is worth mentioning that although the time-varying spectral density possesses a peak around zero frequency, it does not reliably indicate long-range dependence as the LIGO detector strain channel data are not calibrated below 10 Hz during the pre-processing phase (Abbott et al., 2020).

## D.2 Daily PM 2.5 data

PM 2.5 refers to particulate matter with an aerodynamic diameter $\leqslant 2.5$ $\mu$m. It is an air pollutant which causes the air to appear hazy when the concentration levels are elevated. In this subsection, we consider the daily PM 2.5 data of Jilin Province, China, during the 10-year period November 28, 2009 – November 25, 2019. See Geng et al. (2021) for detailed information of the dataset which is continually updated and available at http://tapdata.org.cn. Prior to analysis, we removed the 29th of February in each leap year so that each year has 365 days. We also take the logarithm of the data. The log data are presented in Figure VIII (a). There are strong signs of periodicity in the time series with peaks in winter and troughs in summer. In order to remove the periodic component and (possible) time-varying trend, we use local linear regression with Gaussian kernel as described in Section 6.1 of Zhao (2015). The residuals are shown in Figure VIII (b). The de-trended time series seems to be locally stationary. We choose a thinning factor $i = 1$ and window size $m = 80$ for the BDP-DW approach. The estimated Bayes factor is zero indicating that the time series may not be stationary. Furthermore, the estimated time-varying spectral density (Figure IX) seems to suggest that the non-stationarity comes from the frequency component near zero frequency as the tv-PSD at zero frequency increases with time.

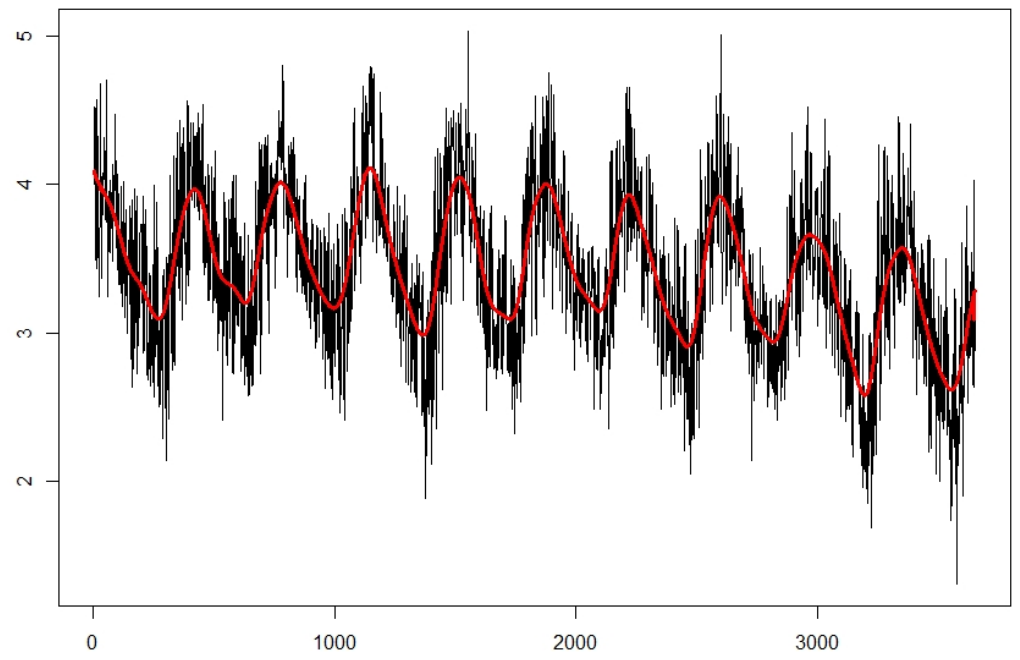

(a) The log PM 2.5 time series (with estimated trend shown in red)

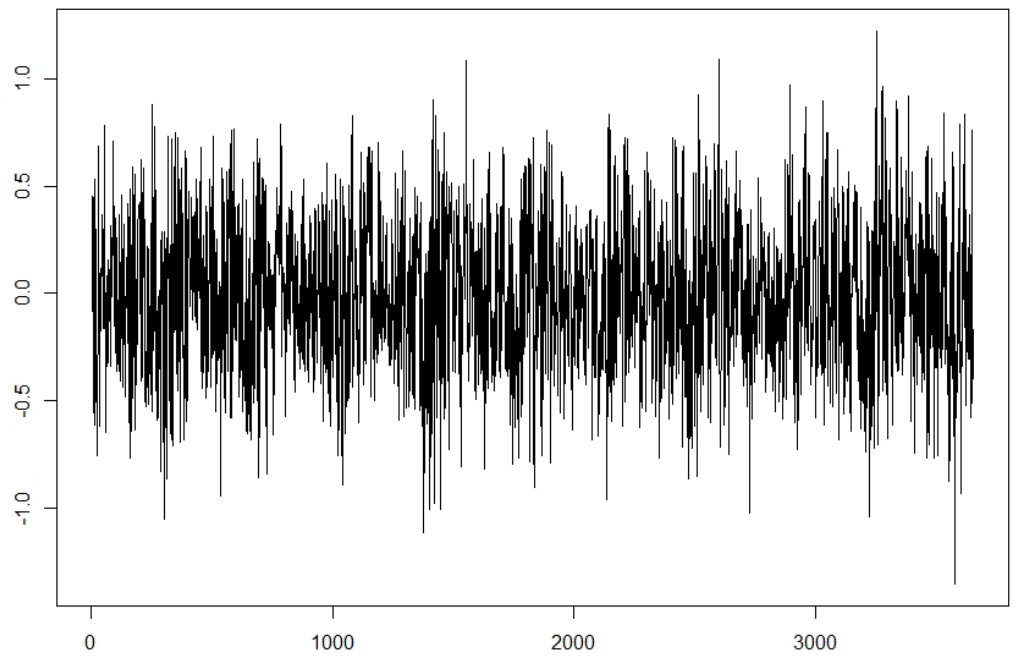

(b) The de-trended log time series

Figure VIII: Time series plots of PM 2.5 data

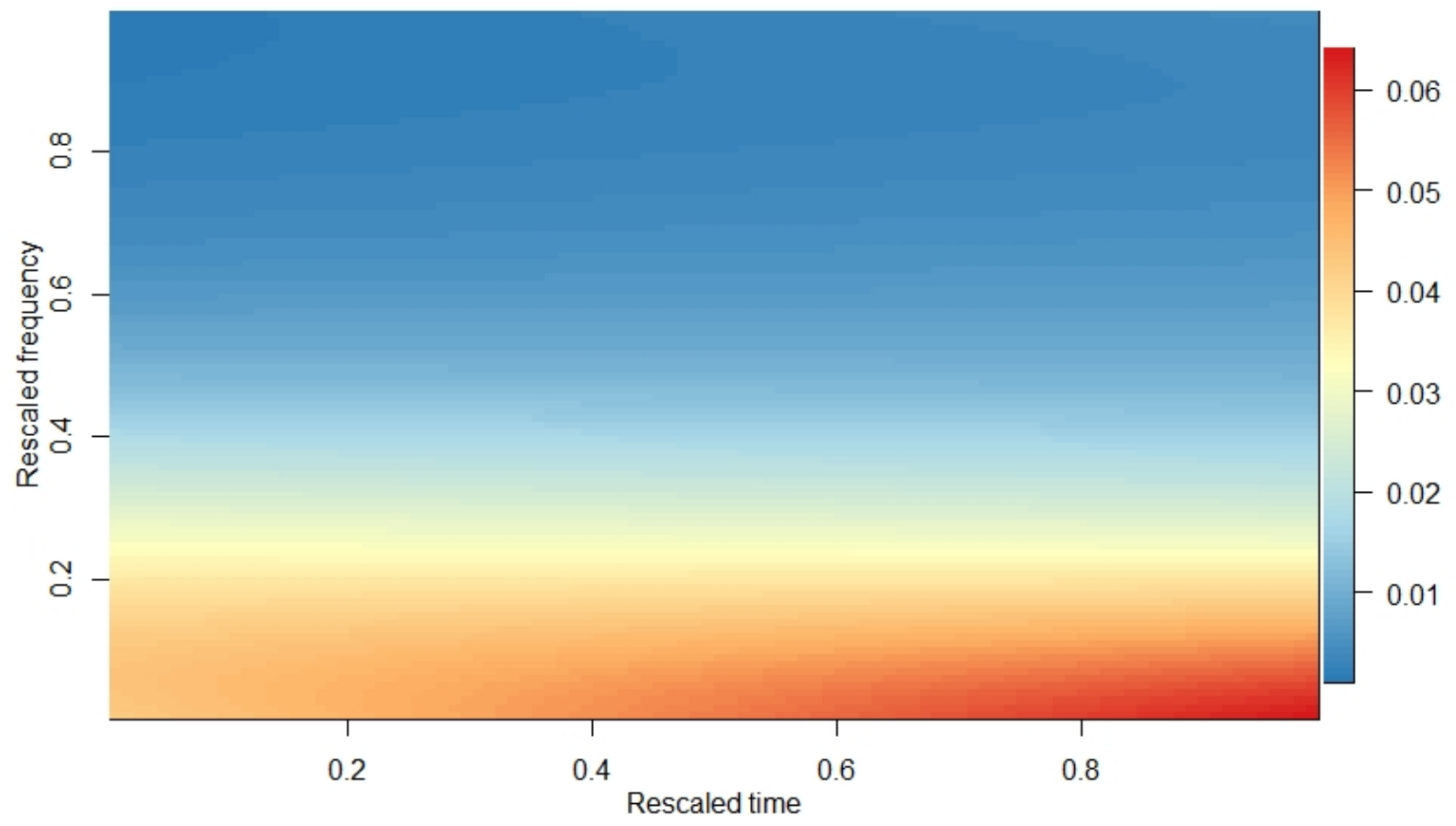

Figure IX: Estimated time-varying spectral density function of PM 2.5 data.

# E Proofs

We will now prove the results on consistency and contraction rates. To this end, we introduce the following decomposition of the likelihood in (2.4) of the main text for $i = 1, 2, 3$:

$$L_{DW}^{(i)}(f) = \exp\left[-mB_i\left(S^{(i)}(f) + D^{(i)}(f) + \Delta(f)\right)\right]\exp\left[mB_i h(f_0)\right], \tag{E.1}$$

where

$$S^{(i)}(f) = \frac{1}{mB_i}\sum_{l=1}^{B_i}\sum_{j=1}^{m}\frac{1}{f(u_{j,l,i},\lambda_j)}\left(\mathrm{MI}_{i(l-1)m+j} - \mathrm{E}\left[\mathrm{MI}_{i(l-1)m+j}\right]\right), \tag{E.2}$$

$$D^{(i)}(f) = \frac{1}{mB_i}\sum_{l=1}^{B_i}\sum_{j=1}^{m}\left(\frac{\mathrm{E}\left[\mathrm{MI}_{i(l-1)m+j}\right]}{f(u_{j,l,i},\lambda_j)} + \ln f(u_{j,l,i},\lambda_j)\right) - h(f), \tag{E.3}$$

$$\Delta(f) = h(f) - h(f_0), \tag{E.4}$$

$$h(f) = \int_0^1\int_0^1\left(\ln f(u,\lambda) + \frac{f_0(u,\lambda)}{f(u,\lambda)}\right)\mathrm{d}u\,\mathrm{d}\lambda, \tag{E.5}$$

$u_{j,l,i} = \frac{i(l-1)m+j}{T}$ is as in Definition 3 of the main text.

In the above decomposition, $S^{(i)}$ effectively captures the stochastic fluctuation of the periodogram ordinates around their expectations. We will show that such a fluctuation will vanish and obtain the corresponding rate when the sample size tends to infinity in Theorem E.2 as the main result of Section E.2. The term $D^{(i)}$ is a deterministic remainder term which will be proved to be asymptotically negligible in Theorem E.1 as a main result in Section E.1. Finally, $\Delta$ is some measure of discrepancy between $f$ and the true underlying time-varying spectral density $f_0$. By Proposition E.1 below this measure of discrepancy is equivalent to the squared $L_2$-distance between $f$ and $f_0$ (in the sense that it can be upper and lower bounded by multiples of the latter).

For better readability, we restrict the detailed discussion to the likelihood approximations as in (2.4) in the main text, where only full frequency sets of moving periodograms are being used. However, if $T/m \notin \mathbb{N}$ and the original dynamic Whittle likelihood in (2.3) in the main text is used or a version of the thinned dynamic Whittle likelihood that also takes all available moving periodograms (not only the complete sets) into account, then the required result as given in Theorem E.2 still holds. This is because by an application of the Markov inequality we easily get that the additional at most $m$ summands in (E.2) are $O_{\mathrm{P}}(m)$, such that their contribution to the whole sum (uniformly over $f$) is at most $O_{\mathrm{P}}(m/T)$ which is of smaller order than the rate given in Theorem E.2. Similarly, the additional factors in (E.3) are asymptotically negligible (compare with Theorem E.1).

To derive asymptotic properties of (E.2) and (E.3), we require the following properties of moving periodograms obtained by Häfner and Kirch (2017).

**Proposition P.1.** Under Assumptions $\mathcal{A}.1$ and $\mathcal{A}.2$ in the main text, it holds:

(a) $\mathrm{MI}_{i(l-1)m+j} = f_0(u_{j,l,i}, \lambda_j)\, \mathrm{MI}^{\varepsilon}_{i(l-1)m+j} + R_{j,l,i}$, where $\mathrm{MI}^{\varepsilon}_t = \frac{1}{2m+1}\left|\sum_{\nu=0}^{2m} \varepsilon_{\nu+t-m} \exp(-i\pi\nu\lambda_{\mathrm{mod}(t)})\right|^2$ is the $2\pi$-scaled moving peridogram of the corresponding i.i.d. sequence $\{\varepsilon_t\}_{t\in\mathbb{Z}}$ in Definition S.1 (a) and $R_{j,l,i}$ satisfies $\sup_{l=1,\ldots,B_i} \sup_{j=1,\cdots,m} \mathrm{E}|R_{j,l,i}|^2 = O\left(m^{-1}\right)$ as $T \to \infty$.

(b) For any $|t_1-t_2| \geqslant 2m+1$, $\mathrm{MI}^{\varepsilon}_{t_1}$ and $\mathrm{MI}^{\varepsilon}_{t_1}$ are independent such that $\mathrm{cov}(\mathrm{MI}^{\varepsilon}_{t_1}, \mathrm{MI}^{\varepsilon}_{t_2}) = 0$. Also, for any $l_1, l_2 = 1, \cdots, \lfloor \frac{T}{m} \rfloor$ and $j = 1, \cdots, m$, $\mathrm{MI}^{\varepsilon}_{(l_1-1)m+j}$ and $\mathrm{MI}^{\varepsilon}_{(l_2-1)m+j}$ have the same distribution.

(c) $\mathrm{E}\left[\mathrm{MI}^{\varepsilon}_{(l-1)m+j}\right] = 1$, $\mathrm{Var}(\mathrm{MI}^{\varepsilon}_{(l-1)m+j}) = 1 + \frac{\mathrm{E}\varepsilon_1^4-3}{2m+1} = O(1)$ for any $l = 1, \cdots, B_1$ and $j = 1, \cdots, m$. Furthermore, $\sup_{l=1,\cdots,B_1}\left|\mathrm{cov}\left(\mathrm{MI}^{\varepsilon}_{(l-1)m+j_1}, \mathrm{MI}^{\varepsilon}_{(l-1)m+j_2}\right)\right| = O\left(\frac{1}{m}\right) + O\left(\frac{1}{|j_1-j_2|^2}\right)$ for $j_1 \neq j_2$.

*Proof.* All assertions follow from various results of Häfner and Kirch (2017): (a) is a direct consequence of Corollary 2.1 (on noting that $2\pi\mathrm{MI}_{i(l-1)m+j} = \mathrm{MI}_{\lfloor u_{j,l,i}T\rfloor}(\lambda_j)$, where the right-hand side is the moving local periodogram defined in (2.3) there), (b) is due to the construction of the moving Fourier coefficients (cf. Definition 2.1 there). Finally, (c) is shown in the proof of Theorem 2.3 (equations (5.10), (5.6), (5.7) and (5.8) there). □

## E.1 Auxiliary deterministic results

In this section we derive several results about functions that will be used in the proofs. As a main result, we obtain that the term $D^{(i)}$ in (E.3) is asymptotically negligible.

We start with the following proposition that shows the equivalence of the measure of discrepancy $\Delta(f)$ as in (E.4) to the squared $L_2$-norm.

**Proposition E.1.** *For any $\varphi, \psi \in C\left([0,1]^2\right)$ which share a common lower bound $a > 0$, we have*

$$\frac{(\varphi-\psi)^2}{2\max\{\|\varphi\|_\infty, \|\psi\|_\infty\}^2} \leqslant \frac{\varphi}{\psi} - \ln\frac{\varphi}{\psi} - 1 \leqslant \frac{1}{a^2}(\varphi-\psi)^2.$$

*In particular, we get for any $f \in \Theta$,*

$$\frac{1}{2\,M_0^2}\,\|f-f_0\|_2^2 \leqslant \Delta(f) \leqslant \frac{1}{\delta^2}\,\|f-f_0\|_2^2 \leqslant \frac{1}{\delta^2}\|f-f_0\|_\infty^2$$

*with $\Delta(f)$ as in (E.4).*

*Proof.* First, observe that

$$\frac{\varphi}{\psi} - \ln\frac{\varphi}{\psi} - 1 = \frac{\varphi-\psi}{\psi} - \ln\left(1 + \frac{\varphi-\psi}{\psi}\right).$$

Since $\ln(1+x) \geqslant \frac{x}{1+x}$ for any $x > -1$, we have

$$\frac{\varphi-\psi}{\psi} - \ln\left(1+\frac{\varphi-\psi}{\psi}\right) \leqslant \frac{\left(\frac{\varphi-\psi}{\psi}\right)^2}{1+\frac{\varphi-\psi}{\psi}} = \frac{(\varphi-\psi)^2}{\varphi\psi} \leqslant \frac{1}{a^2}(\varphi-\psi)^2,$$

proving the second inequality of the proposition. As for the first inequality, we know from Taylor's theorem that $\ln(1+x) = x - \frac{x^2}{2(1+b_x x)^2}$ for some $b_x \in (0,1)$. Therefore, for any $\varphi, \psi$ and $(u,v) \in [0,1]^2$, there exist $b \in (0,1)$ such that

$$\frac{\varphi(u,v)-\psi(u,v)}{\psi(u,v)} - \ln\left(1+\frac{\varphi(u,v)-\psi(u,v)}{\psi(u,v)}\right) = \frac{\left(\varphi(u,v)-\psi(u,v)\right)^2}{2\left[\psi(u,v)+b\left(\varphi(u,v)-\psi(u,v)\right)\right]^2}.$$

From this, the assertion follows by

$$\psi(u,v)+b\left(\varphi(u,v)-\psi(u,v)\right) \geqslant \min\{\varphi(u,v),\psi(u,v)\} \geqslant a > 0$$

and

$$\psi(u,v)+b\left(\varphi(u,v)-\psi(u,v)\right) \leqslant \max\{\varphi(u,v),\psi(u,v)\} \leqslant \max\{\|\varphi\|_\infty,\|\psi\|_\infty\}.$$

□

The next proposition gives a rate of convergence between Riemann sums and the corresponding integral. It is needed in the proof of Theorem E.1 to provide an upper bound for the difference between $h(f)$ and the corresponding discrete sum in $D^{(i)}(f)$ as in (E.3).

**Proposition E.2.** *Let $\varphi \in C^1\left([0,1]^2\right)$ with*

$$\|\varphi\|_\infty + \|\partial_1\varphi\|_\infty + \|\partial_2\varphi\|_\infty \leqslant M$$

*for some $M > 0$. Then, with the notation in Definition 3 of the main text,*

$$\left|\frac{1}{mB_i}\sum_{l=1}^{B_i}\sum_{j=1}^{m}\varphi\left(u_{j,l,i},\lambda_j\right) - \int_0^1\int_0^1 \varphi(x,y)\,\mathrm{d}x\,\mathrm{d}y\right| = O\left(\frac{m}{T}+\frac{1}{m}\right)\cdot M,$$

*where the constants in the $O$-term do not depend on $\varphi$.*

*Proof.* By $u_{1,1,i} = 1/T$, $1-u_{1,B_i,i} = O(m/T)$ and $1-\lambda_m = O(1/m)$, we get with $\lambda_0 = 0$

$$\int_0^1\int_0^1 \varphi(x,y)\,\mathrm{d}x\ \mathrm{d}y = \int_{u_{1,1,i}}^{u_{1,B_i,i}}\int_{\lambda_0}^{\lambda_m}\varphi(x,y)\,\mathrm{d}x\,\mathrm{d}y + O\left(\frac{m}{T}+\frac{1}{m}\right)\cdot M.$$

Because $1/[(u_{1,2,i}-u_{1,1,i})\cdot B_i] = 1+O(m/T)$ and $1/[(\lambda_1-\lambda_0)\,m] = 1+O(1/m)$, we

further get

$$\frac{1}{mB_i}\sum_{l=1}^{B_i}\sum_{j=1}^{m}\varphi\left(u_{j,l,i},\lambda_j\right)=\frac{1}{mB_i}\sum_{l=1}^{B_i-1}\sum_{j=1}^{m}\varphi\left(u_{j,l,i},\lambda_j\right)+O\left(\frac{1}{B_i}\right)\cdot M$$
$$=\left(u_{1,2,i}-u_{1,1,i}\right)\left(\lambda_1-\lambda_0\right)\sum_{l=1}^{B_i-1}\sum_{j=1}^{m}\varphi\left(u_{j,l,i},\lambda_j\right)+O\left(\frac{m}{T}+\frac{1}{m}\right)\cdot M.$$

Consequently, it is sufficient to consider

$$\left|\int_{u_{1,1,i}}^{u_{1,B_i,i}}\int_{\lambda_0}^{\lambda_m}\varphi(x,y)\,\mathrm{d}x\,\mathrm{d}y-\left(u_{1,2,i}-u_{1,1,i}\right)\left(\lambda_1-\lambda_0\right)\sum_{l=1}^{B_i-1}\sum_{j=1}^{m}\varphi\left(u_{j,l,i},\lambda_j\right)\right|$$
$$\leqslant\sum_{l=1}^{B_i-1}\sum_{j=1}^{m}\int_{u_{1,l,i}}^{u_{1,l+1,i}}\int_{\lambda_{j-1}}^{\lambda_j}\left|\varphi(x,y)-\varphi(u_{j,l,i},\lambda_j)\right|\,\mathrm{d}x\,\mathrm{d}y.$$

Furthermore, by $\lambda_j-\lambda_{j-1}\leqslant 1/m$ and $u_{1,l+1,i}-u_{1,l,i}=im/T$ and $m/T=O(1/B_i)$, we get by the mean value theorem that

$$\int_{u_{1,l,i}}^{u_{1,l+1,i}}\int_{\lambda_{j-1}}^{\lambda_j}\left|\varphi(x,y)-\varphi(u_{j,l,i},\lambda_j)\right|\,\mathrm{d}x\,\mathrm{d}y$$
$$\leqslant\left\|\partial_1\varphi\right\|_\infty\left(\lambda_j-\lambda_{j-1}\right)\int_{u_{1,l,i}}^{u_{1,l+1,i}}\left|x-u_{j,l,i}\right|\,\mathrm{d}x$$
$$+\left\|\partial_2\varphi\right\|_\infty\left(u_{1,l+1,i}-u_{1,l,i}\right)\int_{\lambda_{j-1}}^{\lambda_j}\left|y-\lambda_j\right|\,\mathrm{d}y$$
$$\leqslant\left\|\partial_1\varphi\right\|_\infty\left(\lambda_j-\lambda_{j-1}\right)\left(u_{1,l+1,i}-u_{1,l,i}\right)^2+\left\|\partial_2\varphi\right\|_\infty\left(u_{1,l+1,i}-u_{1,l,i}\right)\left(\lambda_j-\lambda_{j-1}\right)^2$$
$$=\frac{1}{mB_i}\,O\left(\frac{m}{T}+\frac{1}{m}\right)\cdot M.$$

Therefore,

$$\sum_{l=1}^{B_i-1}\sum_{j=1}^{m}\int_{u_{1,l,i}}^{u_{1,l+1,i}}\int_{\lambda_{j-1}}^{\lambda_j}\left|\varphi(x,y)-\varphi(u_{j,l,i},\lambda_j)\right|\,\mathrm{d}x\,\mathrm{d}y=O\left(\frac{m}{T}+\frac{1}{m}\right)\cdot M,$$

completing the proof of the assertion. □

We can now conclude that the deterministic remainder term $D^{(i)}(f)$ as in (E.3) is indeed asymptotically negligible.

**Theorem E.1.** *Under Assumptions $\mathcal{A}.1$ and $\mathcal{A}.2$, we have*

$$\sup_{f\in\Theta}\left|D^{(i)}(f)\right| = O\left(\frac{1}{\sqrt{m}}\right),$$

*where $D^{(i)}$ is defined in (E.3).*

*Proof.* First, consider $\varphi(u,\lambda)$ with $\varphi = \ln f + \frac{f_0}{f}$ for $f \in \Theta$ as defined in (3.1) in the main text. Then, there exists $M > 0$ (only depending on $\delta$ and $M_j$, $j = 0,1,2$) such that

$$\|\varphi\|_\infty + \|\partial_1\varphi\|_\infty + \|\partial_2\varphi\|_\infty \leqslant M.$$

Indeed,

$$\left|\ln f(u,\lambda) + \frac{f_0(u,\lambda)}{f(u,\lambda)}\right| \leqslant \max(|\ln M_0|, |\ln\delta|) + \frac{M_0}{\delta},$$

$\partial_1\left(\ln f + \frac{f_0}{f}\right) = \frac{1}{f}\partial_1 f + \frac{1}{f}\partial_1 f_0 - \frac{f_0}{f^2}\partial_1 f$ is bounded in absolute value by $2M_1/\delta + M_0M_1/\delta^2$ and the assertion for the partial derivative with respect to the second variable follows similarly.

Thus by Propositions P.1 and E.2, the definition of $\Theta$ in (3.1) in the main text and the Cauchy-Schwarz inequality, we have

$$\begin{aligned}
&\sup_{f\in\Theta}\left|D^{(i)}(f)\right| \\
&\leqslant \sup_{f\in\Theta}\Bigg|\int_0^1\int_0^1\left(\ln f(u,\lambda) + \frac{f_0(u,\lambda)}{f(u,\lambda)}\right)\mathrm{d}\lambda\,\mathrm{d}u \\
&\qquad\qquad - \frac{1}{mB_i}\sum_{l=1}^{B_i}\sum_{j=1}^{m}\left(\ln f(u_{j,l,i},\lambda_j) + \frac{f_0(u_{j,l,i},\lambda_j)}{f(u_{j,l,i},\lambda_j)}\right)\Bigg| \\
&\quad + \sup_{f\in\Theta}\frac{1}{mB_i}\sum_{l=1}^{B_i}\sum_{j=1}^{m}\frac{\mathrm{E}\,|R_{j,l,i}|}{f(u_{l,i},\lambda_j)} \\
&= O\left(\frac{m}{T} + \frac{1}{m}\right) + O\left(\frac{1}{\sqrt{m}}\right) = O\left(\frac{1}{\sqrt{m}}\right)
\end{aligned}$$

by Assumption $\mathcal{A}.1$. $\square$

The last result of this section is about a compact superset of the parameter space

$\Theta$. This compact set is useful in the next section for proving the measurability as well as the final proof of posterior consistency.

**Proposition E.3.** *(a) We have $\Theta \subset \Gamma$, where $\Theta$ is defined in (3.1) in the main text and*

$$\Gamma := \left\{ f \in \mathrm{Lip}\left([0,1]^2\right) : f \geqslant \delta, \|f\|_\infty \leqslant M_0, L(f) \leqslant \sqrt{2}\max\{M_1, M_2\} \right\}, \quad \text{(E.6)}$$

$$L(f) := \sup_{(x_1,y_1)\neq(x_2,y_2)\in[0,1]^2} \left| \frac{f(x_1,y_1) - f(x_2,y_2)}{\sqrt{(x_1-x_2)^2 + (y_1-y_2)^2}} \right| .$$

*Moreover, when equipped with the uniform metric $d_\infty$, $\Gamma$ is a compact metric space.*

*(b) For any $r > 0$, $\inf_{f\in\Gamma\setminus\mathcal{U}(f_0,r)} \|f - f_0\|_2 > 0$, where*

$$\mathcal{U}(h,r) = \{f \in \Gamma : \|f - h\|_\infty < r\} \quad \text{(E.7)}$$

*and $\|\cdot\|_2$ is the usual $L^2$ norm (see e.g. (3.3) in the main text for a definition).*

*Proof.* First, $\Theta \subset \Gamma$ follows from Proposition 2.2.5 of Cobzaş et al. (2019). The closeness of $\Gamma$ with respect to $d_\infty$ is implied by Proposition 2.4.1 of Cobzaş et al. (2019) and the fact that limits preserve inequalities. The total boundedness of $\Gamma$ with respect to $d_\infty$ is a direct consequence of Theorem 2.7.1 of van der Vaart and Wellner (1996). Since $\Gamma$ is both closed and totally bounded, it is compact.

As for (b), note that $\Gamma \setminus \mathcal{U}(f_0, r) = \Gamma \cap \left\{f \in C\left([0,1]^2\right) : \|f - f_0\|_\infty < r\right\}^c$, where the complement is taken with respect to $C\left([0,1]^2\right)$. This means that $\Gamma \setminus \mathcal{U}(g_0, r)$, as the intersection of two closed sets with respect to $d_\infty$, is again closed. The closeness, combined with part (a), yields the compactness of $\Gamma \setminus \mathcal{U}(f_0, r)$. The continuity of the mapping $f \mapsto \|f - f_0\|_2$ with respect to $d_\infty$ can be seen from the following inequalities:

$$\left|\|f - f_0\|_2 - \|g - f_0\|_2\right| \leqslant \|f - g\|_2 \leqslant d_\infty(f, g),$$

where $f, g \in C\left([0,1]^2\right)$ are arbitrary. Therefore, there exists $\varphi \in \Gamma \setminus \mathcal{U}(f_0, r)$ such that

$$\inf_{f\in\Gamma\setminus\mathcal{U}(f_0,r)} \|f - f_0\|_2 = \|\varphi - f_0\|_2 > 0,$$

where the last inequality is due to the fact that $\varphi \neq f_0$ and $\varphi, f_0 \in C\left([0,1]^2\right)$. $\square$

## E.2 Concentration inequality and related results

The aim of this section is to prove a concentration inequality for the stochastic term (E.2) above. We will now prove several propositions from which we can then derive the concentration inequalities for the stochastic terms (E.2) in Theorem E.2. To this end, in view of the decomposition of the periodogram as in Proposition P.1 (a), we first consider the following factor-6-thinned version of the sum as in $S^{(1)}$ defined in (E.2):

$$A_T(f) := A_T(f; \{X_t\}_{t=-m+1}^{T+m}; \{v_{j,l}\}) := \frac{1}{mB}\sum_{l=1}^{B}\sum_{j=1}^{m}\frac{f_0(v_{j,l},\lambda_j)\mathrm{MI}^{\varepsilon}_{6(l-1)m+j}}{f(v_{j,l},\lambda_j)}, \quad \text{(E.8)}$$

$$R_T(f) := R_T(f; \{X_t\}_{t=-m+1}^{T+m}; \{v_{j,l}\}) := \frac{1}{mB}\sum_{l=1}^{B}\sum_{j=1}^{m}\frac{R_{j,6l-5,1}}{f(v_{j,l},\lambda_j)}, \quad \text{(E.9)}$$

where $B = B_{1,6} = \left\lceil \frac{T-m}{6m} \right\rceil$, $v_{j,l} = v_{j,l,1,6} = \frac{6(l-1)m+j}{T}$. The term $R_{j,6l-5,1}$ is due to $6(l-1)m+j = 1\cdot[(6l-5)-1]m+j$.

We can now first derive some concentration inequality for $A_T(f)$:

**Proposition E.4.** *Under Assumptions $\mathcal{A}$.1 and $\mathcal{A}$.2, for any positive function $f$ on $[0,1]^2$ with lower bound $\delta > 0$ and any $a > 0$, we have*

$$\mathrm{P}\left(|A_T(f) - \mathrm{E}A_T(f)| \geqslant a\right) = O\left(\frac{1}{a^2T}\right),$$

*where the multiplicative constant for the right-hand side depends only on $\delta$ and the underlying tv-PSD $f_0$. Specifically, the above result, in combination with Proposition P.1 (c), yields*

$$|A_T(f_0) - \mathrm{E}A_T(f_0)| = \left|\frac{1}{mB}\sum_{l=1}^{B}\sum_{j=1}^{m}\mathrm{MI}^{\varepsilon}_{6(l-1)m+j} - 1\right| = o_{\mathrm{P}}(1).$$

*Proof.* By the Chebyshev's inequality,

$$\mathrm{P}\left(|A_T(f) - \mathrm{E}A_T(f)| \geqslant a\right) \leqslant \frac{\mathrm{Var}\left(A_T(f)\right)}{a^2}$$
$$\leqslant \frac{1}{a^2}\frac{1}{m^2B^2}\sum_{l_1=1}^{B}\sum_{l_2=1}^{B}\sum_{j_1=1}^{m}\sum_{j_2=1}^{m}\Bigg[$$

$$\frac{f_0(v_{j_1,l_1},\lambda_{j_1})}{f(v_{j_1,l_1},\lambda_{j_1})}\frac{f_0(v_{j_2,l_2},\lambda_{j_2})}{f(v_{j_2,l_2},\lambda_{j_2})}\left|\mathrm{cov}\left(\mathrm{MI}^{\varepsilon}_{6(l_1-1)m+j_1},\mathrm{MI}^{\varepsilon}_{6(l_2-1)m+j_2}\right)\right|\Bigg]$$
$$= O(1)\frac{1}{a^2}\frac{1}{m^2B^2}\sum_{l_1=1}^{B}\sum_{l_2=1}^{B}\sum_{j_1=1}^{m}\sum_{j_2=1}^{m}\left|\mathrm{cov}\left(\mathrm{MI}^{\varepsilon}_{6(l_1-1)m+j_1},\mathrm{MI}^{\varepsilon}_{6(l_2-1)m+j_2}\right)\right|.$$

Note that $|l_1-l_2|\geqslant 1$ implies $|6(l_1-l_2)m+j_1-j_2|>2m+1$, which in turn results in $\mathrm{cov}\left(\mathrm{MI}^{\varepsilon}_{6(l_1-1)m+j_1},\mathrm{MI}^{\varepsilon}_{6(l_2-1)m+j_2}\right)=0$ due to Proposition P.1 (b). This fact, combined with Proposition P.1 (c), yields

$$\frac{1}{m^2B^2}\sum_{l_1=1}^{B}\sum_{l_2=1}^{B}\sum_{j_1=1}^{m}\sum_{j_2=1}^{m}\left|\mathrm{cov}\left(\mathrm{MI}^{\varepsilon}_{6(l_1-1)m+j_1},\mathrm{MI}^{\varepsilon}_{6(l_2-1)m+j_2}\right)\right|$$
$$=\frac{1}{m^2B^2}\sum_{l=1}^{B}\sum_{j_1=1}^{m}\sum_{j_2=1}^{m}\left|\mathrm{cov}\left(\mathrm{MI}^{\varepsilon}_{6(l-1)m+j_1},\mathrm{MI}^{\varepsilon}_{6(l-1)m+j_2}\right)\right|$$
$$=O(1)\frac{1}{m^2B^2}\left(Bm+B\sum_{j_1=1}^{m}\sum_{\substack{j_2=1\\ j_2\neq j_1}}^{m}\frac{1}{|j_1-j_2|^2}\right)=O\left(\frac{1}{mB}\right)=O\left(\frac{1}{T}\right)$$

as

$$\sum_{j_1=1}^{m}\sum_{\substack{j_2=1\\ j_2\neq j_1}}^{m}\frac{1}{|j_1-j_2|^2}=\sum_{|j|=1}^{m-1}\frac{m-|j|}{j^2}=O(m). \tag{E.10}$$

□

We need the following law-of-large-numbers-type result to obtain a uniform concentration inequality in Proposition E.6.

**Proposition E.5.** *Under Assumptions $\mathcal{A}$.1 and $\mathcal{A}$.2, we have as $T\to\infty$*

$$\frac{1}{B}\sum_{l=1}^{B}\left(\frac{1}{m}\sum_{j=1}^{m}\mathrm{MI}^{\varepsilon}_{6(l-1)m+j}\right)^2\to 1 \text{ in P-probability.}$$

*Proof.* Let $Y_{l,B}=\frac{1}{m}\sum_{j=1}^{m}\mathrm{MI}^{\varepsilon}_{6(l-1)m+j}$, $l=1,\cdots,B$. We know from Proposition P.1 (b) that $Y_{1,B},\cdots,Y_{B,B}$ are independent and identically distributed. By Proposition P.1

(c) and direct calculation, we have by (E.10) as $T \to \infty$

$$\mathrm{E}Y_{1,B} = 1, \mathrm{Var}\left(Y_{1,B}\right) = O\left(\frac{1}{m}\right) + O\left(\frac{1}{m^2}\right) \sum_{j_1=1}^{m} \sum_{\substack{j_2=1 \\ j_2 \neq j_1}}^{m} \frac{1}{(j_1 - j_2)^2} = O\left(\frac{1}{m}\right) \to 0.$$

Consequently, $Y_{1,B} \to 1$ in P-probability.

Denote $Z_{l,B} = Y_{l,B}^2$, $l = 1, \cdots, B$. We will employ a general version of the Kolmogorov-Feller weak law of large numbers for triangular arrays (e.g. Theorem 2.2.11 of Durrett (2019)) to show $\frac{1}{B}\sum_{l=1}^{B} Z_{l,B} \to 1$ in P-probability by verifying the two conditions given there. The first condition is verified by using the Chebshev's inequality and the fact that $Z_{l,B}$, $l = 1, \cdots, B$ are identically distributed:

$$\begin{aligned}
&\sum_{l=1}^{B} \mathrm{P}\left(Z_{l,B} > B\right) = B\,\mathrm{P}\left(Z_{1,B} > B\right) \\
&= B\,\mathrm{P}\left(Y_{1,B} > \sqrt{B}\right) \leqslant B\,\mathrm{P}\left(|Y_{1,B} - 1| \geqslant \sqrt{B} - 1\right) \\
&\leqslant \frac{B}{\left(\sqrt{B} - 1\right)^2} \mathrm{Var}\left(Y_{1,B}\right) \to 0.
\end{aligned}$$

Concerning the second condition, let $\tilde{Z}_{l,B} = Z_{l,B}\mathrm{I}\left\{Z_{l,B} \leqslant B\right\}$, where I is the indicator function. Then, for $B > 4$, an application of Lemma 2.2.13 of Durrett (2019) to calculate $\mathrm{E}\tilde{Z}_{1,B}^2$ yields for $T \to \infty$

$$\begin{aligned}
&\frac{1}{B^2} \sum_{l=1}^{B} \mathrm{E}\tilde{Z}_{l,B}^2 = \frac{1}{B}\mathrm{E}\tilde{Z}_{1,B}^2 = \frac{1}{B}\int_0^\infty 2y\mathrm{P}\left(\tilde{Z}_{1,B} > y\right) \mathrm{d}y \leqslant \frac{1}{B}\int_0^B 2y\mathrm{P}\left(Z_{1,B} > y\right) \mathrm{d}y \\
&= \frac{1}{B}\int_0^4 2y\mathrm{P}\left(Z_{1,B} > y\right) \mathrm{d}y + \frac{1}{B}\int_4^B 2y\mathrm{P}\left(Z_{1,B} > y\right) \mathrm{d}y \\
&\leqslant \frac{16}{B} + \frac{1}{B}\int_4^B 2y\mathrm{P}\left(|Y_{1,B} - 1| \geqslant \sqrt{y} - 1\right) \mathrm{d}y \\
&\leqslant \frac{16}{B} + \frac{1}{B}\int_4^B \frac{2y}{(\sqrt{y} - 1)^2} \mathrm{d}y\,\mathrm{Var}\left(Y_{1,B}\right) \leqslant \frac{16}{B} + \frac{8(B-4)}{B}\mathrm{Var}\left(Y_{1,B}\right) \to 0.
\end{aligned}$$

Therefore, it follows from Theorem 2.2.11 of Durrett (2019) that for $T \to \infty$,

$$\frac{1}{B}\sum_{l=1}^{B} Z_{l,B} - \mathrm{E}\tilde{Z}_{1,B} \to 0 \text{ in P-probability.}$$

Finally, for $T \to \infty$,

$$\mathrm{E}Z_{1,B} = \mathrm{E}Y_{1,B}^2 = \mathrm{Var}\left(Y_{1,B}\right) + 1 \to 1$$

and by $Y_{1,B} \to 1$ in P-probability and the dominated convergence theorem

$$\int_0^1 \mathrm{P}\left(Z_{1,B} > y\right) \mathrm{d}y = \int_0^1 \mathrm{P}\left(Y_{1,B} > \sqrt{y}\right) \mathrm{d}y \to 1.$$

The above two equations lead to

$$0 \leqslant \int_B^\infty \mathrm{P}\left(Z_{1,B} > y\right) \mathrm{d}y \leqslant \int_1^\infty \mathrm{P}\left(Z_{1,B} > y\right) \mathrm{d}y = \mathrm{E}Z_{1,B} - \int_0^1 \mathrm{P}\left(Z_{1,B} > y\right) \mathrm{d}y \to 0$$

for $T \to \infty$. Therefore, as $T \to \infty$,

$$\begin{aligned}
&\mathrm{E}\tilde{Z}_{1,B} = \int_0^\infty \mathrm{P}\left(\tilde{Z}_{1,B} > y\right) \mathrm{d}y = \int_0^B \mathrm{P}\left(y < Z_{1,B} \leqslant B\right) \mathrm{d}y \\
&= \int_0^B \mathrm{P}\left(Z_{1,B} > y\right) \mathrm{d}y - B\mathrm{P}\left(Z_{1,B} > B\right) \\
&= \mathrm{E}Z_{1,B} - \int_B^\infty \mathrm{P}\left(Z_{1,B} > y\right) \mathrm{d}y - B\mathrm{P}\left(Z_{1,B} > B\right) \to 1.
\end{aligned}$$

□

The following proposition gives a uniform concentration inequality over $\Gamma$ defined in (E.6). The introduction of $\Gamma$, which is a compact superset of the parameter space $\Theta$ as shown in Proposition E.3 (a), facilitates the verification of measurability.

**Proposition E.6.** *Under Assumptions $\mathcal{A}.1$ and $\mathcal{A}.2$, $\sup_{f\in\Gamma} |A_T(f) - \mathrm{E}A_T(f)|$ is a random variable for each $T$, where $\Gamma$ is as in (E.6). Furthermore, there exist a constant $c > 0$ such that*

$$\lim_{T\to\infty} \mathrm{P}\left(\sup_{f\in\Gamma} |A_T(f) - \mathrm{E}A_T(f)| > cB^{-\frac{1}{4}}\right) = 0.$$

*Proof.* It can be easily verified that $|A_T(\cdot) - \mathrm{E}A_T(\cdot)|$ is a Carathéodory function (Lemma 4.51 of Aliprantis and Border (2006)). The measurability of $\sup_{f\in\Gamma} |A_T(f) - \mathrm{E}A_T(f)|$ can be shown by noting that $\Gamma$ is a compact metric space due to Proposition E.3 (a) and invoking the measurable maximum theorem (Theorem 18.19 of Aliprantis and Border (2006)).

According to Proposition E.4, for any $a > 0$, we have

$$\inf_{f\in\Gamma} \mathrm{P}\left(|A_T(f) - \mathrm{E}A_T(f)| < \frac{a}{2}\right) = 1 - \sup_{f\in\Gamma} \mathrm{P}\left(|A_T(f) - \mathrm{E}A_T(f)| \geqslant \frac{a}{2}\right)$$
$$\geqslant 1 - \frac{c_1}{a^2 T},$$

where $c_1$ is a constant independent of $a$ and $T$. Therefore, as long as $a^2 T \geqslant 2c_1$, we have

$$\inf_{f\in\Gamma} \mathrm{P}\left(|A_T(f) - \mathrm{E}A_T(f)| < \frac{a}{2}\right) \geqslant \frac{1}{2}.$$

As a result of Proposition P.1 (b), the random variables $\sum_{j=1}^{m} \frac{f_0(v_{j,l},\lambda_j)\mathrm{MI}^{\varepsilon}_{6(l-1)m+j}}{f(v_{j,l},\lambda_j)}$, $l = 1, \cdots, B$ are independent. Then, by Lemma 2.3.7 of van der Vaart and Wellner (1996), provided that $a^2 T$ is large enough,

$$\begin{aligned}
&\mathrm{P}\left(\sup_{f\in\Gamma} |A_T(f) - \mathrm{E}A_T(f)| > a\right) \\
&\leqslant \frac{2\mathrm{P}\left(\sup_{f\in\Gamma}\left|\frac{1}{mB}\sum_{l=1}^{B}\xi_l\sum_{j=1}^{m}\frac{f_0(v_{j,l},\lambda_j)\mathrm{MI}^{\varepsilon}_{6(l-1)m+j}}{f(v_{j,l},\lambda_j)}\right| \geqslant \frac{a}{4}\right)}{\inf_{f\in\Gamma}\mathrm{P}\left(|A_T(f) - \mathrm{E}A_T(f)| < \frac{a}{2}\right)} \\
&\leqslant 4\mathrm{P}\left(\sup_{f\in\Gamma}\left|\frac{1}{mB}\sum_{l=1}^{B}\xi_l\sum_{j=1}^{m}\frac{f_0(v_{j,l},\lambda_j)\mathrm{MI}^{\varepsilon}_{6(l-1)m+j}}{f(v_{j,l},\lambda_j)}\right| \geqslant \frac{a}{4}\right),
\end{aligned} \tag{E.11}$$

where $\{\xi_l\}$ are i.i.d. Rademacher random variables independent of $\{\varepsilon_l\}$. Recall that $\mathcal{U}(h, r)$, which is defined in (E.7), is a supremum-norm ball centered at $h \in \Gamma$ with radius $r$. According to Proposition E.3 (a) and Theorems IV and XIII of Kolmogorov and Tikhomirov (1993), for any $r > 0$, there exists integer $\mathcal{N}(r)$ and $\{h_i\}_{i=1}^{\mathcal{N}(r)} \subset \Gamma$ such that $\Gamma \subset \cup_{i=1}^{\mathcal{N}(r)} \mathcal{U}(h_i, r)$ and

$$\ln \mathcal{N}(r) \leqslant \frac{\tilde{c}}{r^2}, \tag{E.12}$$

where $\tilde{c}$ is a constant independent of $r$. We have

$$\begin{aligned}
&\sup_{f\in\mathcal{U}(h_i,r)}\left|\frac{1}{mB}\sum_{l=1}^{B}\xi_l\sum_{j=1}^{m}\frac{f_0(v_{j,l},\lambda_j)\mathrm{MI}^{\varepsilon}_{6(l-1)m+j}}{f(v_{j,l},\lambda_j)}\right| \\
&\leqslant \left|\frac{1}{mB}\sum_{l=1}^{B}\xi_l\sum_{j=1}^{m}\frac{f_0(v_{j,l},\lambda_j)\mathrm{MI}^{\varepsilon}_{6(l-1)m+j}}{h_i(v_{j,l},\lambda_j)}\right|
\end{aligned}$$

$$
\begin{aligned}
&+ \sup_{f\in\mathcal{U}(h_i,r)} \left| \frac{1}{mB}\sum_{l=1}^{B}\xi_l \sum_{j=1}^{m} f_0(v_{j,l},\lambda_j)\mathrm{MI}^{\varepsilon}_{6(l-1)m+j}\left(\frac{1}{f(v_{j,l},\lambda_j)} - \frac{1}{h_i(v_{j,l},\lambda_j)}\right)\right| \\
&\leqslant \left| \frac{1}{mB}\sum_{l=1}^{B}\xi_l \sum_{j=1}^{m} \frac{f_0(v_{j,l},\lambda_j)\mathrm{MI}^{\varepsilon}_{6(l-1)m+j}}{h_i(v_{j,l},\lambda_j)}\right| + \frac{r\,\|f_0\|_\infty}{\delta^2 mB}\sum_{l=1}^{B}\sum_{j=1}^{m}\mathrm{MI}^{\varepsilon}_{6(l-1)m+j}.
\end{aligned}
$$

Therefore,

$$
\begin{aligned}
&\sup_{f\in\Gamma}\left| \frac{1}{mB}\sum_{l=1}^{B}\xi_l \sum_{j=1}^{m} \frac{f_0(v_{j,l},\lambda_j)\mathrm{MI}^{\varepsilon}_{6(l-1)m+j}}{f(v_{j,l},\lambda_j)}\right| \\
&\leqslant \max_{i=1,\cdots,\mathcal{N}(r)} \sup_{f\in\mathcal{U}(h_i,r)}\left| \frac{1}{mB}\sum_{l=1}^{B}\xi_l \sum_{j=1}^{m} \frac{f_0(v_{j,l},\lambda_j)\mathrm{MI}^{\varepsilon}_{6(l-1)m+j}}{f(v_{j,l},\lambda_j)}\right| \\
&\leqslant \max_{i=1,\cdots,\mathcal{N}(r)} \left| \frac{1}{mB}\sum_{l=1}^{B}\xi_l \sum_{j=1}^{m} \frac{f_0(v_{j,l},\lambda_j)\mathrm{MI}^{\varepsilon}_{6(l-1)m+j}}{h_i(v_{j,l},\lambda_j)}\right| + \frac{r\,\|f_0\|_\infty}{\delta^2 mB}\sum_{l=1}^{B}\sum_{j=1}^{m}\mathrm{MI}^{\varepsilon}_{6(l-1)m+j}.
\end{aligned}
$$

In conjunction with (E.11), it follows that for any $a > 0$,

$$
\begin{aligned}
&\mathrm{P}\left(\sup_{f\in\Gamma}|A_T(f) - \mathrm{E}A_T(f)| > a\right) \leqslant 4\mathrm{P}\left(\frac{1}{mB}\sum_{l=1}^{B}\sum_{j=1}^{m}\mathrm{MI}^{\varepsilon}_{6(l-1)m+j} \geqslant \frac{\delta^2 a}{8r\,\|f_0\|_\infty}\right) \\
&+ 4\mathrm{P}\left(\max_{i=1,\cdots,\mathcal{N}(r)}\left| \frac{1}{mB}\sum_{l=1}^{B}\xi_l \sum_{j=1}^{m} \frac{f_0(v_{j,l},\lambda_j)\mathrm{MI}^{\varepsilon}_{6(l-1)m+j}}{h_i(v_{j,l},\lambda_j)}\right| \geqslant \frac{a}{8}\right). \qquad \text{(E.13)}
\end{aligned}
$$

By the Hoeffding's inequality (see e.g. Lemma 2.2.7 of van der Vaart and Wellner (1996)), we have

$$
\begin{aligned}
&\mathrm{P}\left(\max_{i=1,\cdots,\mathcal{N}(r)}\left| \frac{1}{mB}\sum_{l=1}^{B}\xi_l \sum_{j=1}^{m} \frac{f_0(v_{j,l},\lambda_j)\mathrm{MI}^{\varepsilon}_{6(l-1)m+j}}{h_i(v_{j,l},\lambda_j)}\right| \geqslant \frac{a}{8}\,\middle|\, \varepsilon\right) \\
&\leqslant \min\left\{1, \sum_{i=1}^{\mathcal{N}(r)}\mathrm{P}\left(\left| \frac{1}{mB}\sum_{l=1}^{B}\xi_l \sum_{j=1}^{m} \frac{f_0(v_{j,l},\lambda_j)\mathrm{MI}^{\varepsilon}_{6(l-1)m+j}}{h_i(v_{j,l},\lambda_j)}\right| \geqslant \frac{a}{8}\,\middle|\, \varepsilon\right)\right\} \\
&\leqslant \min\left\{1, 2\mathcal{N}(r)\exp\left(-\frac{a^2}{128\max_{i=1,\cdots,\mathcal{N}(r)}\sum_{l=1}^{B}\left(\frac{1}{mB}\sum_{j=1}^{m}\frac{f_0(v_{j,l},\lambda_j)\mathrm{MI}^{\varepsilon}_{6(l-1)m+j}}{h_i(v_{j,l},\lambda_j)}\right)^2}\right)\right\}
\end{aligned}
$$

$$\leqslant \min\left\{1, 2\mathcal{N}(r)\exp\left(-\frac{Ba^2}{128\frac{\|f_0\|_\infty^2}{\delta^2}\frac{1}{B}\sum_{l=1}^B\left(\frac{1}{m}\sum_{j=1}^m \mathrm{MI}_{6(l-1)m+j}^\varepsilon\right)^2}\right)\right\}$$

$$\leqslant 2\mathcal{N}(r)\exp\left(-\frac{Ba^2}{128\frac{\|f_0\|_\infty^2}{\delta^2}b}\right) + \mathrm{I}\left\{\frac{1}{B}\sum_{l=1}^B\left(\frac{1}{m}\sum_{j=1}^m \mathrm{MI}_{6(l-1)m+j}^\varepsilon\right)^2 \geqslant b\right\}.$$

for any $b > 0$, where I is the indicator function. Choosing $c > \tilde{c}^{\frac{1}{4}}$, $a = a_T = cB^{-\frac{1}{4}}$, $r = r_T = \frac{16\|f_0\|_\infty}{\delta}a_T$ and $b = 2$, we have $a_T^2 T \to \infty$ due to Assumption $\mathcal{A}.1$. Moreover, by inequality (E.12), Propositions E.4 and E.5 together with (E.13), we get

$$\mathrm{P}\left(\sup_{f\in\Gamma}|A_T(f) - \mathrm{E}A_T(f)| > cB^{-\frac{1}{4}}\right) \leqslant 8\mathcal{N}(r_T)\exp\left(-\frac{Ba_T^2}{128\frac{\|f_0\|_\infty^2}{\delta^2}b_T}\right)$$
$$+ 4\left[\mathrm{P}\left(\frac{1}{mB}\sum_{l=1}^B\sum_{j=1}^m \mathrm{MI}_{6(l-1)m+j}^\varepsilon \geqslant 2\right) + \mathrm{P}\left(\frac{1}{B}\sum_{l=1}^B\left(\frac{1}{m}\sum_{j=1}^m \mathrm{MI}_{6(l-1)m+j}^\varepsilon\right)^2 \geqslant 2\right)\right]$$
$$\leqslant 8\exp\left(\frac{\tilde{c}}{r_T^2}\right)\exp\left(-\frac{Ba_T^2}{256\frac{\|f_0\|_\infty^2}{\delta^2}}\right) + o(1) = 8\exp\left(\frac{1}{256\frac{\|f_0\|_\infty^2}{\delta^2}}\left(\frac{\tilde{c}}{a_T^2} - Ba_T^2\right)\right) + o(1)$$
$$= 8\exp\left(-\check{c}\sqrt{B}\right) + o(1) = o(1)$$

as $T \to \infty$, where $\check{c} = \frac{1}{256\frac{\|f_0\|_\infty^2}{\delta^2}}\left(c^2 - \frac{\tilde{c}}{c^2}\right) > 0$. □

It is worth noting that the sequence $\{v_{j,l}\}$ does not have any impact on the proof. Namely, the above proof is valid even if we replace $A_T(f)$ by $A_T(f; \{X_t\}_{t=-m+1}^{T+m}; \{x_{j,l}\})$ (see (E.8) for a definition) where $\{x_{j,l}\} \subset [0, 1]$ is an arbitrary sequence.

Finally, we get a uniform concentration equality for the (centered) remainder term:

**Proposition E.7.** *Under Assumptions $\mathcal{A}.1$ and $\mathcal{A}.2$, $\sup_{f\in\Gamma}|R_T(f) - \mathrm{E}R_T(f)|$ is a random variable for each $T$. Moreover, for any $a > 0$,*

$$\mathrm{P}\left(\sup_{f\in\Gamma}|R_T(f) - \mathrm{E}R_T(f)| \geqslant a\right) = O\left(\frac{1}{a\sqrt{m}}\right).$$

*Proof.* By invoking the same arguments in Proposition E.6, we obtain the measurability of $\sup_{f\in\Gamma}|R_T(f)-\mathrm{E}R_T(f)|$. As for the inequality, note that

$$\sup_{f\in\Gamma}|R_T(f)-\mathrm{E}R_T(f)|\leqslant\frac{1}{\delta mB}\sum_{l=1}^{B}\sum_{j=1}^{m}|R_{j,6l-5,1}-\mathrm{E}R_{j,6l-5,1}|\,.$$

Therefore, by Markov's inequality, Cauchy-Schwarz inequality and Proposition P.1 (b), it follows that

$$\begin{aligned}&\mathrm{P}\left(\sup_{f\in\Gamma}|R_T(f)-\mathrm{E}R_T(f)|\geqslant a\right)\leqslant\mathrm{P}\left(\frac{1}{\delta mB}\sum_{l=1}^{B}\sum_{j=1}^{m}|R_{j,6l-5,1}-\mathrm{E}R_{j,6l-5,1}|\geqslant a\right)\\&\leqslant\frac{1}{\delta amB}\mathrm{E}\sum_{l=1}^{B}\sum_{j=1}^{m}|R_{j,6l-5,1}-\mathrm{E}R_{j,6l-5,1}|\leqslant\frac{2}{\delta amB}\sum_{l=1}^{B}\sum_{j=1}^{m}\mathrm{E}\,|R_{j,6l-5,1}|\\&\leqslant\frac{2}{\delta a}\sqrt{\sup_{l=1,2,\cdots,B}\sup_{j=1,\cdots,m}\mathrm{E}|R_{j,6l-5,1}|^2}=O\left(\frac{1}{a\sqrt{m}}\right).\end{aligned}$$

□

Putting Propositions E.6 and E.7 together, we obtain the following concentration inequality for the stochastic term of all 3 likelihood approximations.

**Theorem E.2.** *Under Assumptions $\mathcal{A}.1$ and $\mathcal{A}.2$ in the main text, for any positive divergent sequence $M_T\to\infty$,*

$$\sup_{f\in\Gamma}\left|S^{(i)}(f)\right|=o_{\mathrm{P}}\left(M_T\max\left\{\left(\frac{m}{T}\right)^{\frac{1}{4}},m^{-\frac{1}{2}}\right\}\right),$$

*where $S^{(i)}$ and $\Gamma$ are defined in (E.2) and (E.6), respectively.*

*Proof.* First, by Propositions E.6 and E.7 it holds for the factor-6-thinned version considered so far in this section that

$$\begin{aligned}&\frac{\sup_{f\in\Gamma}|A_T(f)+R_T(f)-\mathrm{E}\left[A_T(f)+R_T(f)\right]|}{M_T\max\left\{\left(\frac{m}{T}\right)^{\frac{1}{4}},m^{-\frac{1}{2}}\right\}}\\&\leqslant O(1)\frac{\sup_{f\in\Gamma}|A_T(f)-\mathrm{E}A_T(f)|}{M_T\,B^{-\frac{1}{4}}}+\frac{\sup_{f\in\Gamma}|R_T(f)-\mathrm{E}R_T(f)|}{M_T\,m^{-\frac{1}{2}}}=o_{\mathrm{P}}(1).\end{aligned}$$

From this, we now get the assertion for the actual likelihoods as follows:

First, consider the following shifted factor-6-thinned-versions,

$$A_T^{(k,6)}(f) = \frac{1}{mB_{k,6}} \sum_{l=1}^{B_{k,6}} \sum_{j=1}^{m} \frac{f_0(v_{j,l,k,6}, \lambda_j) \mathrm{MI}^{\varepsilon}_{[6(l-1)+(k-1)]m+j}}{f(v_{j,l,k,6}, \lambda_j)},$$

$$R_T^{(k,6)}(f) = \frac{1}{mB_{k,6}} \sum_{l=1}^{B_{k,6}} \sum_{j=1}^{m} \frac{R_{j,6l-6+k,1}}{f(v_{j,l,k,6}, \lambda_j)},$$

where $B_{k,6} = \left\lceil \frac{T-km}{6m} \right\rceil$ and $v_{j,l,k,6} = \frac{[6(l-1)+(k-1)]m+j}{T}$, $k = 1, \cdots, 6$. Note that we may express $A_T^{(k,6)}(f) = A_T(f; \{X_t\}_{t=-m+1+(k-1)m}^{T+m}; \{v_{j,l,k,6}\})$ and similarly $R_T^{(k,6)}(f) = R_T(f; \{X_t\}_{t=-m+1+(k-1)m}^{T+m}; \{v_{j,l,k,6}\})$ due to (E.8) and (E.9), respectively. Therefore, because $\{X_t : t \geqslant -m+1+(k-1)m\}$ is also a locally stationary time series fulfilling all assumptions with length $T$ replaced by $T-(k-1)m$ where $m/(T-(k-1)m) = m/T\,(1+o(1)) = (6\,B_{k,6})^{-1}\,(1+o(1))$. Hence, Proposition E.6 still holds if we replace $A_T(f)$ with $A_T^{(k,6)}(f)$ and $B$ with $B_{k,6}$. Similarly, Proposition E.7 remains true if we substitute $R_T^{(k,6)}(f)$ for $R_T(f)$.

Finally, this implies the desired result for all 3 cases as

$$S^{(i)}(f) := \begin{cases} \sum_{k=1}^{6} \frac{B_{k,6}}{B_1} \left[ A_T^{(k,6)}(f) + R_T^{(k,6)}(f) - \mathrm{E}\left( A_T^{(k,6)}(f) + R_T^{(k,6)}(f) \right) \right] & i{=}1, \\ \sum_{k=1,3,5} \frac{B_{k,6}}{B_2} \left[ A_T^{(k,6)}(f) + R_T^{(k,6)}(f) - \mathrm{E}\left( A_T^{(k,6)}(f) + R_T^{(k,6)}(f) \right) \right] & i{=}2, \\ \sum_{k=1,4} \frac{B_{k,6}}{B_3} \left[ A_T^{(k,6)}(f) + R_T^{(k,6)}(f) - \mathrm{E}\left( A_T^{(k,6)}(f) + R_T^{(k,6)}(f) \right) \right] & i{=}3 \end{cases}$$

due to (E.2), Proposition P.1 (a) and $B_{k,6} = O(B_i)$ for any $k = 1, \cdots, 6$ and $i = 1, 2, 3$.

□

*Remark.* It is worth noting that the conclusion of Theorem E.2 can be re-written as

$$\sup_{f \in \Gamma} \left| S^{(i)}(f) \right| = O_{\mathrm{P}} \left( \max \left\{ \left( \frac{m}{T} \right)^{\frac{1}{4}}, m^{-\frac{1}{2}} \right\} \right)$$

according to Lemma C.1 of Kirch and Reckruehm (2022).

## E.3 Proof of the prior positivity as given in Theorem 3.1 in the main text

Let $b(\cdot; k_1, k_2, G)$ the Bernstein polynomial associated with $G$ which is first introduced in Section 3.1 of the main text. Suppose $G$ possesses probability density function $g$, then

$b(\cdot; k_1, k_2, G)$ is an approximation of $g$ in terms of uniform metric. In fact, we have

**Proposition E.8.** *For any probability density function* $g \in C^1([0,1]^2)$*, let* $G$ *denote both the corresponding probability measure and the cumulative distribution function associated with* $g$*. We have*

*(a)*

$$\|g - b(\cdot; k_1, k_2, G)\|_\infty \leqslant \frac{4 \max\{\|\partial_1 g\|_\infty, \|\partial_2 g\|_\infty\}}{\min\{\sqrt{k_1}, \sqrt{k_2}\}}.$$

*(b)*

$$\|b(\cdot; k_1, k_2, G)\|_\infty \leqslant \|g\|_\infty, \quad \|\partial_j b(\cdot; k_1, k_2, G)\|_\infty \leqslant \|\partial_j g\|_\infty, \; j = 1, 2.$$

*Proof.* The proof of (a) can be found in the proof of Lemma 2.5 of Kruijer (2008). Note that the Lipschitz constant of $g$ is bounded from above by $\sqrt{2}\max\{\|\partial_1 g\|_\infty, \|\partial_2 g\|_\infty\}$ according to the proof of Proposition 2.2.5 of Cobzaş et al. (2019). As for (b), let

$$B(x, y; k_1, k_2, G) = \sum_{j_1=0}^{k_1}\sum_{j_2=0}^{k_2}\binom{k_1}{j_1}\binom{k_2}{j_2} x^{j_1}(1-x)^{k_1-j_1} y^{j_2}(1-y)^{k_2-j_2} G\left(\frac{j_1}{k}, \frac{j_2}{k}\right).$$

Then, $B(\cdot; k_1, k_2, G)$ is the classical Bernstein polynomial associated with $G$ and $\partial_{12} B(\cdot; k_1, k_2, G) = b(\cdot; k_1, k_2, G)$, where $\partial_{12} f$ denotes the mixed partial derivative of function $f$ with respect to the first and the second variable. Note also $\partial_{12} G = g$. Thus, the inequalities follow from (17) of Foupouagnigni and Wouodjié (2020).

□

We also need a proposition about the linear combination of tensor-product beta densities which will be useful for providing a lower bound for the prior probability.

**Proposition E.9.** *For any*

$$f(x, y) = \sum_{j_1=1}^{k}\sum_{j_2=1}^{k} w_{j_1,j_2} \beta(x; j_1, k - j_1 + 1)\, \beta(y; j_2, k - j_2 + 1), (x, y) \in [0,1]^2,$$

*we have*

$$\inf_{(x,y)\in[0,1]^2} f(x, y) \geqslant k^2 \min_{j_1, j_2 \in \{1, \cdots, k\}} w_{j_1, j_2},$$

$$\|f\|_\infty \leqslant k^2 \max_{j_1,j_2\in\{1,\cdots,k\}} |w_{j_1,j_2}|\,,$$

$$\|\partial_i f\|_\infty \leqslant 2k^3 \max_{j_1,j_2\in\{1,\cdots,k\}} |w_{j_1,j_2}|\,,\ i=1,2.$$

*Proof.* The first and second inequalities can be deduced by noticing

$$\sum_{j=1}^{k} \beta\,(x; j, k-j+1) = k\sum_{j=1}^{k}\binom{k-1}{j-1} x^{j-1}(1-x)^{k-j} = k$$

for any $x \in [0,1]$ due to the binomial formula. For the third inequality, by direct calculation, we have

$$\begin{aligned}
\partial_1 f &= \sum_{j_2=1}^{k} \beta\,(y; j_2, k-j_2+1) \sum_{j_1=1}^{k} w_{j_1,j_2}\partial_1\beta\,(x; j_1, k-j_1+1)\\
&= k\sum_{j_2=1}^{k} \beta\,(y; j_2, k-j_2+1)\Bigg[-w_{1,j_2}\beta\,(x;1,k-1)\\
&+\sum_{j_1=2}^{k-1} w_{j_1,j_2}\left(\beta\,(x; j_1-1, k-j_1+1) - \beta\,(x; j_1, k-j_1)\right) + w_{k,j_2}\beta\,(x; k-1, 1)\Bigg]\\
&= k\sum_{j_1=1}^{k-1}\sum_{j_2=1}^{k} \left(w_{j_1+1,j_2} - w_{j_1,j_2}\right)\beta\,(x; j_1, k-j_1)\,\beta\,(y; j_2, k-j_2+1)
\end{aligned}$$

with the convention $\sum_{j=1}^{0} = 0$. Since $\sum_{j=1}^{k-1} \beta\,(\cdot; j, k-j) = k-1$, we have

$$\begin{aligned}
\|\partial_1 f\|_\infty &\leqslant k\sum_{j_1=1}^{k-1}\sum_{j_2=1}^{k} |w_{j_1+1,j_2} - w_{j_1,j_2}|\,\beta\,(x; j_1, k-j_1)\,\beta\,(y; j_2, k-j_2+1)\\
&\leqslant k \max_{\substack{j_1\in\{1,\cdots,k-1\}\\ j_2\in\{1,\cdots,k\}}} |w_{j_1+1,j_2} - w_{j_1,j_2}| \sum_{j_1=1}^{k-1}\beta\,(x; j_1, k-j_1)\sum_{j_2=1}^{k}\beta\,(y; j_2, k-j_2+1)\\
&= k^2(k-1)\max_{\substack{j_1\in\{1,\cdots,k-1\}\\ j_2\in\{1,\cdots,k\}}} |w_{j_1+1,j_2} - w_{j_1,j_2}| \leqslant 2k^3 \max_{j_1,j_2\in\{1,\cdots,k\}} |w_{j_1,j_2}|
\end{aligned}$$

The case $i=2$ can be deduced similarly. □

Now we can proceed to proving prior positivity as stated in Theorem 3.1 of the main text.

*Proof of Theorem 3.1.* Let $\tau_0 = \int_0^1 \int_0^1 f_0(x,y)\,\mathrm{d}x\,\mathrm{d}y$ and $q_0 := \frac{f_0}{\tau_0}$. $Q_0$ denotes both the probability measure and the cumulative distribution function corresponding to $q_0$. First, observe that $b(\cdot;k,k,Q_0) - b(\cdot;k,k,G)$ is also a Bernstein polynomial with weights $w_{k,k}(Q_0)(j_1,j_2) - w_{k,k}(G)(j_1,j_2)$ such that all bounds as in Proposition E.9 also hold for this difference. We will make repeated use of this fact throughout the proof.

Let

$$A_\tau = \left\{\tau : \left|1 - \frac{\tau}{\tau_0}\right| \leqslant \epsilon_\tau\right\},\ A_{k,G} = \left\{G : \left\|w_{k,k}(G) - w_{k,k}(Q_0)\right\|_1 \leqslant \epsilon_G\right\},$$

where $w_{k,k}(G) = (w_{k,k}(G)(j_1,j_2))_{j_1,j_2=1}^k$ is a vector of weights in the Bernstein polynomial and $\|\cdot\|_1$ is the usual $l^1$-norm on $\mathbb{R}^{k^2}$. Our goal is to find suitable $k$, $\epsilon_\tau$ and $\epsilon_G$ (depending on $r$) such that $A = \{\tau b(\cdot;k,k,G) : \tau \in A_\tau, G \in A_{k,G}\}$ is a subset of $U_\infty(r)$, i.e., that it fulfills $A \subset \Theta$ as well as $\|f - f_0\|_\infty < r$. To this end, define $k = \left\lceil \frac{256 \max\{\|\partial_1 f_0\|_\infty^2, \|\partial_2 f_0\|_\infty^2\}}{r^2} \right\rceil$. By assumption $\|f_0\|_\infty < M_0$, $\|\partial_1 f_0\|_\infty < M_1$, $\|\partial_2 f_0\|_\infty < M_2$ and $\delta_0 = \min_{(x,y)\in[0,1]^2} f_0(x,y) > \delta$, we may define

$$\begin{aligned}\epsilon_\tau = \min\Bigg\{&\frac{r}{4\|f_0\|_\infty}, \frac{1}{2}\left(\frac{M_0}{\|f_0\|_\infty} - 1\right), \frac{1}{2}\left(\frac{M_1}{\|\partial_1 f_0\|_\infty} - 1\right), \frac{1}{2}\left(\frac{M_2}{\|\partial_2 f_0\|_\infty} - 1\right),\\ &\frac{1}{2}\left(1 - \frac{\delta}{\delta_0}\right)\Bigg\} > 0.\end{aligned}$$

For these $\epsilon_\tau$, positivity of the following choice for $\epsilon_G$ is guaranteed, i.e.,

$$\begin{aligned}\epsilon_G = \min\Bigg\{&\frac{r}{4k^2\tau_0(1+\epsilon_\tau)}, \frac{\frac{M_0}{1+\epsilon_\tau} - \|f_0\|_\infty}{k^2\tau_0}, \frac{\frac{M_1}{1+\epsilon_\tau} - \|\partial_1 f_0\|_\infty}{2k^3\tau_0}, \frac{\frac{M_2}{1+\epsilon_\tau} - \|\partial_2 f_0\|_\infty}{2k^3\tau_0},\\ &\frac{\delta_0 - \frac{\delta}{1-\epsilon_\tau}}{k^2\tau_0}\Bigg\} > 0.\end{aligned}$$

We will first start with condition $\|f - f_0\|_\infty < r$: For any $\tau b(\cdot;k,k,G) \in A$, we have by the triangular inequality, Propositions E.8 and E.9 that

$$\begin{aligned}&\|f_0 - \tau b\,(\cdot;k,k,G)\,\|_\infty\\ &\leqslant \tau_0\|q_0 - b\,(\cdot;k,k,Q_0)\,\|_\infty + |\tau_0 - \tau|\,\|b\,(\cdot;k,k,Q_0)\,\|_\infty + \tau\|b(\cdot;k,k,Q_0) - b(\cdot;k,k,G)\|_\infty\\ &\leqslant \tau_0\|q_0 - b\,(\cdot;k,k,Q_0)\,\|_\infty + \tau_0\left|1 - \frac{\tau}{\tau_0}\right|\|q_0\|_\infty\\ &\qquad + \tau k^2 \max_{j_1,j_2\in\{1,\cdots,k\}} |w_{k,k}(Q_0)(j_1,j_2) - w_{k,k}(G)(j_1,j_2)|\end{aligned}$$

$$\leqslant \frac{4\max\left\{\|\partial_1 f_0\|_\infty, \|\partial_2 f_0\|_\infty\right\}}{\sqrt{k}} + \epsilon_\tau \|f_0\|_\infty + \tau_0\left(1+\epsilon_\tau\right)k^2\epsilon_G < r.$$

Next, we will achieve $A \subset \Theta$. Indeed, by very similar arguments as above we have

$$\begin{aligned}
&\|\tau b\left(\cdot;k,k,G\right)\|_\infty \leqslant \tau\left(\|(b\left(\cdot;k,k,G\right) - b(\cdot;k,k,Q_0))\|_\infty + \|b\left(\cdot;k,k,Q_0\right)\|_\infty\right) \\
&\leqslant \left(1+\epsilon_\tau\right)\left(k^2\tau_0\epsilon_G + \|f_0\|_\infty\right) \leqslant M_0.
\end{aligned}$$

Similar arguments where now the bounds for the partial derivatives are being used yield

$$\begin{aligned}
&\|\partial_1\left(\tau b\left(\cdot;k,k,G\right)\right)\|_\infty \leqslant \tau\left(\|\partial_1\left(b\left(\cdot;k,k,G\right) - b(\cdot;k,k,Q_0)\right)\|_\infty + \|\partial_1 b\left(\cdot;k,k,Q_0\right)\|_\infty\right) \\
&\leqslant \tau_0(1+\epsilon_\tau)\left(2k^3 \max_{j_1,j_2\in\{1,\cdots,k\}} |w_{k,k}(Q_0)(j_1,j_2) - w_{k,k}(G)(j_1,j_2)| + \|\partial_1 q_0\|_\infty\right) \\
&\leqslant \left(1+\epsilon_\tau\right)\left(2k^3\tau_0\epsilon_G + \|\partial_1 f_0\|_\infty\right) \leqslant M_1.
\end{aligned}$$

Analogously,

$$\|\partial_2\left(\tau b\left(\cdot;k,k,G\right)\right)\|_\infty \leqslant \left(1+\epsilon_\tau\right)\left(2k^3\tau_0\epsilon_G + \|\partial_2 f_0\|_\infty\right) \leqslant M_2.$$

Finally,

$$\begin{aligned}
&\tau b\left(\cdot;k,k,G\right) \geqslant \tau_0(1-\epsilon_\tau)\left(b\left(\cdot;k,k,G\right) - b\left(\cdot;k,k,Q_0\right) + b\left(\cdot;k,k,Q_0\right)\right) \\
&\geqslant \tau_0(1-\epsilon_\tau)\left(k^2 \min_{j_1,j_2}\left\{w_{k,k}(G)(j_1,j_2) - w_{k,k}(Q_0)(j_1,j_2)\right\} + \delta_0/\tau_0\right) \\
&\geqslant \tau_0(1-\epsilon_\tau)\left(-k^2\epsilon_G + \frac{f_0(\cdot)}{\tau_0}\right) = \left(1-\epsilon_\tau\right)\left(-k^2\tau_0\epsilon_G + \delta_0\right) \geqslant \delta.
\end{aligned}$$

Thus, for these choices of $k$ and $\epsilon_\tau, \epsilon_G$ we get $A \subset U_\infty(r)$. Then, with $\Pi_D$ denoting the prior on $G$,

$$\Pi_{BD}\left(U_\infty(r)\right) \geqslant \Pi_{BD}\left(A\right) = \rho_1(k)\rho_2(k)\Pi_D\left(A_{k,G}\right)\int_{A_\tau} \pi_\tau(u)\,\mathrm{d}u > 0$$

due to the fact that the distributions of $G$, $k_1$, $k_2$ and $\tau$ all have full support.

Furthermore, for $r$ and thus $\epsilon_G$ small enough, we have by Lemma A.1 of Ghosal (2001) that

$$-\ln\Pi_D\left(A_{k,G}\right) = O\left(-k^2\ln\epsilon_G\right) = O\left(-\frac{\ln r}{r^4}\right),$$

because $k^2$ is of order $1/r^2$ while $\epsilon_G$ is of polynomial in $r$ order. Similarly, it follows from Assumption $\mathcal{A}.3$ in the main text that

$$-\ln \rho_i(k) = O\left(k^2 \ln k\right) = O\left(-\frac{\ln r}{r^4}\right), \quad i = 1, 2$$

and for $r$ small enough such that $\epsilon_\tau \leqslant 1/2$, we get by the definition of $A_\tau$ that

$$-\ln \int_{A_\tau} \pi_\tau(u)\, \mathrm{d}u \leqslant -\ln\left(\left[\inf_{\frac{1}{2}\tau_0 \leqslant u \leqslant \frac{3}{2}\tau_0} \pi_\tau(u)\right] 2\tau_0 \epsilon_\tau\right) = O\left(-\ln r\right)$$

completing the proof.

□

## E.4 Proof of posterior consistency and contraction rates as given in Theorem 3.2 in the main text

In view of (E.1), we have the following expression of the posterior,

$$\Pi_T^{(i)}(A) = \frac{\int_A \exp\left[-mB_i\left(S^{(i)}(f) + D^{(i)}(f) + \Delta(f)\right)\right] \Pi_0(\,\mathrm{d}f)}{\int_\Theta \exp\left[-mB_i\left(S^{(i)}(f) + D^{(i)}(f) + \Delta(f)\right)\right] \Pi_0(\,\mathrm{d}f)}$$

for any $A \in \mathcal{B}\left(C\left([0,1]^2\right)\right) \cap \Theta$ and $i = 1, 2, 3$. By Theorem E.1, there exists a positive constant $C$ such that $\sup_{f\in\Theta} |D^{(i)}(f)| \leqslant Cm^{-1/2}$. Also note that $\Theta \subset \Gamma$ due to Proposition E.3 (a). Then for any $a_T$, $b_T > 0$, when event $\left\{\sup_{f\in\Gamma} \left|S^{(i)}(f)\right| \leqslant a_T\right\}$ occurs, it follows that

$$\begin{aligned}
&\int_\Theta \exp\left[-mB_i\left(S^{(i)}(f) + D^{(i)}(f) + \Delta(f)\right)\right] \Pi_0(\,\mathrm{d}f) \\
&\geqslant \int_\Theta \exp\left[-mB_i\left(a_T + \frac{C}{\sqrt{m}} + \Delta(f)\right)\right] \Pi_0(\,\mathrm{d}f) \\
&\geqslant \int_{\Delta(f)\leqslant b_T} \exp\left[-mB_i\left(a_T + \frac{C}{\sqrt{m}} + b_T\right)\right] \Pi_0(\,\mathrm{d}f) \\
&\geqslant \exp\left[-mB_i\left(a_T + \frac{C}{\sqrt{m}} + b_T\right)\right] \Pi_0\left(U_\infty(\delta\sqrt{b_T})\right),
\end{aligned}$$

where the last inequality is due to Proposition E.1. Similarly, for any $A \in \mathcal{B}\left(C\left([0,1]^2\right)\right) \cap \Theta$, we have

$$\begin{aligned}
&\int_A \exp\left[-mB_i\left(S^{(i)}(f)+D^{(i)}(f)+\Delta(f)\right)\right]\Pi_0(\,\mathrm{d}f) \\
&\leqslant \exp\left[mB_i\left(a_T+\frac{C}{\sqrt{m}}-\inf_{f\in A}\Delta(f)\right)\right].
\end{aligned}$$

The above inequalities, in conjunction with the positivity of $\Pi_0\left(U_\infty(\delta\sqrt{b_T})\right)$ due to Theorem 3.1, yield

$$\Pi_T^{(i)}(A) \leqslant \exp\left[mB_i\left(2a_T+\frac{2C}{\sqrt{m}}+b_T-\frac{1}{mB_i}\ln\Pi_0\left(U_\infty(\delta\sqrt{b_T})\right)-\inf_{f\in A}\Delta(f)\right)\right]. \tag{E.14}$$

In order to prove posterior consistency in the $d_\infty$ metric, we let $A = U_\infty^c(r)$ for any fixed $r > 0$. Then with Propositions E.1 and E.3, we get

$$\inf_{f\in A}\Delta(f) \geqslant \frac{\inf_{f\in U_\infty^c(r)}\|f-f_0\|_2^2}{2M_0^2} \geqslant \frac{\inf_{f\in\Gamma\setminus\mathcal{U}(f_0,r)}\|f-f_0\|_2^2}{2M_0^2} > 0.$$

Therefore, let $b_T$ be a constant satisfying $b_T = b < \inf_{f\in A}\Delta(f)$ and $a_T = \ln m \cdot \max\left\{m^{-\frac{1}{2}}, (m/T)^{\frac{1}{4}}\right\}$ such that $a_T = o(1)$ due to Assumption $\mathcal{A}$.1, we obtain that the right-hand side of (E.14) goes to 0 when $T \to \infty$.

Since $\mathrm{P}\left(\sup_{f\in\Gamma}\left|S^{(i)}(f)\right| < a_T\right) \to 1$ as $T \to \infty$ according to Theorem E.2, we obtain $\Pi_T^{(i)}\left(U_\infty^c(r)\right) = o_{\mathrm{P}}(1)$.

In the following, we will use $c$ to denote a generic positive constants that may differ from one line to the next. To derive a posterior contraction rate in the $L_2$-norm, let $A_T = U_2^c\left(M_T\max\left\{m^{-\frac{1}{4}}, (m/T)^{\frac{1}{8}}\right\}\right)$, then by Proposition E.1, we have

$$\inf_{f\in A_T}\Delta(f) \geqslant \frac{\inf_{f\in A_T}\|f-f_0\|_2^2}{2M_0^2} = c\left(M_T^2\max\left\{m^{-\frac{1}{2}}, \left(\frac{m}{T}\right)^{\frac{1}{4}}\right\}\right),$$

By choosing $a_T = M_T\max\left\{m^{-\frac{1}{2}}, (m/T)^{\frac{1}{4}}\right\}$, $b_T = \frac{1}{\delta^2}(m/T)^{\frac{1}{4}}$, for sufficiently large $T$, we have by Theorem 3.1 that

$$2a_T+\frac{2C}{\sqrt{m}}+b_T-\frac{1}{mB_i}\ln\Pi_0\left(U_\infty(\delta\sqrt{b_T})\right)-\inf_{g\in A}\Delta(g)$$

$$
\begin{aligned}
&\leqslant 2a_T + \frac{2C}{\sqrt{m}} + b_T - c\frac{1}{mB_i}\frac{\ln \delta^2 b_T}{\delta^4 b_T^2} - c\left(M_T^2 \max\left\{m^{-\frac{1}{2}}, \left(\frac{m}{T}\right)^{\frac{1}{4}}\right\}\right)\\
&\leqslant c\left(M_T \max\left\{m^{-\frac{1}{2}}, \left(\frac{m}{T}\right)^{\frac{1}{4}}\right\} + m^{-\frac{1}{2}} + \left(\frac{m}{T}\right)^{\frac{1}{4}}\right.\\
&\qquad \left. + \frac{\ln(T/m)}{m\sqrt{T/m}} - M_T^2 \max\left\{m^{-\frac{1}{2}}, \left(\frac{m}{T}\right)^{\frac{1}{4}}\right\}\right)\\
&\leqslant c\left(-M_T^2 \max\left\{m^{-\frac{1}{2}}, \left(\frac{m}{T}\right)^{\frac{1}{4}}\right\}\right).
\end{aligned}
$$

By Assumption $\mathcal{A}.1$ this shows that the right hand side of (E.14) converges to 0. Because $\mathrm{P}\left(\sup_{f\in\Gamma}\left|S^{(i)}(f)\right| < a_T\right) \to 1$ as $T \to \infty$ according to Theorem E.2, we conclude that $\Pi_T^{(i)}\left(U_2^c\left(M_T \max\left\{m^{-\frac{1}{4}}, (m/T)^{\frac{1}{8}}\right\}\right)\right) = o_{\mathrm{P}}(1)$, which is the assertion.

# F R code demonstration of the BDP-DW procedure implemented in the R package `beyondWhittle`

In this section, we present a chunk of R code demonstrating how to use `beyondWhittle` R package to estimate the time-varying spectral density of DGP LS2c defined in Section 4.2 of the main text.

```
#===================================================================
#  load required packages
#===================================================================
if("beyondWhittle" %in% rownames(installed.packages()) == FALSE){
install.packages("beyondWhittle")
}
library(beyondWhittle)
#===================================================================
#  (I) Generating the simulated data for DGP LS2c
#===================================================================
#set seed
set.seed(200)

#set the length of time series
len_d <- 1500
```

```
#generate data from DGP LS2c defined in Section 4.2.2,
#type ?sim_tvarma12 for more info
sim_data <- sim_tvarma12(len_d = 1500,
                         dgp = "LS2",
                         innov_distribution = "c")
#====================================================================
#  (II) Plotting the time-varying spectral density of DGP LS2c
#====================================================================
#specify time grid at which the tv-PSD is evaluated
res_time <- seq(0, 1, by = 0.005)

#specify frequency grid at which the tv-PSD is evaluated
freq <- pi * seq(0, 1, by = 0.01)

#calculate the true tv-PSD of DGP LS2c at the pre-specified grid
#type ?psd_tvarma12 for more info
true_tvPSD <- psd_tvarma12(rescaled_time = res_time,
                           freq = freq,
                           dgp = "LS2")

#plot the true tv-PSD
#type ?plot.bdp_dw_tv_psd for more info
plot(true_tvPSD)
#====================================================================
#  (III) Applying the BDP-DW approach to the simulated data
#        with m = 50 and thinning factor 2
#====================================================================
#type ?gibbs_bdp_dw for more info
result <- gibbs_bdp_dw(data = sim_data,
                       m = 50,
                       likelihood_thinning = 2,
                       res_time = res_time,
                       freq = freq)
#====================================================================
#  (IV) Summary of the posterior samples and estimated Bayes factor
#====================================================================
```

```
summary(result)
bayes_factor(result)
#====================================================================
#  (V) Comparison of the posterior mean, the posterior median and
#      the true time-varying spectral density (without boundary
#      modification)
#====================================================================
#type ?plot.bdp_dw_result for more info
par(mfrow = c(1,3))

plot(result, which = 1,
     zlim = range(result$tvpsd.mean,
                  result$tvpsd.median,
                  true_tvPSD$tv_psd))

plot(result, which = 2,
     zlim = range(result$tvpsd.mean,
                  result$tvpsd.median,
                  true_tvPSD$tv_psd))

plot(true_tvPSD,
     zlim = range(result$tvpsd.mean,
                  result$tvpsd.median,
                  true_tvPSD$tv_psd)),
     main = "true tv-PSD")

par(mfrow = c(1,1))
#====================================================================
# (VI) Plotting the pointwise 90 percent credible region
#      (without boundary modification)
#====================================================================
par(mfrow = c(1,3))
plot(result, which = 3,
     zlim = range(result$tvpsd.p05,
                  result$tvpsd.mean,
                  result$tvpsd.p95))
```

```
plot(result, which = 1,
     zlim = range(result$tvpsd.p05,
                  result$tvpsd.mean,
                  result$tvpsd.p95))

plot(result, which = 4,
     zlim = range(result$tvpsd.p05,
                  result$tvpsd.mean,
                  result$tvpsd.p95))
par(mfrow = c(1,1))
```